\documentclass[preprint, 3p, authoryear]{elsarticle}
\usepackage{lipsum}
\usepackage[english]{babel}
\usepackage[mathcal]{eucal}
\usepackage[utf8]{inputenc}
\usepackage[colorinlistoftodos, color=green!40, prependcaption]{todonotes}
\usepackage[pdftex,
            pdftitle={Combinatorial_plastic_slip_FCC},
            pdfauthor={Afonso D. M. Barroso}]{hyperref} 
\usepackage{amssymb}
\usepackage{amsmath}
\usepackage{xcolor}
\usepackage{microtype}
\usepackage{caption}
\usepackage{subcaption}
\usepackage{graphicx, lipsum} 
\usepackage{soul}
\usepackage{comment}
\usepackage{upgreek}
\usepackage{gensymb}
\usepackage{physics}
\usepackage{multirow, array}
\usepackage{threeparttable}
\renewcommand{\arraystretch}{1.2}

\graphicspath{ {./Figures/} }

\begin{document} 

\title{Plastic deformation as a phase transition: a combinatorial model of plastic flow in copper single crystals 
}

\author[1]{Afonso D. M. Barroso\corref{cor1}}
\ead{afonso.barroso@manchester.ac.uk}
\author[1]{Elijah Borodin}
\ead{elijah.borodin@manchester.ac.uk}
\author[1]{Andrey P. Jivkov}
\ead{andrey.jivkov@manchester.ac.uk}
\cortext[cor1]{Corresponding author}

\affiliation[1]{organization={School of Engineering, The University of Manchester},
postcode={M13 9PL}, city={Manchester}, country={UK}}

\date{\today}

\begin{abstract}
Continuum models of plasticity fail to capture the richness of microstructural evolution because the continuum is a homogeneous construction. The present study shows that an alternative way is available at the mesoscale in the form of truly discrete constructions and in the discrete exterior calculus. A pre-existing continuum mean-field model with two parameters is rewritten in the language of the latter to model the properties of a network of plastic slip events in a perfect copper single crystal under uniaxial tension. The behaviour of the system is simulated in a triangular 2D mesh in 3D space employing a Metropolis-Hastings algorithm. Phases of distinct character emerge and both first-order and second-order phase transitions are observed. The phases represent arrangements of the plastic slip network with different combinations of collinear, coplanar, non-collinear and non-coplanar active slip systems. Furthermore, some of these phases can be interpreted as representing crystallographic phenomena like activation of secondary slip systems, strain localisation and fracture or amorphisation. The first-order transitions mostly occur as functions of the applied stress, while the second-order transitions occur exclusively as functions of the mean-field coupling parameter. The former are reminiscent of transitions in other statistical-mechanical models, while the latter find parallels in experimental observations. \hfill
\break \hfill \break
\end{abstract}


\maketitle


\section{Introduction} \label{seq: Introduction}

Since Brenner's experiments in the 1950s \citep{Brenner1956, Brenner1957, Brenner1958}, metal specimens of very small dimensions have put into question commonly held assumptions about the nature of plasticity and the conditions at its onset, such as the maximum yield strength attainable by metals \citep{Hou2022}, the critical resolved shear stress and the effect of normal stresses on slip planes \citep{Tschopp2007}. Unlike bulk-sized crystals and polycrystals, these microscopic single crystals, usually in the shape of microwhiskers (single crystals that are much longer than they are wide), have almost an absence of microstructure and can even be dislocation-free to the limit of image resolution \citep{Kamada1974}. As increasing magnitudes of stress are applied to a perfect crystal, very high levels of potential strain energy are accumulated until a dislocation is nucleated, thereby providing a localised and propagating release mechanism that leads to the nucleation of more dislocations \citep{Cao2008}. Hence, there is a parallel between the nucleation of that first dislocation and the initiation of a plastic front. In fact, experimental studies have observed that many microwhiskers suffer a sudden and sharp drop in yield stress due to the initiation of a Lüders band that propagates up and down the material \citep{Charsley1960, Kamada1974}. Furthermore, the yield strength of micrometre-sized single crystals varies with size, orientation and strain rate \citep{Peng2012, Dupont2012}.

It is plausible that the punctuated behaviour of small-scale crystals reflects one or more phase transitions occurring in the microstructure. Second-order phase transitions have been associated with the percolation of ordered regions surrounding a dislocation network \citep{Dubrovskii1979}. A connection has also been made between single slip in a forest of dislocations (forest hardening) and the second-order phase transition of an Ising model \citep{Ortiz1999}. Furthermore, a second-order phase transition has been experimentally observed in the change from stage III (characterised by a linearly decreasing work-hardening coefficient) to stage IV in the plastic deformation of copper (characterised by a constant work-hardening coefficient) \citep{Schafler2005} and it has been noted that plastic slip avalanches follow scaling and universality laws that are typical of phase transitions \citep{Friedman2012, Sole2011}. The Ising model \citep{Ising1925} (and other such statistical-mechanical models) is famous for its description of phase transitions in ferromagnetism, as well as other physical applications (e.g. see \cite{Wang2024}). A parallel between electromagnetism and plasticity mechanics has been remarked upon previously by \cite{Hollander1960}.

Nevertheless, the continuum models fail to describe the strange behaviour observed at the scale limit of the very small, because the continuum itself is an inappropriate approximation. \cite{Kroner1995,Kroner2001} recognised that, since defects can only exist in contrast with a surrounding ordered structure, they cannot be explicitly described by a continuum – because a continuum is, by definition, a homogeneous structure. Continuum models carry with them the implicit assumption of the permanence of points or volume elements and of a topology stationary in time. An expression of the kind $x = x(X,t)$ assumes that the particles described by coordinates $x$ in the current configuration and by $X$ in a reference configuration are immutable, but the movement of dislocations displaces atoms inhomogeneously, which causes $X \rightarrow x$ to not be a bijective function. This produces a discontinuous displacement field, which cannot be treated appropriately in a continuum theory. Similarly, dislocations modify volume elements, so that the current configuration cannot be mapped back to the reference configuration by a bijective function. Therefore, a crystal undergoing plastic deformation has neither point permanence nor a stationary topology.

One renowned alternative is that of discrete dislocation dynamics (DDD). However, this approach has a high computational cost, in part because it attempts to mix discrete methods with continuous methods, namely in the sense that dislocations exist as geometric lines in a continuum but their stress fields are functions of continuous coordinates (cf. \cite{Gurrutxaga-Lerma2013}). Furthermore, it has been shown that the discretisation of the slip planes, rather than the dislocation lines, is a more appropriate procedure when compromising averaging errors with computational cost \citep{Roy2008, Sandfeld2013}.

Then, the question arises as to whether it is possible to have a fully discrete model of plasticity, with discrete dislocation lines and discrete planes as well as discrete stress and strain. To achieve this vision, one would have to abandon the traditional vector calculus altogether. Most prominently, the use of exterior calculus has gained traction over the past couple of decades, due to its close relationship with differential geometry and its natural partition of the continuum into the direct sum of spaces of objects with different geometric dimensions ($k$-forms) --- e.g. see \cite{Rashad2023} for non-linear elasticity and see \cite{Hochrainer2005} and \cite{Sozio2023} for dislocation plasticity. Still, the continuum assumption is present in these approaches, because the underlying mathematical space is a continuum. The crucial point is in making the stress and strain fields discrete as well -- that is, functions of discrete spaces rather than continua.

It is important to note that dislocation motion is fundamentally a two-dimensional phenomenon. Two volume elements slide over each other along a two-dimensional interface which determines the discontinuity, and the boundary of that interface is the dislocation line. This has been previously recognised in the literature \citep{Xia2015}. \cite{Wijnen2021} successfully implemented a model of plastic slip where only the slipping of material along slip planes --- and not the movement of individual dislocations --- is considered explicitly in the case of compression of a microwhisker.

One requires then a discrete mathematical framework incorporating geometry and topology and allowing for the `observer' to move from one geometrical object to its boundary. The apparatus of discrete exterior calculus (DEC) lends itself well to these requirements (e.g. see \cite{Chen2017}). It allows for the definition of inherently discrete models of physics and mechanics. By `inherently discrete' we mean a bottom-up approach that formulates the conservation laws and physical processes as functions of discrete objects, rather than a top-down approach, such as the finite element method, which discretises a differential equation to obtain a numerical solution. The inherently discrete method is related to these other discretisation methods, but is ultimately different in its approach \citep{Berbatov2023}. The argument that one should skip the integral and differential formulations and go directly from the laws of conservation to a discrete formulation has been under development since the start of the century \citep{Tonti2001, Tonti2014, Ferretti2015}, although the application of discrete mathematics, such as graph theory, is not new to Physics \citep[pp. 3--4]{Grady&Polimeni2010}.

Researchers have turned their attention to discrete formulations of mechanics over a decade ago and elasticity has been the prime target of these \citep{Dassios2018, Seruga2020, Boom2022}. However, there is an effort to extend this framework to all mechanical phenomena involving the microstructure of materials \citep{Zhu2023, Jivkov2023}. An attempt at modelling non-linear elasticity and plasticity using solely discrete objects was recently made with some success \citep{Dassios2024}. Regarding dislocation-mediated plasticity, some examples of these efforts include, but are not limited to: \cite{Ariza2005}, who used the concept of a discrete Gauss linking number to quantify dislocation entanglements in a cell complex; \cite{Lyu2017}, who formulated a multi-scale approach to plasticity on a primal-dual construction of cell complexes by using $n$-order atomistically informed constitutive equations on $n$-order geometrical objects; and \cite{Srinivasa2021}, who partitioned a material manifold into localised Riemannian manifolds to describe regions of a material enclosed by dislocations.

In the present study, we describe the kinetics of microscopic plastic slips in micrometre-sized single crystals by joining together combinatorial mathematics and statistical methods. Specifically, we make use of the works of Naǐmark, Bayandin and colleagues, wherein is developed a statistical mean-field model of microscopic slip events based on the assumption of the self-similarity (fractal character) of the probability distribution of those defects \citep{Naimark1992, Naimark1998, Naimark2016, Naimark2017, Bayandin2019}. Crucially, the model is of microscopic instances of plastic slip, not individual dislocation lines. This idea was also taken up by \cite{Kroner1990}, who included in his theory the ``elementary point stacking fault''. In Kröner's words, ``(…) as far as length measurement is concerned, it makes no essential difference whether the elementary point stacking which builds up the surface stacking fault really forms a surface fault or is distributed, e.g., at random in the volume of the crystal''. Therefore, we take it as (sufficiently) true that the random distribution of microscopic shearing events in a material volume is globally equivalent to the presence of macroscopic shears (stacking faults or, in our case, the slipped areas that end in dislocation lines). This stochastic model is best suited to describe plastic waves or any kind of cascade-like deformation \citep{Bayandin2019}, as occurs in microwhiskers \citep{Charsley1960, Kamada1974}. In the present work, this continuum model is rewritten into a discrete stochastic formulation using a combinatorial structure to describe the material and a version of discrete exterior calculus on such a structure to calculate energies associated with shears and applied stresses. As proof of concept, the method is applied to the case of micrometre-sized perfect single copper crystals under uniaxial tension. While the stochastic model proposed by Naǐmark and colleagues relies on a mean-field approach, which is known to not describe accurately short- and long-range interactions between dislocations, the use of a mean-field is nonetheless justified here (or, at least, it is less inappropriate) owing to the very small size of the specimen simulated.

Section \ref{seq: Theory} provides an overview of the continuum model that serves as a basis for the discrete model and of the theory of cell complexes and discrete exterior calculus needed to understand the proposed framework. It concludes with the formulation of the discrete framework. The computational methodology is described in Section \ref{seq: Methodology}. In Section \ref{seq: Results}, the results of multiple simulations applied to the test case of micrometre-sized perfect copper single crystals under uniaxial tension are shown and the evolution of the slip network in the material is explored. The most interesting result is the emergence of distinct phases in this network and of phase transitions, which are discussed in more depth in Section \ref{seq: Discussion}. Section \ref{seq: Conclusion} closes the article with a summary of the findings and with concluding remarks, including commentary on the limitations of the approach developed here.
\section{Theory: towards a combinatorial mean-field model of microscopic slips} \label{seq: Theory}

\subsection{Continuum mean-field model of microscopic slip events} \label{subseq: Model}


Consider a three-dimensional linear isotropic material body $\mathcal{B}$ with boundary $\partial \mathcal{B}$ and volume $V_{\mathcal{B}}$ subject to a continuous displacement field $\vec{u}(x_i) \equiv u_k(x_i), \: i \in \{1,2,3\}$. The local (point-wise) gradient of the displacement field, referred to as the local distortion, is given by $\boldsymbol{\upbeta}^{\mathrm{local}} = u_{i,k}$, and the linearised local strain is given by $\boldsymbol{\upvarepsilon}^{\mathrm{local}} = (u_{i,k} + u_{k,i})/2$, where the subscript $_{,k}$ indicates partial differentiation with respect to $x_k$. Suppose that a dislocation inside $\mathcal{B}$ glides along its slip plane and sweeps out an area $\delta A$ in the direction $\vec{b}$. If this process is accounted for in $\vec{u}$, the displacement field becomes discontinuous, hence the local distortion and local strain are not well-defined at the area element $\delta A$. By cutting the body along the slip plane into two parts, thereby extending $\partial \mathcal{B}$ and removing the problematic element $\delta A$ from $\mathcal{B}$, it can be shown that
\begin{equation}
    \frac{1}{V_{\mathcal{B}}} \int_{\partial \mathcal{B}} \vec{b} \otimes \vec{n} \: dA =
    \frac{1}{V_{\mathcal{B}}} \int_{\mathcal{B}} \boldsymbol{\upbeta}^{\mathrm{local}} \: dV,
    \label{eq: global_strain}
\end{equation}
where $\vec{n}$ is the unit normal to $\partial \mathcal{B}$. Since the right-hand side of Eq. \ref{eq: global_strain} is a measure of the global distortion experienced by the body (the average of the local distortions), it is possible to quantify the shear deformation experienced by the surface element $\delta A$ as \citep[pp. 14--17]{Kocks1975}
\begin{equation}
\boldsymbol{\upbeta}^{\mathrm{surface}} = \frac{\delta A}{V_{\mathcal{B}}} \: \vec{b} \otimes \vec{n}.
    \label{eq: surface_distortion}
\end{equation}
Naǐmark used the symmetric part of this tensor to define a ``micro-defect tensor'' for microscopic slip events \citep{Naimark1998},
\begin{equation}
    \vb{s} = \frac{s}{2}(\vec{n} \otimes \vec{\ell} + \vec{\ell} \otimes \vec{n}),
    \label{eq: microdefect_tensor}
\end{equation}
where $s\vec{\ell}$ gives the unit direction $\vec{\ell}$ and magnitude $s$ of slip and $\vec{n}$ is a unit vector defining the orientation of the slip plane --- but not the plane itself. The tensor $\vec{n} \otimes \vec{\ell} + \vec{\ell} \otimes \vec{n}$ is the generalised Schmid factor \citep{Kocks1982}. The distribution of these microscopic slips (microslips) can be assumed to possess a fractal character, which allows for the assumption of the self-similarity of their probability distribution $W(s, \vec{\ell}, \vec{n})$ \citep{Hahner1998, Chen2013}. This distribution is given by an expression of the kind $W = Z^{-1} \mathrm{exp}(-E/Q)$, where $Z$ is a partition function, $E$ is the energy density available for the creation of the defect, and $Q$ is an energetic relaxation parameter \citep{Naimark1998}. The macroscopic defect tensor $p_{ik}$ is obtained from the average of $s_{ik}$ over the ensemble,
\begin{equation}
    \vb{p} = \rho \int \vb{s} \: W(s,\vec{n},\vec{\ell}) \: d \vb{s},
    \label{eq: macrodefect_tensor}
\end{equation}
where $\rho$ is the volumetric concentration of defects in the entire material body (dimensions $\mathrm{L}^{-3}$). The driving force for the process is determined by the energy available for the creation of an individual defect,
\begin{equation}
    E = E_0 +(\alpha \vb{s} - \boldsymbol{\upsigma} - \lambda \vb{p}) : \vb{s},
    \label{eq: continuum_lagrangian}
\end{equation}
where $E_0$ is a ground state energy density, $\alpha$ and $\lambda$ are material parameters, $\boldsymbol{\upsigma}$ is the externally applied Cauchy stress tensor, and $:$ is the double contraction of tensors \citep{Naimark2017}.

The difference $E-E_0$ in Eq. \eqref{eq: continuum_lagrangian} quantifies the balance between the work done on a region of the material that is prone to slip and that region's resistance to slip. Therefore, a negative value of $E-E_0$ quantifies the (positive) energy surplus in the system upon the creation of a microslip \citep[pp. 12]{Kocks1975}. By ``surplus'', we mean the energy that is not used to produce the defect itself, which, in a more comprehensive model, would be converted into the stress field of the defect, or elastic relaxation, or converted into dissipated heat.
The term $\alpha \vb{s}:\vb{s}$ is the self-energy of the defect and stands as an energetic barrier to its formation, while the terms $\boldsymbol{\upsigma} : \vb{s}$ and $\lambda \vb{p} : \vb{s}$ quantify the work done to the region by the externally applied stress and by a mean-field of the defect ensemble, respectively. Therefore, the parameter $\lambda$ is the strength of the coupling of the mean-field to the orientation of the defect being potentially created.

\subsection{Cell complexes as the basis for modelling microscopic slips} \label{subseq: Cell Complexes}


Our aim is to rewrite the model described in the preceding section in a fully discrete way as discussed in the Introduction. Firstly, it is necessary to introduce the topological space for modelling microslips --- the cell complex. Cell complexes are the natural extension of planar graphs to dimensions higher than two; while a graph is characterised by a set of nodes and a topology-defining set of edges, a $d$-dimensional cell complex is composed of nodes, edges and higher-dimensional geometric objects, each defined as a set of constituent nodes \citep[pp. 7-35]{Kozlov2008}. Each $p$-dimensional object in a $d$-dimensional cell complex ($d \geq p \geq 0$) is called a $p$-cell; it is homeomorphic to a ball in $\mathbb{R}^p$, and is composed of at least $p+1$ nodes (0-cells). A $p$-cell composed of exactly $p+1$ nodes is called a simplex. Furthermore, the boundary of each $p$-cell with $p > 0$ is composed of the union of $(p-1)$-cells and the intersection of any two $p$-cells is either a $(p-1)$-cell on the boundary of both or the empty set \citep[pp. 38--62]{Grady&Polimeni2010}. 

The structures built on cell complexes are chains and cochains. A $p$-chain $c_p$ is a formal linear combination of $p$-cells $x_p$, i.e., $c_p=\sum \xi_p x_p$, with $\xi_p \in \mathbb{R}$. The $p$-chains form a vector space with individual $p$-cells as basis vectors, hence by abuse of notation, $x_p$ denotes a basis $p$-chain. A $p$-cochain $c^p$ is a linear functional over $p$-chains, i.e., an assignment of a value (scalar, vector, covector, tensor) to each $p$-chain. The $p$-cochains form a vector space with basis vectors $x^p$ defined by $x^p(x_p)=1$.


Analysis on cell complexes uses discrete differential forms, which are analogues of the continuous differential forms from the smooth exterior calculus \citep{Flanders1989}. One option for defining discrete $p$-forms is as functions on $p$-chains, i.e., as $p$-cochains. This is the approach taken in the discrete exterior calculus \citep{Hirani2003}. In this case, the calculus uses a second cell complex dual to the primal one with several ad hoc assumptions for the operations. A second option for defining discrete $p$-forms is as functions on relations between cells with a difference of $p$ in their dimension \citep{Forman2002}. These are referred to as combinatorial differential forms and the analysis is performed on an extended complex whose space of $p$-cochains is isomorphic to the space of combinatorial $p$-forms \citep{Berbatov2022}. Nevertheless, the development in the present work does not require a choice of option, as we directly consider $p$-cochains as an assignment of values to $p$-chains in a simplicial complex and perform only operations that do not require a dual complex as in \cite{Hirani2003}, nor an extended complex as in \cite{Berbatov2022}.

When the cell complex is embedded in a metric space by a function that assigns each 0-cell to a tuple of coordinate values, each $p$-cell $x_p$ with $p>0$ acquires a measure (length, area, volume), denoted by $\mu(x_p)$. In our investigation, we deal exclusively with 3-dimensional simplicial complexes, which are composed of nodes/vertices/points (0-cells), edges/lines (1-cells), faces/triangles (2-cells) and volumes/tetrahedra (3-cells). Our considerations do not involve all operations with discrete differential forms, only the inner product of cochains. This has been developed by \cite{Berbatov2022} following the Riemannian formalism. If $C^p$ is the space of all $p$-cochains, then, for a simplicial complex, the inner product of $p$-cochains $\langle \cdot, \cdot \rangle : C^p \times C^p \rightarrow \mathbb{R}$, is explicitly given for two $p$-cochains $\chi^p$ and $\omega^p$ by,
\begin{equation}
    \langle \chi^p , \omega^p \rangle =
    \sum_{x_p} \frac{\chi^p(x_p) \cdot \omega^p(x_p)}{(d+1)(p+1)\mu(x_p)^2}
    \sum_{n_0 \preccurlyeq x_p} \kappa(n_0)
    \sum_{v_d \succcurlyeq n_0} \mu(v_d) = \chi^p(x_p) \cdot \omega^p(x_p) \: \langle x^p , x^p \rangle,
    \label{eq: cochain_inner_product}
\end{equation}
where $d$ is the top dimension of the complex, $v_d$ is a top-dimensional cell, $x_p \succcurlyeq y_q$ encodes the property that the $p$-cell $x_p$ and the $q$-cell $y_q$ intersect, with $p \geq q$ (if $p > q$, $x_p \succ y_q$ is used), and $\kappa(n_0)$ is a weight assigned to the node $n_0$ which accounts for the curvature of the boundary of the cell complex. The first sum is taken over all nodes that are part of $x_p$ and the third sum is taken over all $d$-cells that have $n_0$ on their boudnary. For a cell complex in the shape of a parallelepiped, $\kappa(n_0)=8$ if $n_0$ is a corner of the complex, $\kappa(n_0)=4$ if $n_0$ is on an external edge of the complex, $\kappa(n_0)=2$ if $n_0$ is on an external face, and $\kappa(n_0)=1$ if $n_0$ is an interior node of the complex \citep{Berbatov2022}. The inner product is non-zero only if there is at least one basis $p$-chain where both $p$-cochains take values. It is worth noting that the inner product is a global operation which returns one single scalar value, unlike the double contraction of tensors, which is computed at each point of the continuum. Furthermore, the inner product of $p$-cochains carries with it a physical dimension of $\mathrm{L}^{d-2p}$.

While Eq. \eqref{eq: cochain_inner_product} was developed for the inner product of scalar-valued (ordinary) cochains \citep{Berbatov2022}, it is extended here to the inner product of vector- or covector-valued cochains. The vector or covector part of these cochains is expressed with respect to the coordinate of the embedding space, i.e., it is not intrinsic to the cell complex, while the cochain part is intrinsic. It is possible to formulate combinatorial vector- or covector-valued cochains independent from the embedding space, but this is not necessary for the current considerations. Therefore, the inner product of vector- or covector-valued cochains is given by Eq. \eqref{eq: cochain_inner_product}, where $\cdot$ represents the inner product of vectors or covectors.

\subsection{Combinatorial mean-field model of microscopic slip events} \label{subseq: Combinatorial_Model}

Consider a 3-dimensional simplicial cell complex embedded in Euclidean space $\mathbb{R}^3$ in which a subset of 2-cells cuts the material body along slip planes. Firstly, the microdefect tensor given by Eq. \eqref{eq: microdefect_tensor} is the symmetrised tensor product of a polar vector $s\vec{\ell}$ and an axial vector $\vec{n}$. Axial vectors are a construction of vector calculus and are so called because they represent the axis of a 2-dimensional object, specifically an oriented area. However, in the more general and complete framework of algebraic geometry, axial vectors are Hodge-dual to bivectors \citep[p. 220]{Doran2003}. The unit axial vector $\vec{n}$ is an intensive quantity and it only describes the orientation of a plane, not the plane itself. To achieve our discrete formulation, we require an extensive quantity that is internal to the cell complex \citep{Jivkov2023}. Therefore, the role played by $\vec{n}$ in Eq. \eqref{eq: microdefect_tensor} is instead played by the oriented area it represents, which is a 2-cell. Thus, the cochain $Ac^2$ is used, where $A$ is the area of the 2-cell in question, which can be labelled as $c_2$, and $c^2$ is a unit 2-cochain that assigns a value of 1 to $c_2$. Due to the way $\vb{s}$ was defined in Eq. \eqref{eq: microdefect_tensor}, candidates for $c_2$ are those 2-cells that cut the material body along slip planes. The tensor $\vb{s}$ can then be expressed as a vector-valued 2-cochain
\begin{equation}
    S = sA \vec{\ell} \: c^2.
    \label{eq: microdefect_cochain}
\end{equation}
Although $S$ is a 2-cochain, the superscript $^2$ is omitted in order to declutter the notation. Similarly for the stress 2-cochain and mean-field 2-cochain discussed below.

Due to the construction of the 2-cells as boundaries along slip planes, the similarity of Eq. \eqref{eq: microdefect_cochain} with Eq. \eqref{eq: surface_distortion} is straightforward. Hence, the self-energy of a microslip occurring on the 2-cell $c_2$ with magnitude $s$ in the direction $\vec{\ell}$ is given by $\alpha \langle S, S \rangle$ with the inner product given by Eq. \eqref{eq: cochain_inner_product}. If we represent the physical dimensions of $s$ as $[s]$, the cochain $S$ has dimensions $[s] \mathrm{L}^2$, which would make $\langle S, S \rangle$ have dimensions of $[s]^2 \mathrm{L}^3$. Since the combinatorial inner product gives a global value, $\alpha \langle S, S \rangle$ must have dimensions of energy rather than energy density (cf. \citep{Jivkov2023}). Therefore, $s$ is chosen to be dimensionless and $\alpha$ to have dimensions of stress (energy density). This is correspondent with Eq. \eqref{eq: surface_distortion}, where $s = b \cdot\delta A/V_{\mathcal{B}}$, but not with Eq. \eqref{eq: microdefect_tensor}, where $s = b \cdot \delta A$ \citep{Naimark2017}. However, whereas the volume $V_{\mathcal{B}}$ used in Eq. \eqref{eq: surface_distortion} is the volume of the whole material body, in the discrete setting we can take advantage of the inhomogeneous topology and discontinuities in the material body. Recall that it was possible to define the cochain $S$ in Eq. \eqref{eq: microdefect_cochain} by partitioning the material continuum into 3-cells whose boundaries include the slip planes.
From the perspective of a 2-cell where a slip event occurs, the `original material body' only extends as far as the closest 2-cells, since cuts have been made there too. This is the same thought process laid out at the start of Section \ref{subseq: Model} that allows one to remove $\delta A$ from the body and then considers $V_{\mathcal{B}}$ as the total volume of the body, except here the boundary formed by the cut is the 2-cell in-between two 3-cells. In this way, the material body is partitioned into volume elements of what \cite{Frank1951} called ``good crystal''. The deformation state of the material body is thus determined by the composition of all local deformations set on the 2-cells of the cell complex. Therefore, in defining $s$ in Eq. \eqref{eq: microdefect_cochain}, the volume used is the sum of the volumes of the 3-cells incident on the 2-cell where a slip event occurs. Referring to this volume simply as $V$, we obtain $s=bA/V$ in Eq. \eqref{eq: microdefect_cochain}, where $A$ is the (finite) area of the 2-cell under consideration. In this sense, $s$ is a local-global strain measure: from the perspective of the 2-cell, it is global, but from the perspective of the material body, it is local.


The Cauchy stress tensor field is best described as a covector-valued 2-form \citep{Kanso2007}. This is achieved by considering the stress field as a tensor-valued 0-cochain and `flattening' the tensor-value at each node into vector-values at all adjacent 2-cells. In essence, we obtain the traction force covector on a 2-cell by summing the stress tensors at each of its nodes and multiplying by the area vector of the 2-cell (the area vector is the unit normal scaled by the area). We obtain thus a covector-valued 2-cochain $\Sigma$ and the work done by the externally applied stress field is computed by the inner product $\langle \Sigma, S \rangle$.


The mean-field tensor $\vb{p}$ cannot be represented in the cell complex the same way, for it would require a long and computationally expensive process of converting quantities from the discrete to the continuum and back to the discrete. Instead, suppose that $n-1$ microslip events have aleady occurred, each characterised by the microslip 2-cochain $S_i$ defined on 2-cell $x_{2,i}$, with $i \in \{1, ..., n-1\}$. Then, the effect of this ensemble on the creation of an $n$-th microslip occurring on 2-cell $x_{2,n}$ in the direction $\vec{\ell}$ is given by
\begin{equation}
    \vec{p} = \frac{1}{N} \sum_{i=1}^{n-1} S_i(c_{p,i})
    = \frac{n-1}{N} \sum_{i=1}^{n-1} \left( S_i(c_{p,i}) W(s, \gamma) \right),
    \label{eq: macrodefect_cochain}
\end{equation}
where $N$ is the total number of microslip events that can occur and $W(s,\gamma)$ is the probability of the slip event described by the vector $S_i(c_{p,i})$ occurring with magnitude $s$ on the slip system labelled by $\gamma$ -- compare this with the continuum probability distribution $W(s, \vec{n}, \vec{\ell})$.

Then, a mean-field vector-valued 2-cochain $P$ is constructed on the 2-cell $x_{2,n}$ where the new microslip might form, $P = \vec{p} \: x^2_n$ (where $x^2_n$ is the unit cochain on $x_{2,n}$). Unlike $\vb{p}$ and $\vb{s}$ in Eq. \eqref{eq: macrodefect_tensor}, where $s=b \cdot \delta A$ because $\vb{p}$ is dimensionless, $P$ has the same physical dimensions as $S$ (namely, $\mathrm{L}^2$). The value of the norm $|\vec{p}|$ corresponds to the net shear strain in the whole material multiplied by the net area of the oriented slip interfaces (slip 2-cells). A measure of the global/net plastic strain can be obtained by multiplying $|\vec{p}|$ by the area of the cross-section between a plane perpendicular to $\vec{p}$ and the material body. Noteworthy as well is the observation that $\vec{p}$ is perpendicular to the plane where there is a net reduction in area. The work done by the discrete microdefect ensemble to create the $n$-th microslip is given by $\lambda \langle P, S_n \rangle$.

Thus, the total energy available for the creation of a new defect is given generally by
\begin{equation}
    \mathcal{H} = \mathcal{H}_0 + \langle \alpha S - \Sigma - \lambda P , \:S \rangle.
    \label{eq: discrete_lagrangian}
\end{equation}
As with Eq. \eqref{eq: continuum_lagrangian}, the difference $\mathcal{H}-\mathcal{H}_0$ quantifies the difference between the energy needed to create a new microslip and the mechanical energy input into the system by the externally applied stress and the mean-field. As before, we call the negative of this difference the energy surplus. Additionally, we henceforth set $\mathcal{H}_0=0$.

The discrete model described is, in a general sense, a classical mean-field Potts model with 49 tensor-valued `spins' on a 2-dimensional triangular lattice in 3-dimensional space, with some restrictions on which sites can have which `spin' values (the 49 values correspond to 48 possible combinations of positive and negative plane slip normals and positive and negative slip directions, plus the zero state) \citep{Potts1952, Wu1982}. The `spin' values are characterised by a directional local-global strain at volume element interfaces, the externally applied field is the applied mechanical stress, and the self-energy term is included to account for some of the characteristic non-equilibrium kinetics of plastic phenomena. Nevertheless, our model differs substantially from the Potts model of magnetism, because of the particular use of the inner product \eqref{eq: cochain_inner_product} and the special topology we have formulated.

\subsection{Parameter calibration} \label{subseq: Calibration}

It is necessary to calibrate the model parameters $\alpha$ and $\lambda$ present in Eq. \eqref{eq: discrete_lagrangian}. Determining $\alpha$ is straightforward, since $\alpha \langle S, S \rangle$ is the energy barrier that must be overcome by $\langle \Sigma + \lambda P, S \rangle$ at yield. In this context, $\alpha$ accounts both for the creation of a defect and its movement/glide because we consider the material to be free of defects at the start of the simulation. At the onset of plastic behaviour, $P \equiv 0$ and one can estimate
\begin{equation}
    \alpha = \frac{\langle \Sigma, S \rangle}{\langle S, S \rangle} = \frac{(\boldsymbol{\upsigma} \: \vec{n}) \cdot \vec{b} \: V}{\vec{b} \cdot \vec{b} \: A},
    \label{eq: alpha}
\end{equation}
when the magnitude of $\boldsymbol{\upsigma}$ is comparable to the assumed yield strength of the material. The second equality in Eq. \eqref{eq: alpha} is due to $S$ taking a value on one and only one 2-cell, where $A$, $V$ and $\vec{n}$ are the area of the 2-cell, the sum of the volumes of the 3-cells incident on it and the unit normal of the 2-cell, respectively. The value of $\alpha$ will be different for different slip systems due to the directionality of $\vec{b}$ and $\sigma_{ik} n_k$, but also for different 2-cells with different areas and different volumes of the incident 3-cells. This creates a hierarchy of energy barriers that makes it intrinsically easier for slip to occur on some slip systems on some 2-cells. In a previous study, we investigated how the applied stress tensor was `deconstructed', via Eq. \eqref{eq: alpha}, into a self-energy level for each slip system --- some triaxiality states revealed up to 12 different energy levels \citep{Borodin2024}. This splitting of the self-energies is consequential for the activation of different slip systems at different moments of the deformation process, which would impact the trajectory of the deformation state in phase-space.

The other parameter, the mean-field coupling strength $\lambda$, can be estimated by a mechanics-based reasoning as well. The very definition of plastic deformation posits that at the end of the deformation process, once the applied stress is removed, the system should not relax back to its initial state or deform further.
So, assuming the system has reached an equilibrium, whatever mean-field is left at the end of the simulation must be strong enough to sustain itself, that is, it must be strong enough to equal the self-energy of microslips:
\begin{equation}
    \lambda = \frac{\alpha \: \langle S, S \rangle}{\langle P, S \rangle} = \frac{\alpha \: bA^2}{\vec{p} \cdot \vec{\ell} \: V}.
    \label{eq: lambda}
\end{equation}
Although $\lambda$ depends on the orientation $\vec{\ell}$ of the prospective slip, this is merely a reflection of the fact that $\langle P, S \rangle$ is itself a function of this orientation. If the mean-field vector $\vec{p}$ is perpendicular to $\vec{\ell}$, then $\lambda$ explodes to infinity, but in this case $\langle P, S \rangle$ would be null and so $\lambda$ would not even be present in Eq. \eqref{eq: discrete_lagrangian}.

\section{Computational methodology} \label{seq: Methodology}


%
\begin{figure}[t]
     \centering
     \begin{subfigure}{0.3\textwidth}
         \centering
         \includegraphics[height=0.18\textheight]{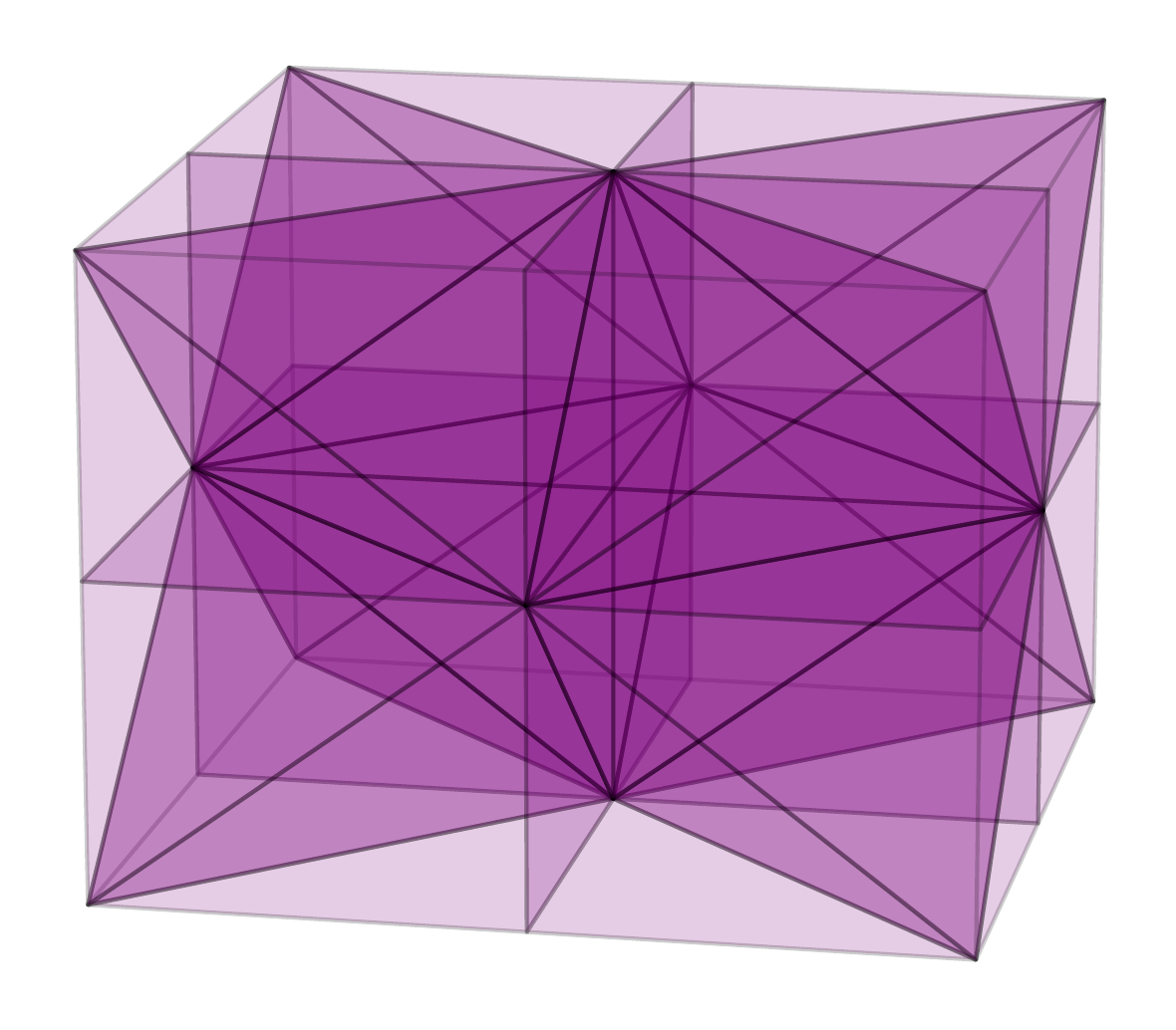}
         \caption{}
         \label{subfig: FCC_nodes}
     \end{subfigure}
     \hfill
     \begin{subfigure}{0.3\textwidth}
         \centering
         \includegraphics[height=0.18\textheight]{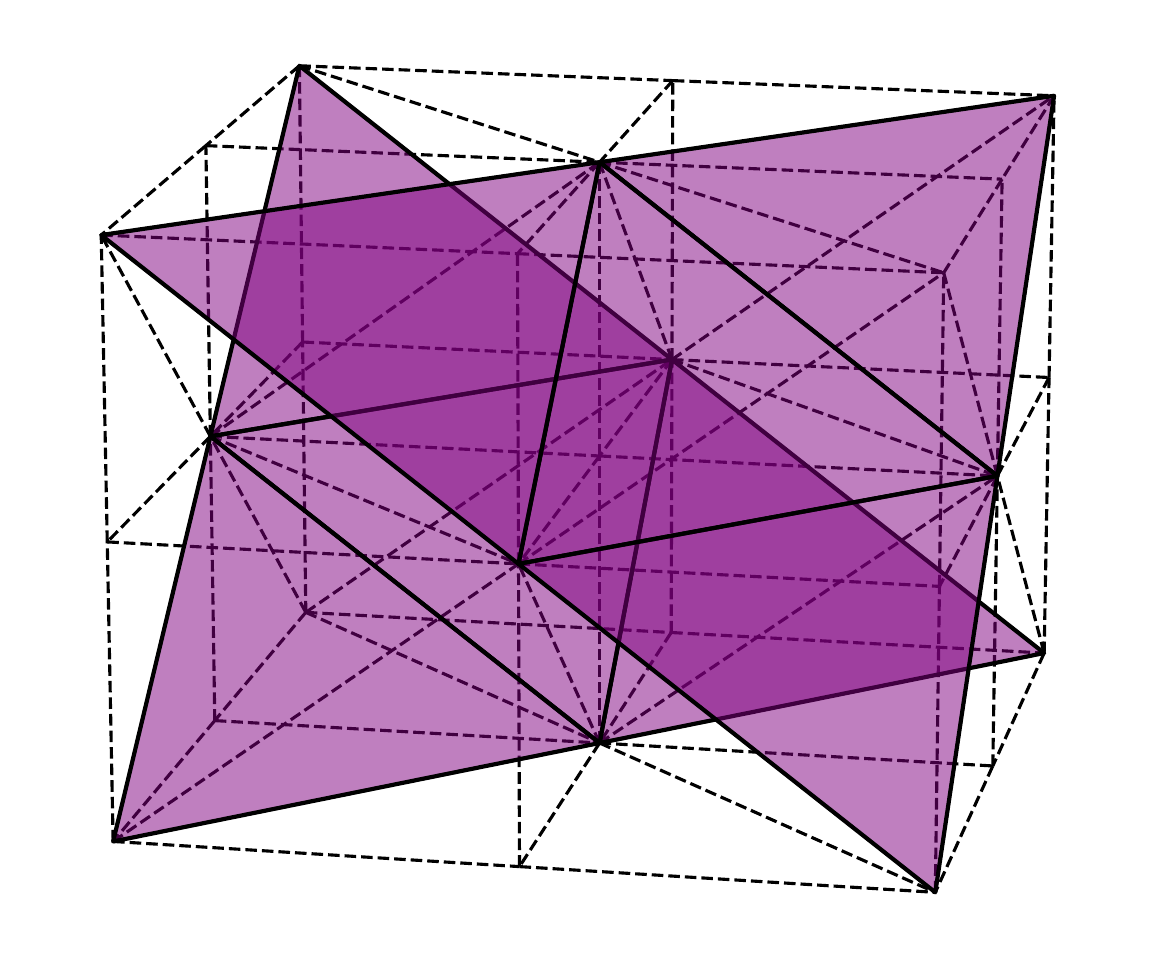}
         \caption{}
         \label{subfig: FCC_slipplanes}
     \end{subfigure}
     \hfill
     \begin{subfigure}{0.3\textwidth}
         \centering
         \includegraphics[height=0.18\textheight]{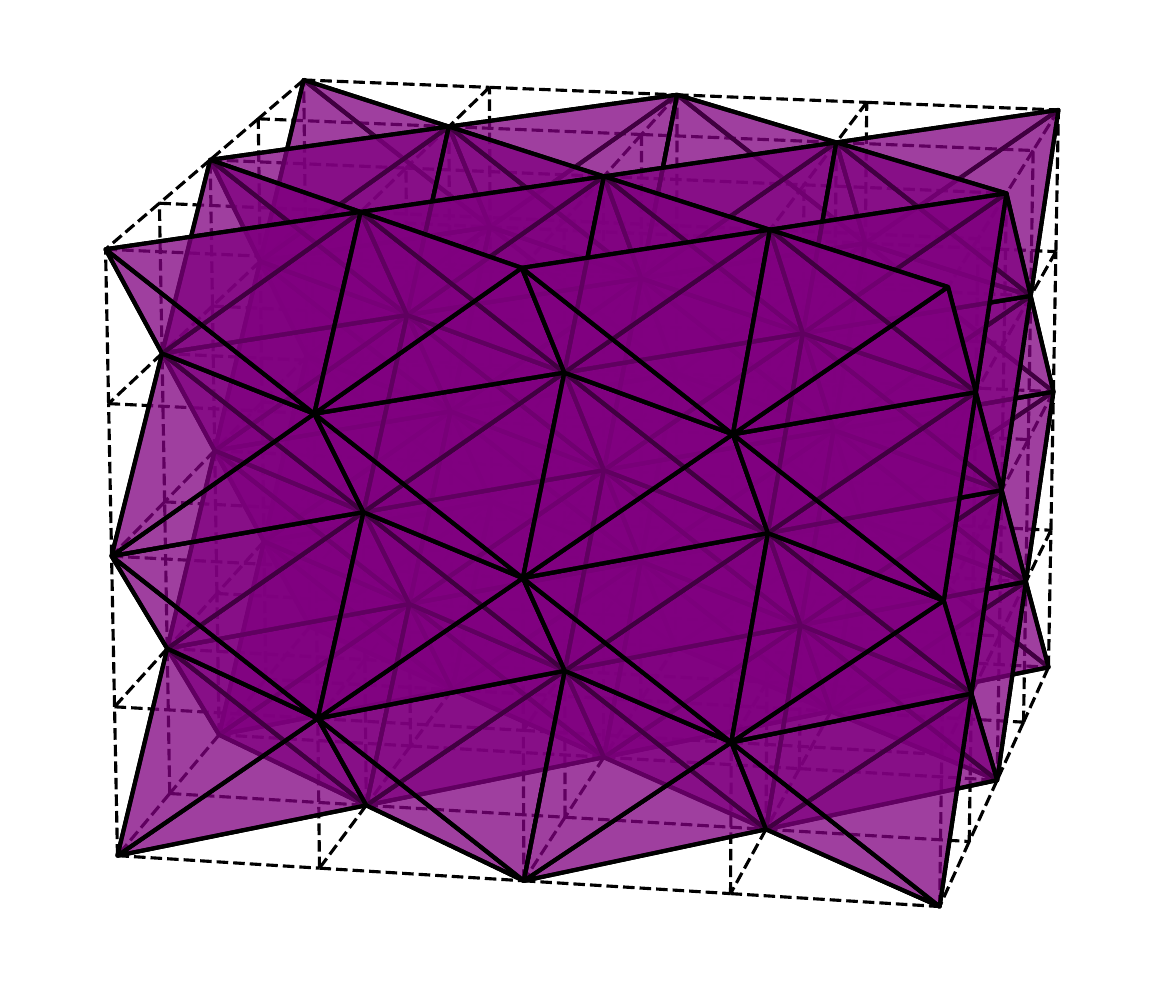}
         \caption{}
         \label{subfig: FCC_polytopes}
     \end{subfigure}
        \caption{
(a) The computational unit cell for a face-centred cubic (FCC) structure; (b) The $(111)$ slip planes in the computational unit cell, each composed of four 2-cells; (c) A $2 \times 2 \times 2$ construction for the FCC structure, with all slip 2-cells highlighted.
}
        \label{fig: FCC_complex}
\end{figure}
In order to simulate the principal slip systems in face-centred cubic (FCC) metals, a 3-dimensional cell complex was built with the Python programming language where a subset of 2-cells and 1-cells replicate the 12 \{111\}$\langle 101 \rangle$ slip systems. This code and the code for what follows are freely available on GitHub \citep{Plastic-Slip-Simulator}. The cell complex is built with computational unit cells (UCs) resembling FCC crystallographic UCs with additional body-centred  and edge-centred vertices necessary to divide the cube into tetrahedra and thus create a simplicial complex \citep[pp. 7-35]{Kozlov2008}. These additional vertices also function to create a homogeneous distribution of 3-cells, both in terms of size and connectivity, which is important to replicate the assumption that the material is homogeneous isotropic. The full simulation space is constructed by layering the computational UCs like bricks. Examples are shown in Fig. \ref{fig: FCC_complex}. This construction results in each computational UC having 27 0-cells, 90 1-cells, 104 2-cells and 40 3-cells. Of course, when computational UCs are placed next to each other and topologically glued, the 0-, 1- and 2-cells on the boundaries will be shared between two or more UCs. Nevertheless, it is worth noting that the 2-cells corresponding to slip planes (of which there are 32) lie inside the UC and therefore do not get shared. In the length scale limit where the nodes on the vertices of the computational UC and on the centres of the faces of the computational UC, as shown in Fig. \ref{subfig: FCC_nodes}, correspond exactly to atomic positions, the length of the edges of the computational UC will be the corresponding lattice parameter(s) and all real crystallographic slip planes will be represented by the 2-cells. For larger length scales, each 2-cell corresponding to a slip plane represents a structure encompassing several real crystallographic slip planes, e.g. by averaging. The value of the slip magnitude $s$ (in Eqs. \eqref{eq: microdefect_tensor} and \eqref{eq: microdefect_cochain}) can be adjusted accordingly. 

This construction has its limitations, namely that, for simulations of macroscopic specimens, the size of the complex quickly becomes unwieldy unless one is willing to compromise with the size of the representative 2-cells. At these scales, it is possible to consider plastic slip caused not by dislocations but by superdislocations, as explored by \cite{Zbib1998} and \cite{Hussein2017}, or to consider averaging procedures over slip planes as explored by \cite{Monavari2013}, \cite{Sandfeld2013} and \cite{Sandfeld2015}; however, we will not develop this further in the present study. Another limitation of this construction is that only the principal slip systems \{111\}$\langle 101 \rangle$ are modelled. A more intricately designed cell complex might be able to reproduce other slip systems that are more rarely activated. Nevertheless, it is worth noting that while the cell complex imposes restrictions on the plasticity process --- some of them desired and some of them due to approximations as mentioned above --- it does not by any means dictate the plasticity process itself. These restrictions are purely of a topological and, when the complex is embedded in a metric space, geometrical nature.


The plasticity process is simulated by employing a Metropolis-Hastings algorithm \citep{Metropolis1953, Hastings1970} to find an equilibrium state of the network of microslips assuming that the defect distribution function $W \propto \mathrm{exp}(-\mathcal{H}/Q)$ follows a Boltzmann distribution. Taking the slipping motion to be of the order of the Burgers vector $\vec{b}$, the relaxation parameter is estimated to be of the order $Q = k_B T$, where $k_B$ is Boltzmann's constant and $T$ is the (constant and homogeneous) temperature of the system.

At each step of the algorithm, a 2-cell $f_2$ corresponding to a slip plane and one of its edges $e_1 \preccurlyeq f_2$ are selected at random. The pairs ($f_2$, $\: \pm e_1$) define two possible events -- slips in the direction of the positive and negative orientation of the edge $e_1$. As explained after Eq. \eqref{eq: microdefect_tensor} and in the lead-up to Eq. \eqref{eq: microdefect_cochain}, the 2-cell $f_2$ determines the plane and position of the microslip event, and the oriented 1-cell $\: \pm e_1$ determines the slip direction by a unit vector parallel to it. If the pair ($f_2$, $\: \pm e_1$) has not been visited in any previous step, then an event with arbitrary sign is generated. The microdefect 2-cochain $S$ is obtained according to Eq. \eqref{eq: microdefect_cochain} by setting $s = |\vec{b}|A/V$, where $A$ is the area of the 2-cell and $V$ is the sum of the volumes of the 3-cells incident on $f_2$. If the pair ($f_2$, $\: e_1$) has been visited in a previous step with event ($f_2$, $\: \pm e_1$), then an attempt is made to create the reverse event, i.e. ($f_2$, $\: \mp e_1$). The effect of the mean-field is computed according to Eq. \eqref{eq: macrodefect_cochain}. The stress 2-cochain is computed \textit{a priori} from the loading context of the simulated specimen by computing the traction forces on each 2-cell, as explained in Section \ref{subseq: Model}. Hence, the energy $\mathcal{H}$ available for this particular microslip event is obtained from Eq. \eqref{eq: discrete_lagrangian} 
. According to the Metropolis-Hastings algorithm, if the change in surplus energy $\Delta \mathcal{H}$ caused by the change in the system (slip or unslip event) is negative, then the event is kept and the algorithm moves on to a new microslip chosen at random. On the other hand, if $\Delta \mathcal{H} \geq 0$, then a (uniformly) random number $p_{\mathrm{acc}} \in [0, 1]$ is generated. If $\mathrm{exp}(-\Delta \mathcal{H}/Q) \geq p_{\mathrm{acc}}$, the new event (slip or unslip) is kept; if $\mathrm{exp}(-\Delta \mathcal{H}/Q) < p_{\mathrm{acc}}$, then the new event (slip or unslip) is discarded. In either case, the algorithm moves on to a new random event pair. The employment of the Metropolis-Hastings algorithm is pertinent in this investigation because it removes the need to compute explicitly the partition function $Z$ by satisfying identically the reversibility condition of probability distributions \citep{Chib1995}. The algorithm converges towards an equilibrium state which is a local minimum of the surplus energy $\mathcal{H}$.

All simulations were carried out on a cube-shaped cell complex of volume 1 $\upmu \mathrm{m}^3$ with 10 computational UCs in each Cartesian direction, giving 40,000 3-cells and 82,400 2-cells of which 32,000 corresponded to slip planes. This results in a total of 96,000 possible slip events --- one for each possible Burgers vector direction on a slip plane 2-cell. The external edges of the cell complex were aligned with the global Cartesian $x$, $y$ and $z$ coordinates (the `lab frame'), which makes the simulated crystal oriented along $\langle 001 \rangle$. A uniaxial stress was applied in the $z$ direction uniformly throughout the cell complex. The temperature $T$ was set to $293 \: \mathrm{K}$. To model copper, the Burgers vector magnitude $|\vec{b}|$ was taken to be $2.556 \: \mathrm{\AA}$ \citep{Simon1992}.
\section{Simulation results} \label{seq: Results}

\subsection{The self-energy of microslips} \label{subseq: Results_alpha}

\begin{table}[t!]
\caption{Schmid's and Boas's notation for the most common slip systems in a face-centred cubic crystal \citep{Schmid1950}}
\centering
    \begin{tabular}{| m{1.4cm} | m{0.9cm} m{0.9cm} m{0.9cm} m{0.9cm} | m{0.9cm} m{0.9cm} m{0.9cm} m{0.9cm} m{0.9cm} m{0.9cm}|  }
    \hline
    & \multicolumn{4}{|c|}{\textbf{Slip planes}} & \multicolumn{6}{|c|}{\textbf{Slip directions}}\\
    \hline
    \textbf{Label} & A & B & C & D & 1 & 2 & 3 & 4 & 5 & 6 \\
    \hline
    \textbf{Miller indices} & $(\bar{1}11)$ & $(111)$ & $(\bar{1} \bar{1} 1)$ & $(1\bar{1}1)$ & $[011]$ & $[0\bar{1}1]$ & $[101]$ & $[\bar{1}01]$ & $[\bar{1}10]$ & $[110]$ \\
    \hline
    \end{tabular}
\label{table: Schmid&Boas}
\end{table}
\begin{figure}[t!]
    \centering
    \begin{subfigure}{0.7\textwidth}
        \centering
        \includegraphics[width=\textwidth]{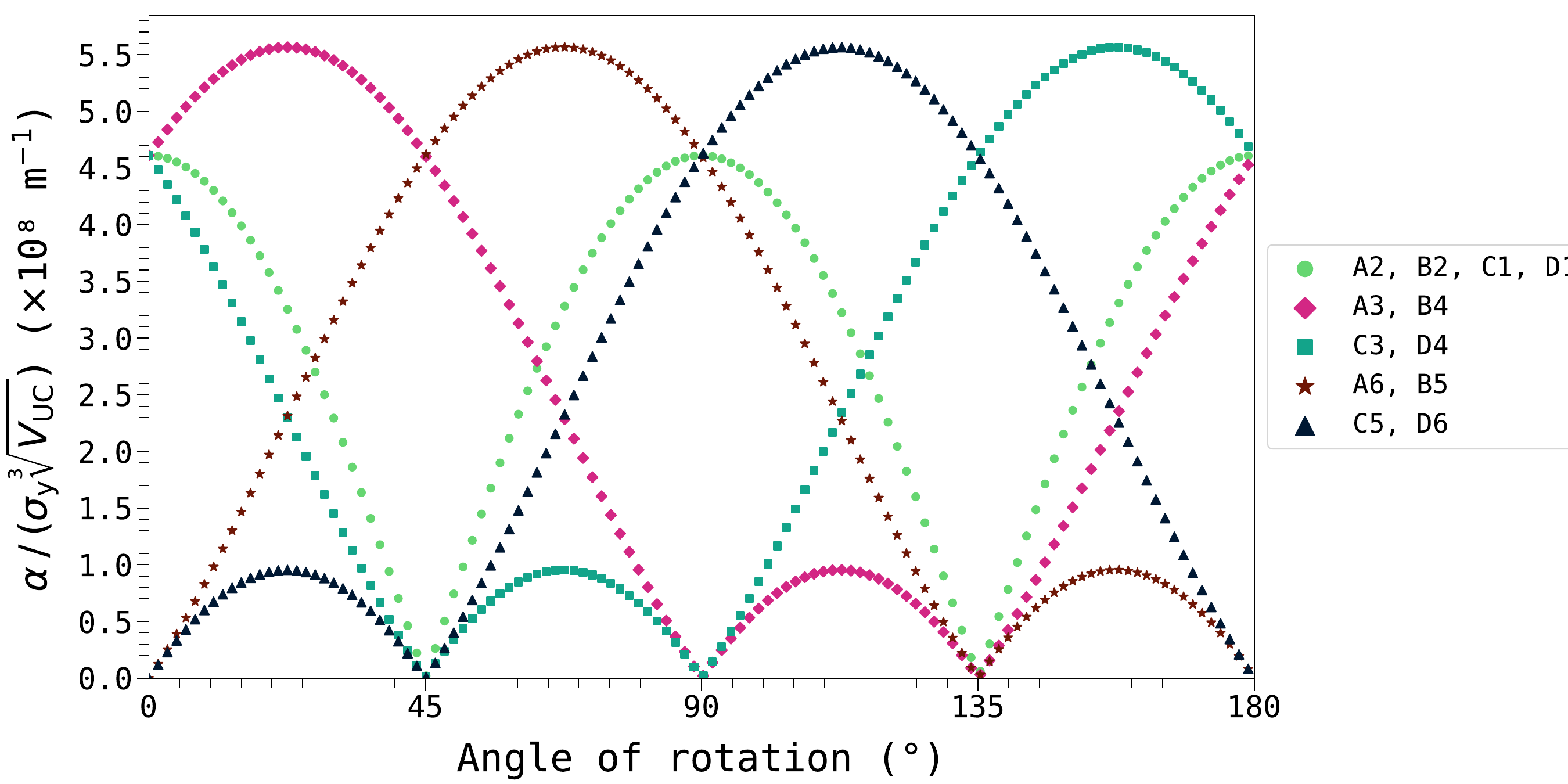}
    \end{subfigure}
    \hfill
    \begin{subfigure}{0.29\textwidth}
        \centering
        \includegraphics[width=0.8\textwidth]{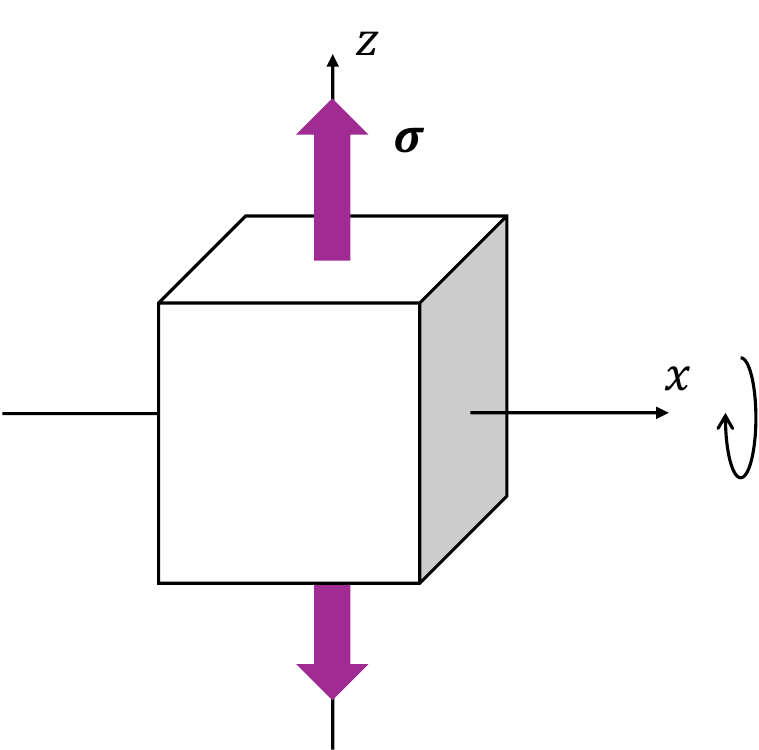}
    \end{subfigure}
    \caption{
Values of the self-energy $\alpha$ for different slip systems as the stress tensor of uniaxial tension in the $z$ direction is rotated about the $x$ axis, as shown on the right. The values were obtained from Eq. \eqref{eq: alpha}. The values are scaled by the magnitude of the applied stress at yield $\sigma_y$ and the cubic root of the volume of the computational UC. The slip systems are labelled using Schmid's and Boas's notation, shown in Table \ref{table: Schmid&Boas}.}
    \label{fig: alpha}
\end{figure}

By inspection, we note that all 2-cells corresponding to slip planes have the same area of approximately $0.2165 \: V_{\mathrm{UC}}^{\: 2/3}$ and the same sum of the volumes of their incident 3-cells, namely $V_{\mathrm{UC}}/16$, where $V_\mathrm{UC}$ is the volume of one computational UC. From Eq. \eqref{eq: alpha} one can then expect the energy barriers $\alpha$ for each slip system to depend only on the orientation of the applied stress. Fig. \ref{fig: alpha} shows the positive values of $\alpha$ for the case of uniaxial tension in the $z$ direction rotated about the $x$ axis (according to Eq. \eqref{eq: alpha}, a negative value of $\alpha$ indicates that the traction $\boldsymbol{\upsigma} \: \vec{n}$ acts in the opposite direction to the tendentious shear $\vec{b}$ and therefore cannot produce the slip event). When the applied tension is in the $z$ direction ($0 \degree$ of rotation), there are two values for the self-energy $\alpha$ of different slip systems. Scaled by the imposed yield stress magnitude and the edge length of each computational UC, these are, approximately, $4.61 \times 10^8 \: \mathrm{m}^{-1}$ and $0 \: \mathrm{m}^{-1}$. Note that this value of 0 m$^{-1}$ does not mean that slip on the A6, B5, C5 and D6 systems is energetically free , but rather that the applied stress is orthogonal to the tendentious slip direction $\vec{b}$ and therefore these slip systems only experience normal stresses (we use Schmid's and Boas's notation for the slip systems in FCC --- see Table \ref{table: Schmid&Boas} \citep{Schmid1950}). As the stress is rotated about the $z$ axis, we observe five classes of periodic curves emerging. For example, the slip systems A3 and B4 behave according to a class of curves that takes the value 0 only at $90 \degree$ and $135 \degree$ within the domain shown, $[0 \degree, 180 \degree]$. These classes of curves reflect the symmetry properties of each slip system relative to the applied stress and are reminiscent of the behaviour of the conventional Schmid factor under rotation.
\subsection{Individual runs of the Metropolis-Hastings algorithm} \label{subseq: Results_individualsimulations}

As stated at the end of Section \ref{seq: Methodology}, we conducted simulations on a domain of cubic shape with a volume of $1 \: \upmu \mathrm{m}^3$ divided into $10 \times 10 \times 10$ computational UCs. Let us henceforth assume that the material is pure copper and, being perfectly free of defects (including dislocation sources), yields at an applied uniaxial tension of $\sigma_y = 10.09$ GPa along $\langle 001 \rangle$ \citep{Liu2009}. Although this value is never observed experimentally, it is pertinent to the case study at hand, since it pertains to pure perfect single crystals. If one takes $\alpha$ from Fig. \ref{fig: alpha} and writes $\lambda$ as a multiple of $\alpha$, then Eq. \ref{eq: discrete_lagrangian} is linear in $\sigma_y$. If the assumed yield stress is lower, then slip events that are less energetically favourable will occur more often, reflecting an easier plastic flow.

\begin{figure}[!p]
     \centering
     \begin{subfigure}{0.329\textwidth}
         \centering
         \includegraphics[width=\textwidth]{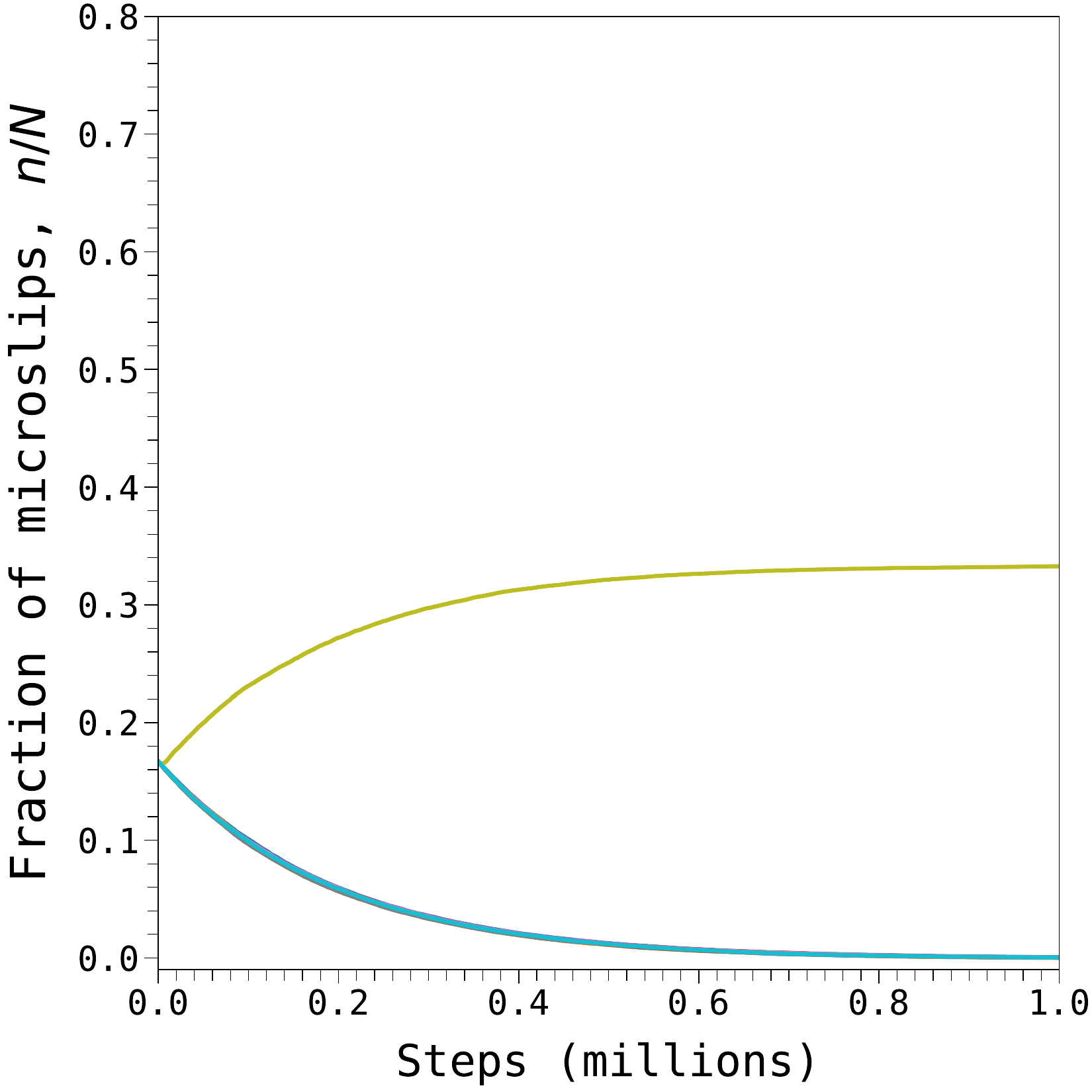}
         \caption{at $0.99 \: \sigma_y$ (9.99 GPa)}
         \label{subfig: low_fractions}
     \end{subfigure}
     \hfill
     \begin{subfigure}{0.329\textwidth}
         \centering
         \includegraphics[width=\textwidth]{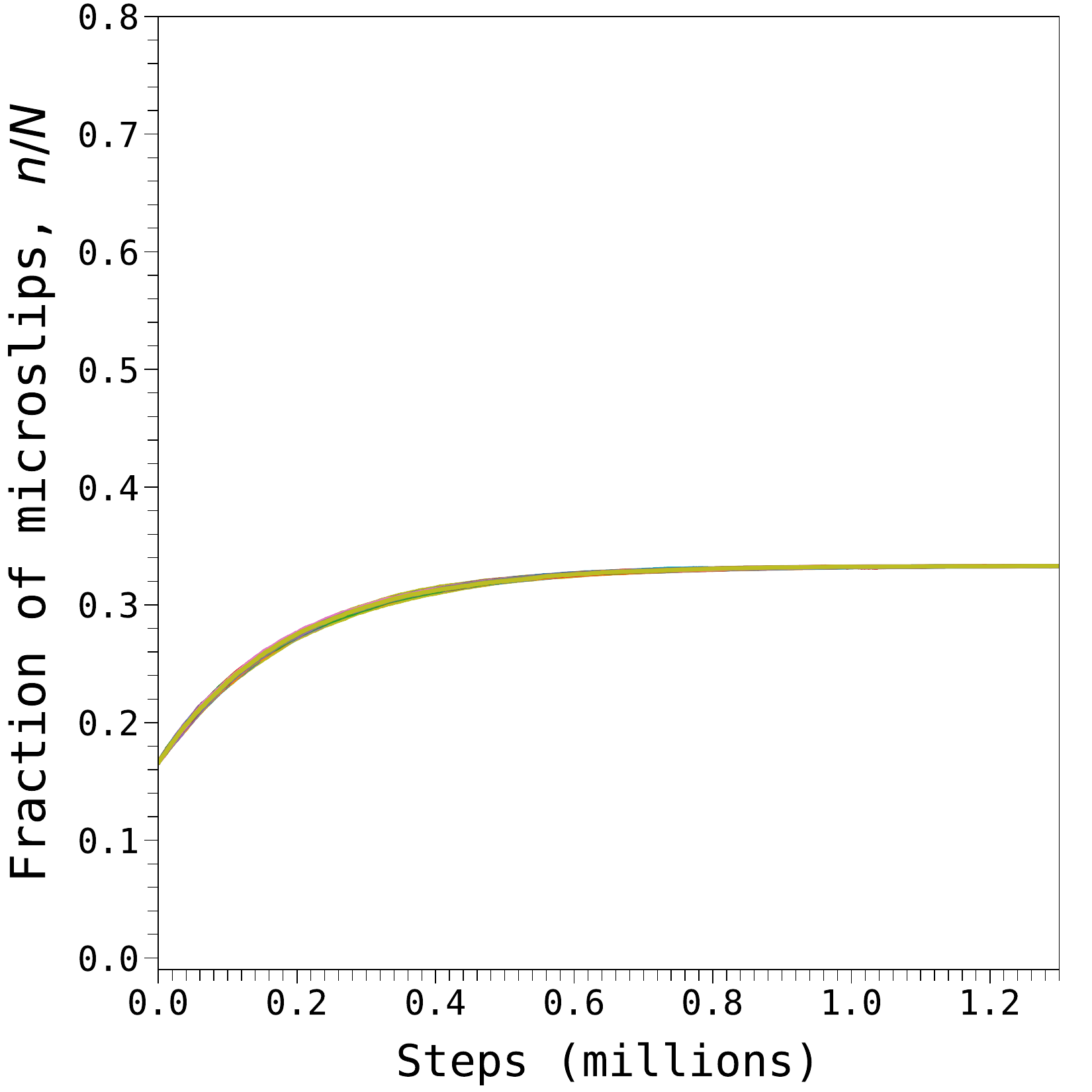}
         \caption{at $\sigma_y$ (10.09 GPa)}
         \label{subfig: med_fractions}
     \end{subfigure}
     \hfill
     \begin{subfigure}{0.329\textwidth}
         \centering
         \includegraphics[width=\textwidth]{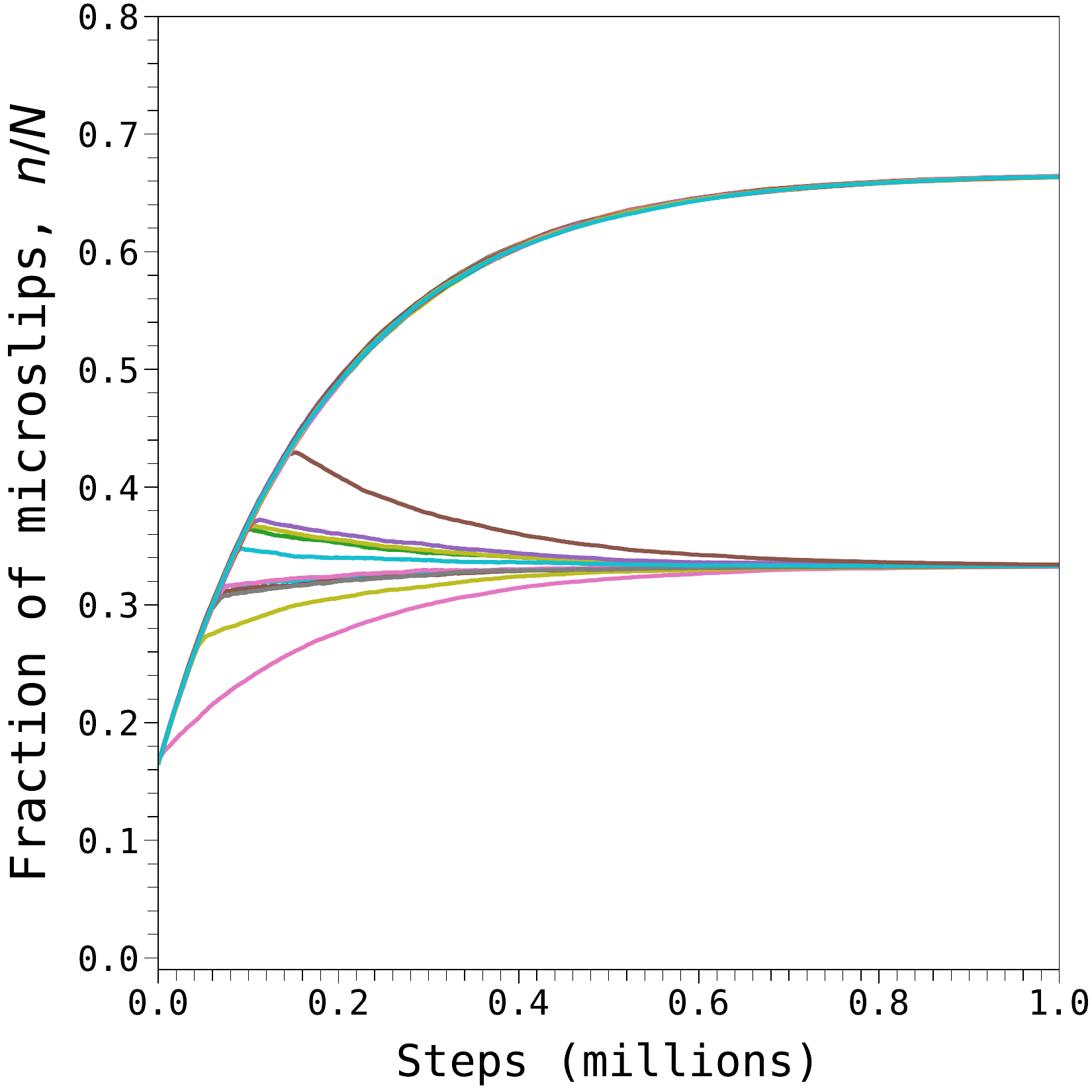}
         \caption{at $1.01 \: \sigma_y$ (10.20 GPa)}
         \label{subfig: hig_fractions}
     \end{subfigure}
     \vfill
     \centering
     \begin{subfigure}{0.329\textwidth}
         \centering
         \includegraphics[width=\textwidth]{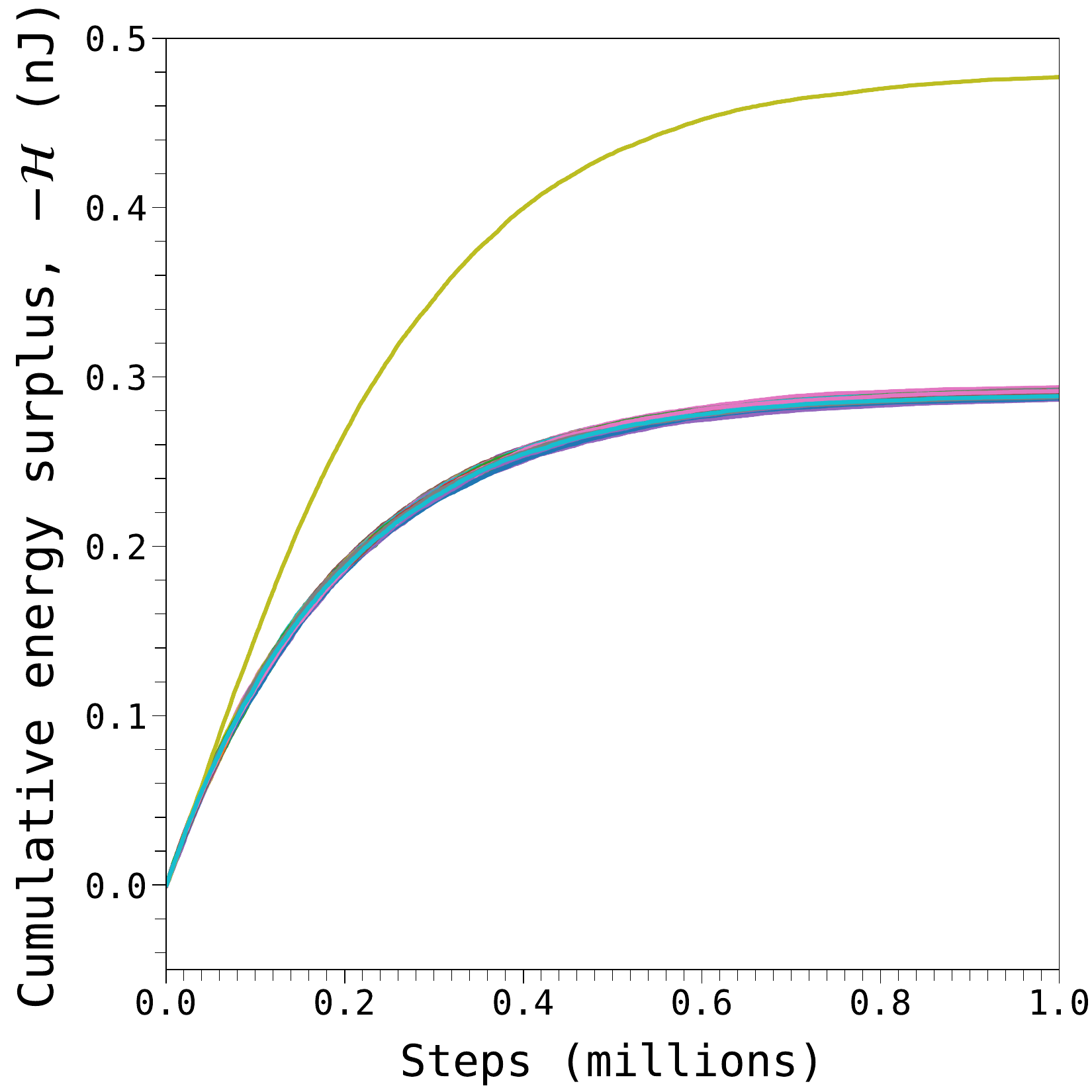}
         \caption{at $0.99 \: \sigma_y$ (9.99 GPa)}
         \label{subfig: low_energies}
     \end{subfigure}
     \hfill
     \begin{subfigure}{0.329\textwidth}
         \centering
         \includegraphics[width=\textwidth]{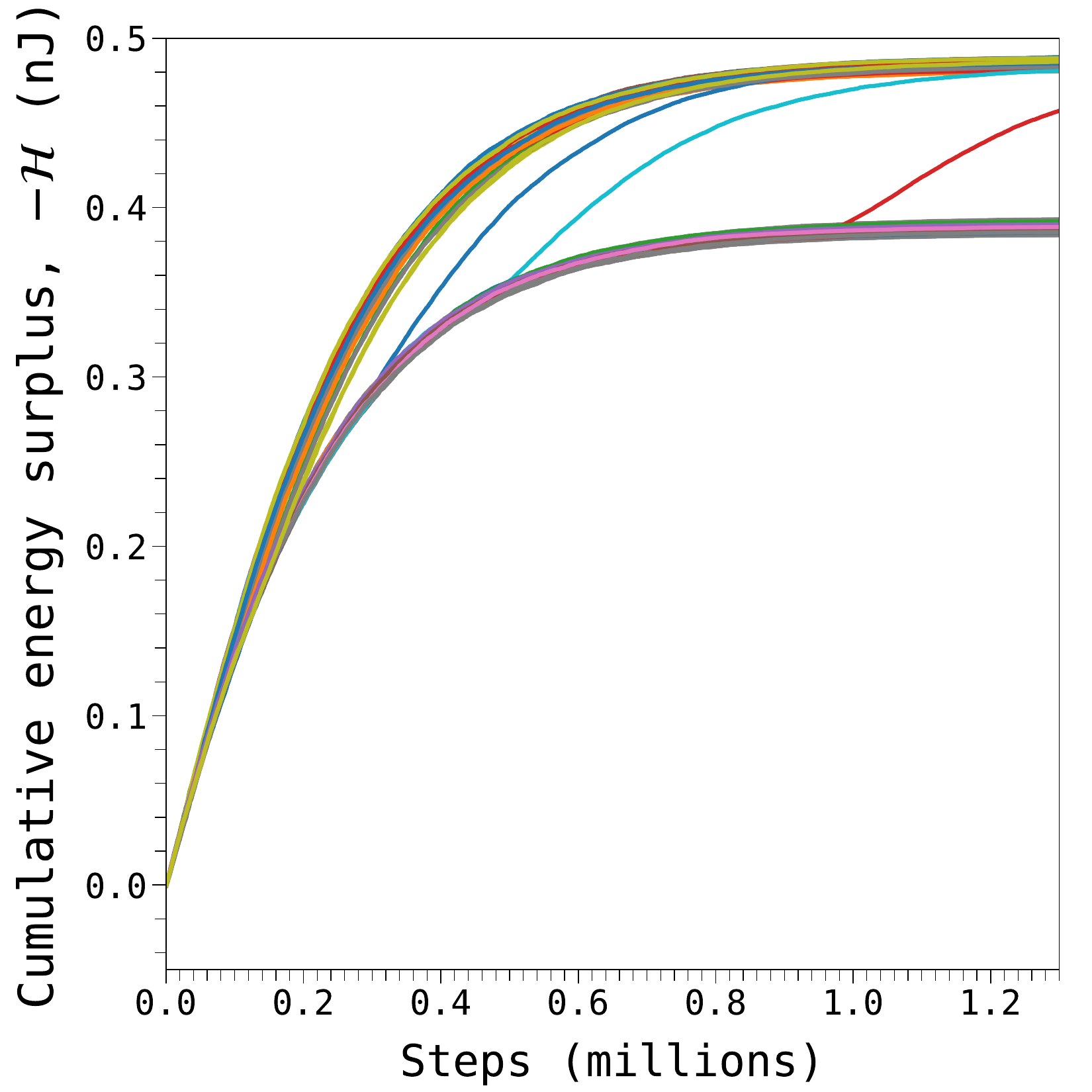}
         \caption{at $\sigma_y$ (10.09 GPa)}
         \label{subfig: med_energies}
     \end{subfigure}
     \hfill
     \begin{subfigure}{0.329\textwidth}
         \centering
         \includegraphics[width=\textwidth]{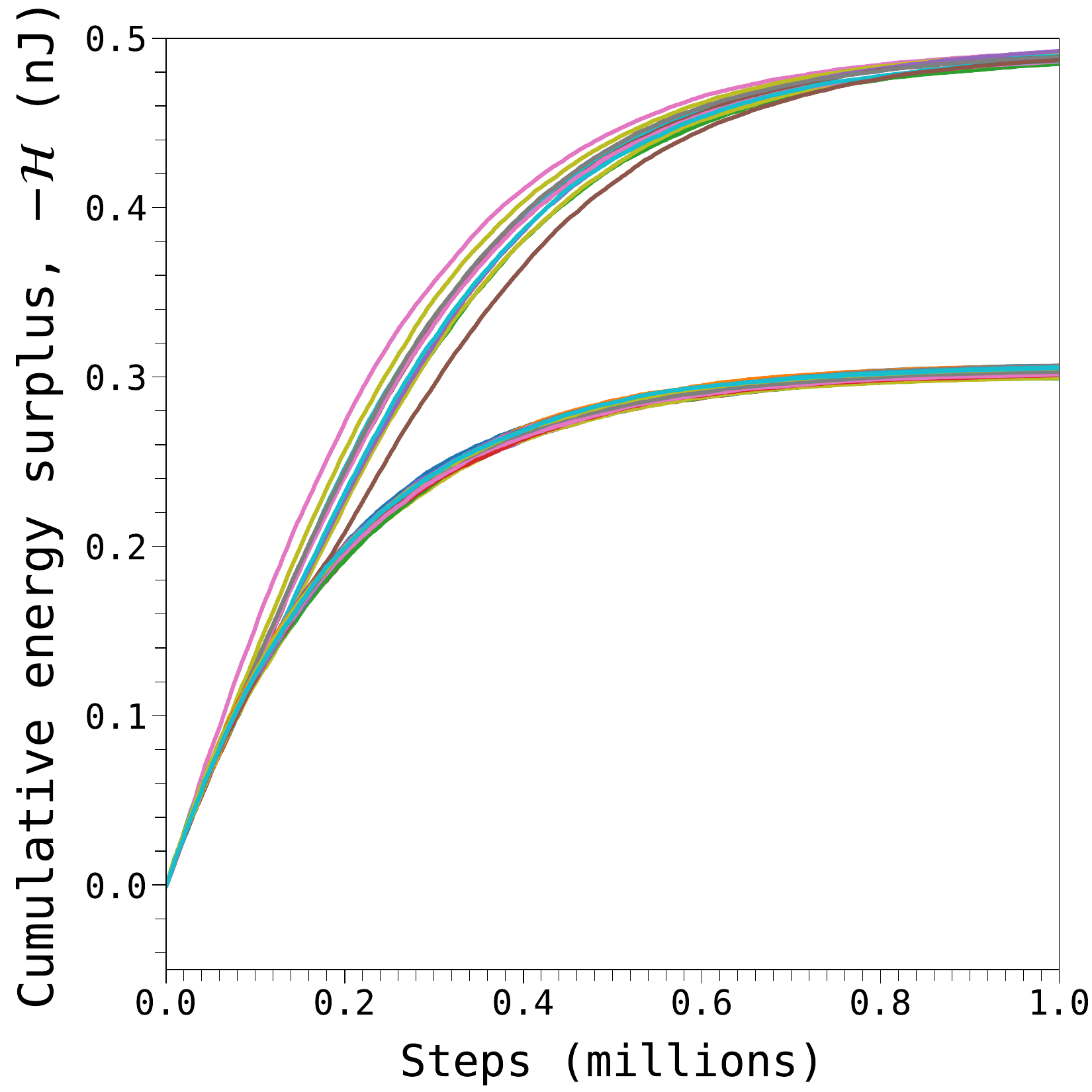}
         \caption{at $1.01 \: \sigma_y$ (10.20 GPa)}
         \label{subfig: hig_energies}
     \end{subfigure}
     \vfill
     \centering
     \begin{subfigure}{0.329\textwidth}
         \centering
         \includegraphics[width=\textwidth]{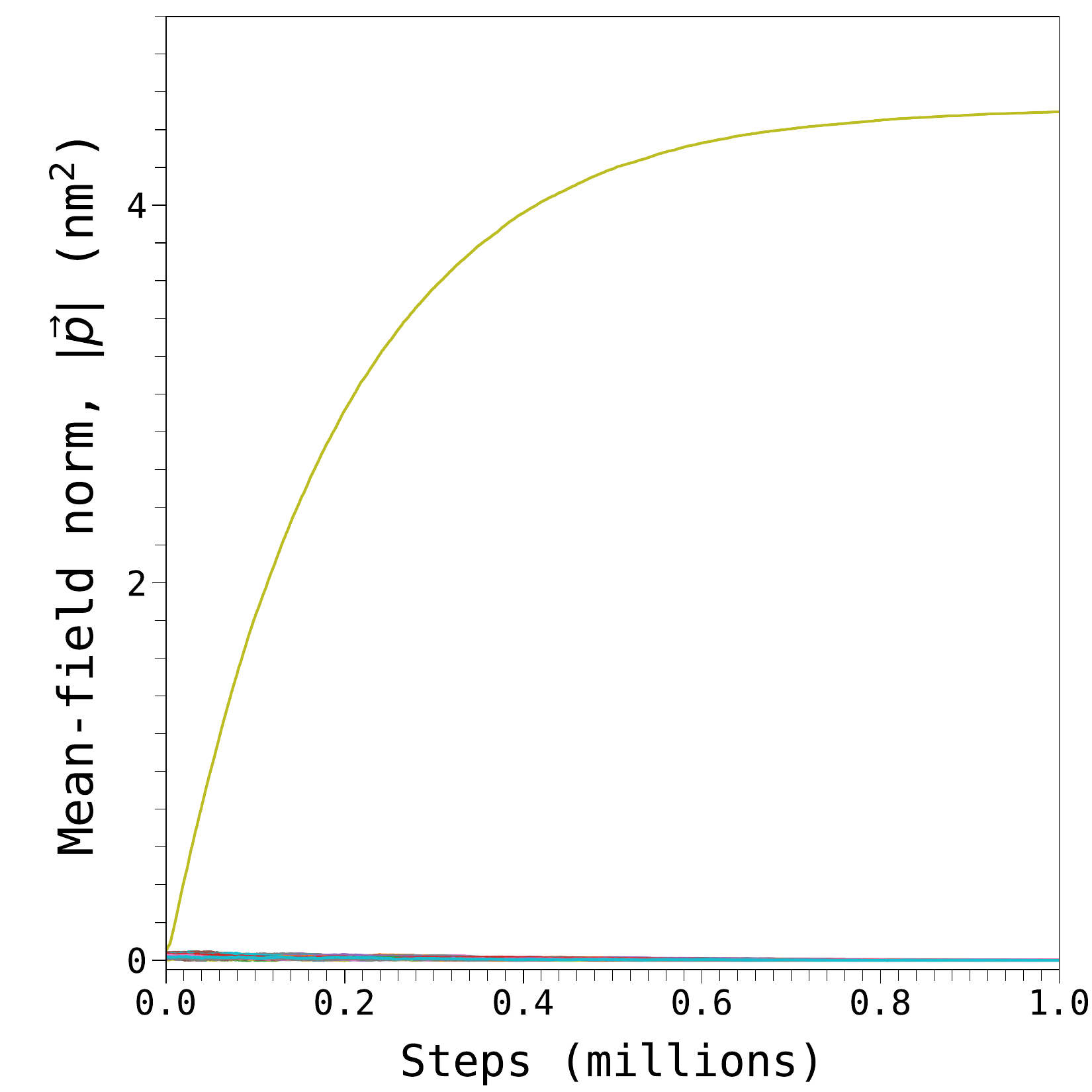}
         \caption{at $0.99 \: \sigma_y$ (9.99 GPa)}
         \label{subfig: low_meanfields}
     \end{subfigure}
     \hfill
     \begin{subfigure}{0.329\textwidth}
         \centering
         \includegraphics[width=\textwidth]{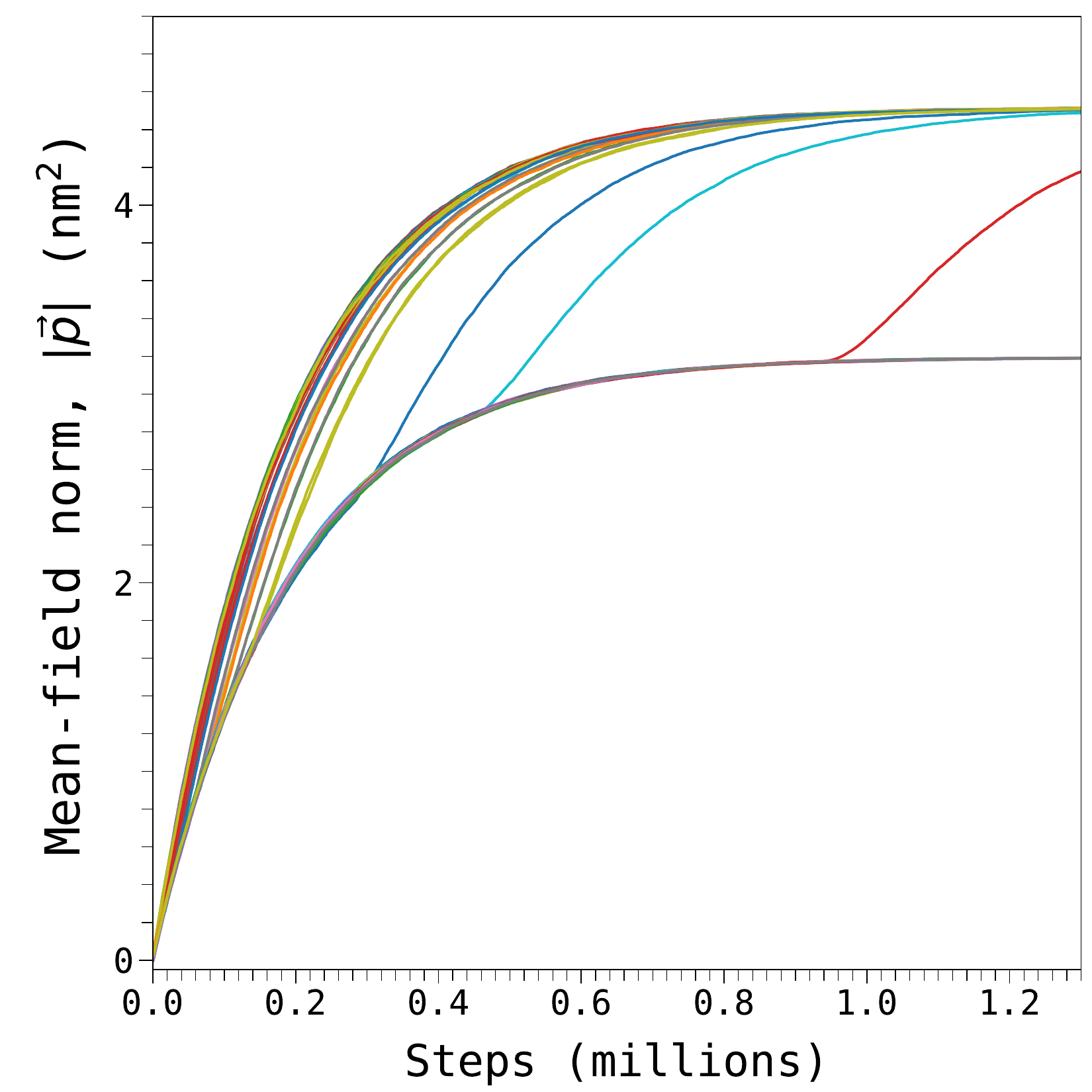}
         \caption{at $\sigma_y$ (10.09 GPa)}
         \label{subfig: med_meanfields}
     \end{subfigure}
     \hfill
     \begin{subfigure}{0.329\textwidth}
         \centering
         \includegraphics[width=\textwidth]{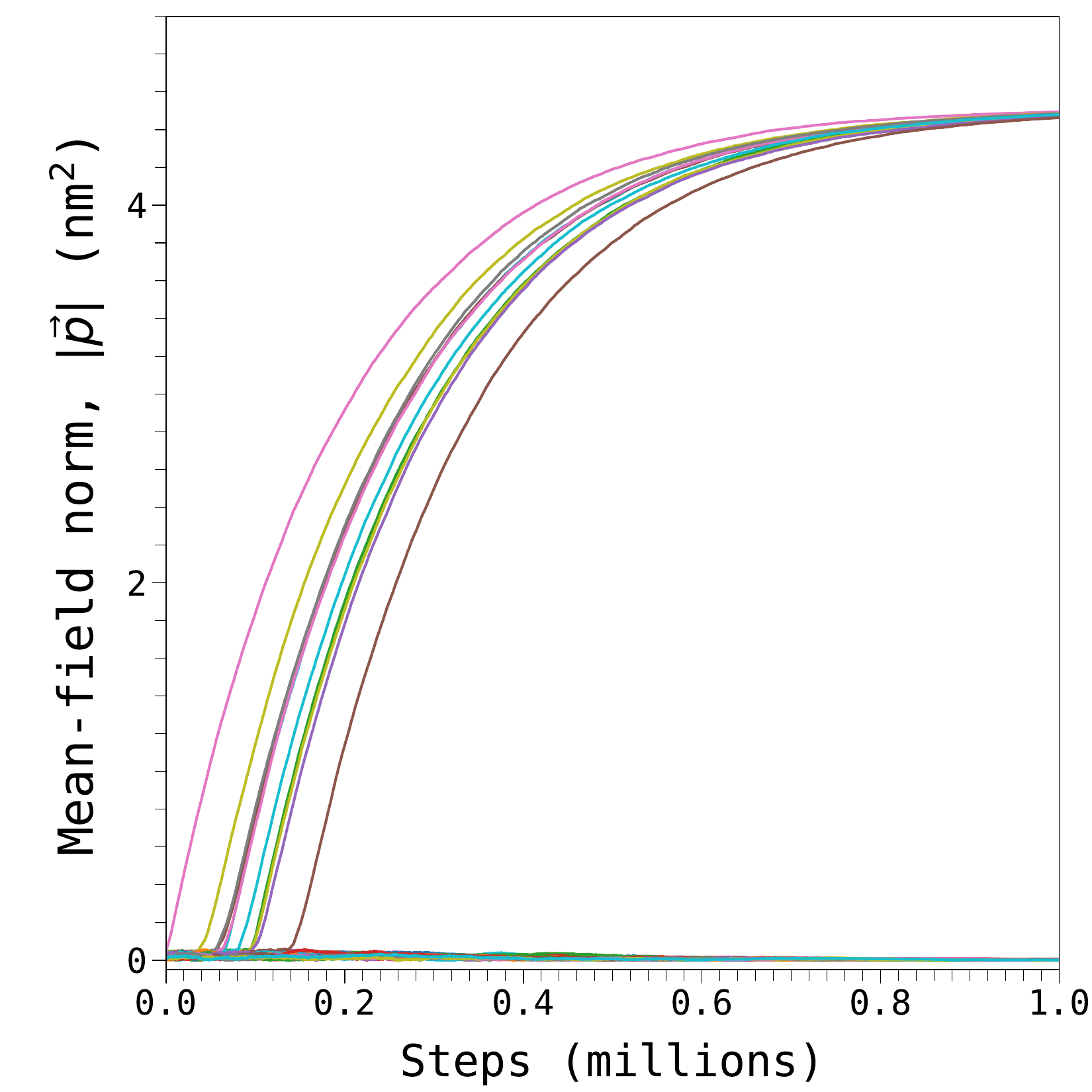}
         \caption{at $1.01 \: \sigma_y$ (10.20 GPa)}
         \label{subfig: hig_meanfields}
     \end{subfigure}
        \caption{
One hundred single runs of the Metropolis-Hastings algorithm at different magnitudes of the externally applied stress. (a)--(c) Fraction of microslip events. (d)--(f) Energy surplus, i.e. the cumulative $-\mathcal{H}$ \eqref{eq: discrete_lagrangian}. (g)--(i) Norm of the mean-field vector $\vec{p}$ as in Eq. \eqref{eq: macrodefect_cochain} at each step of the algorithm. The self-energy parameter $\alpha$ was fixed at 465.15 GPa and the mean-field coupling strength was set to $\lambda =4\: \alpha$.
}
        \label{fig: individual_simulations}
\end{figure}

Using Fig. \ref{fig: alpha}, we set $\alpha = 4.61 \times 10^8 \: \mathrm{m}^{-1} \times \sigma_y \times 10^{-6} \: \mathrm{m} \:/ \:10 = 465.15$ GPa. Fig. \ref{fig: individual_simulations} shows the results of the output of the Metropolis-Hastings algorithm for 100 runs for each of three values of the applied stress: $0.99 \: \sigma_y$ (9.99 GPa), $\sigma_y$ (10.09 GPa) and $1.01 \: \sigma_y$ (10.20 GPa), with $\lambda =4 \: \alpha$. Since we do not know what $\vec{p}$ is, Eq. \eqref{eq: lambda} cannot be used yet. The value of $4\: \alpha$ was chosen through trial and error as representative of the general behaviour of the system. Since the Metropolis-Hastings algorithm does not simulate a process but merely finds an optimal state (that of maximum cumulative energy surplus, $-\mathcal{H}$), we set the initial state of the system to have a uniformly random distribution of slips, assigning one slip event to half of all slip plane 2-cells. The averages and standard deviations of the values attained at the end of the runs are summarised in Table \ref{table: individual_simulation_results}.

The state of ``lowest energy'' (from the perspective of the Metropolis-Hastings algorithm) is the state of highest energy surplus, because the positive energy surplus is the negative of $\mathcal{H}$ (Eq. \eqref{eq: discrete_lagrangian}). Henceforth, setting $\mathcal{H}_0=0$, we call the cumulative $-\mathcal{H}$ the ``energy surplus''. Figs. \ref{subfig: low_fractions}--\ref{subfig: hig_energies} suggest that there are four (meta)stable states in the evolution of the system, which behave as attractors of the system's trajectory with strengths that are functions of the externally applied stress. Curiously, two of the states are degenerate in $n/N$, as all 100 runs in Fig. \ref{subfig: med_fractions} behave the same but there are two distinct energy states in Fig. \ref{subfig: med_energies}. Figs. \ref{subfig: low_meanfields}--\ref{subfig: hig_meanfields} show the norm of the mean-field vector $\vec{p}$ defined in Eq. \eqref{eq: macrodefect_cochain} as a function of the simulation step. For a relatively low applied stress ($0.99 \: \sigma_y$, Fig. \ref{subfig: low_meanfields}), the tension exerted on the material is not strong enough to produce plastic slip -- the fraction $n/N$ goes to zero -- and so the mean-field norm decays to zero. For a relatively large applied stress ($1.01 \: \sigma_y$, Fig. \ref{subfig: hig_meanfields}), the tension exerted is so strong that every possible microslip is produced -- the fraction $n/N$ reaches the allowed maximum of 0.66 -- and so the mean-field norm also decays to zero because the contributions from all microslips produce a spherical effect and cancel each other out.

\begin{table}
\centering
\begin{threeparttable}[t]
\caption{Averages and sample standard deviations of the fraction of microslips, cumulative energy surplus and mean-field norm attained at the end of the simulations in Fig. \ref{fig: individual_simulations}\tnote{a}.}
\begin{tabular}{| m{4.3cm} | m{3.2cm} | m{3.2cm} |  m{3.2cm} |}
    \hline
    \textbf{Quantity} & \textbf{At 0.99} $\sigma_y$ & \textbf{At} $\sigma_y$ & \textbf{At 1.01} $\sigma_y$ \\
    \hline
    Fr. of microslips, $n/N$ & $0.0009 \pm 0.0001$ or $0.3326$\tnote{1} & $0.333$\tnote{2} & $0.333$ or $0.664$\tnote{2} \\
    \hline
    Cum. energy sur., $-\mathcal{H}$ (nJ) & $0.290 \pm 0.002$ or $0.477$\tnote{1} & $0.388 \pm 0.002$ or $0.485 \pm 0.004$ & $0.303 \pm 0.002$ or $0.488 \pm 0.002$ \\
    \hline
    Mean-field norm, $|\vec{p}|$ (nm$^2$) & $0.0017 \pm 0.0007$ or $4.50$\tnote{1} & $3.19$ or $4.51$\tnote{2} & $0.0038 \pm 0.0016$ or $4.48 \pm 0.01$ \\
    \hline
\end{tabular}
\begin{tablenotes}
    \footnotesize
    \item [a] The simulation seen in Figs. \ref{subfig: med_energies} and \ref{subfig: med_meanfields} in red which breaks off from the lower branch at around step number 880,000 was considered an outlier and removed from consideration, since it did not finish converging by the end of the run. The simulation seen in Figs. \ref{subfig: low_fractions}, \ref{subfig: low_energies} and \ref{subfig: low_meanfields} in green, although unique in its behaviour in those figures, was not considered an outlier due to later observations: see Fig. \ref{fig: stress_range} and Table \ref{table: phases}.
    \item [1] Given without sample standard deviation because the sample is of one value only.
    \item [2] Given without sample standard deviation because it was at least four orders of magnitude lower than the average.
\end{tablenotes}
\label{table: individual_simulation_results}
\end{threeparttable}
\end{table}

\begin{figure}[t]
     \centering
     \includegraphics[width=0.6\textwidth]{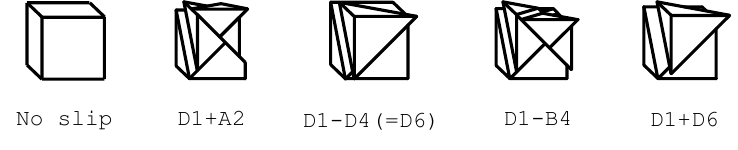}
\caption{
The material simulated in Fig. \ref{fig: individual_simulations} could have different deformation states which are characterised in part by the mean-field vector $\vec{p}$, as noted in the text after Eq. \eqref{eq: macrodefect_cochain}. From left to right: $\vec{p}=(0,0,0)$, $\vec{p} \propto (0,0,1)$, $\vec{p} \propto (1,1,0)$, $\vec{p} \propto (1,1,0)$, and $\vec{p} \propto (1,2,1)$. Diagrams are not up to scale.
}
    \label{fig: necking_shearing}
\end{figure}

It is at yield $\sigma_y$ that the mean-field takes on a predominant role in shaping the behaviour of the system. The applied stress is strong enough to initiate some microslip events, but each subsequent event is conditioned by the orientation of the growing mean-field. At the yield point, the final state with the lowest energy surplus also had the lowest mean-field norm (Fig. \ref{subfig: med_meanfields}) as well as a $z$-component of the mean-field close to zero. The $x$- and $y$-components were identical in absolute value. (This is identified as phase S(1) in Table \ref{table: phases} in Section \ref{subseq: Results_stressranges}.) The end state with the highest energy surplus had the largest mean-field norm and a maximal $z$-component of the mean-field, with vanishing $x$- and $y$-components. (This is identified as phase N(1) in Table \ref{table: phases} in Section \ref{subseq: Results_stressranges}.) Because the mean-field vector is the average of all the $s\vec{\ell}$ vectors in Eq. \eqref{eq: microdefect_cochain}, it indicates the average direction in which slip occurred. Whether the mean-field vector $\vec{p}$ points in one or the other direction depends on the concerted motion of the induced slips. By drawing a diagram of possible situations, as is done in Fig. \ref{fig: necking_shearing}, one sees that different orientations of $\vec{p}$ indicate different geometrical `deformation states', although, due to the averaging feature of employing a mean-field, it is impossible to tell exactly how the simulated specimen would look like.

To understand the origin of the degeneracy observed at yield, we investigated the distribution of the microslip events on the slip systems. All 100 runs had comparable distributions in terms of the slip systems A2, A3, B2, B4, C1, C3, D1 and D4 (see Table \ref{table: Schmid&Boas}), with all systems being equally activated by the end of each run. However, if we consider the positive and negative slip directions on each slip plane as different slip systems, labelling them as A2, -A2, A3, -A3, ..., D4 and -D4, then we see that the runs where the system evolution ends at a higher energy surplus activated all slip systems of the same sign equally, while the runs where the system achieved a lower energy surplus activated four positive slip systems and four negative slip systems equally (systems sharing the same letter and number, e.g. A2 and -A2, were never activated in the same run). This difference is in turn reflected on the mean-field, leading to different `deformation states'. Some possibilities are shown in Fig. \ref{fig: necking_shearing}.

Finally, both states with $n/N \approx 0.33$ had the highest energy surplus, which suggested that, under the assumptions made for these simulations, a lack of plastic slip ($n/N \approx 0$) and a saturation with slips ($n/N \approx 0.66$) were both unfavourable configurations for a plastically deforming microscopic single crystal. This is not surprising, since a slip is the physical manifestation of the release of accumulated elastic strain energy, but too many moving dislocations in a body can induce very high interaction energies.

\subsection{Phase transitions as a function of the applied stress} \label{subseq: Results_stressranges}

\begin{figure}[!p]
     \centering
     \begin{subfigure}{0.329\textwidth}
         \centering
         \includegraphics[width=\textwidth]{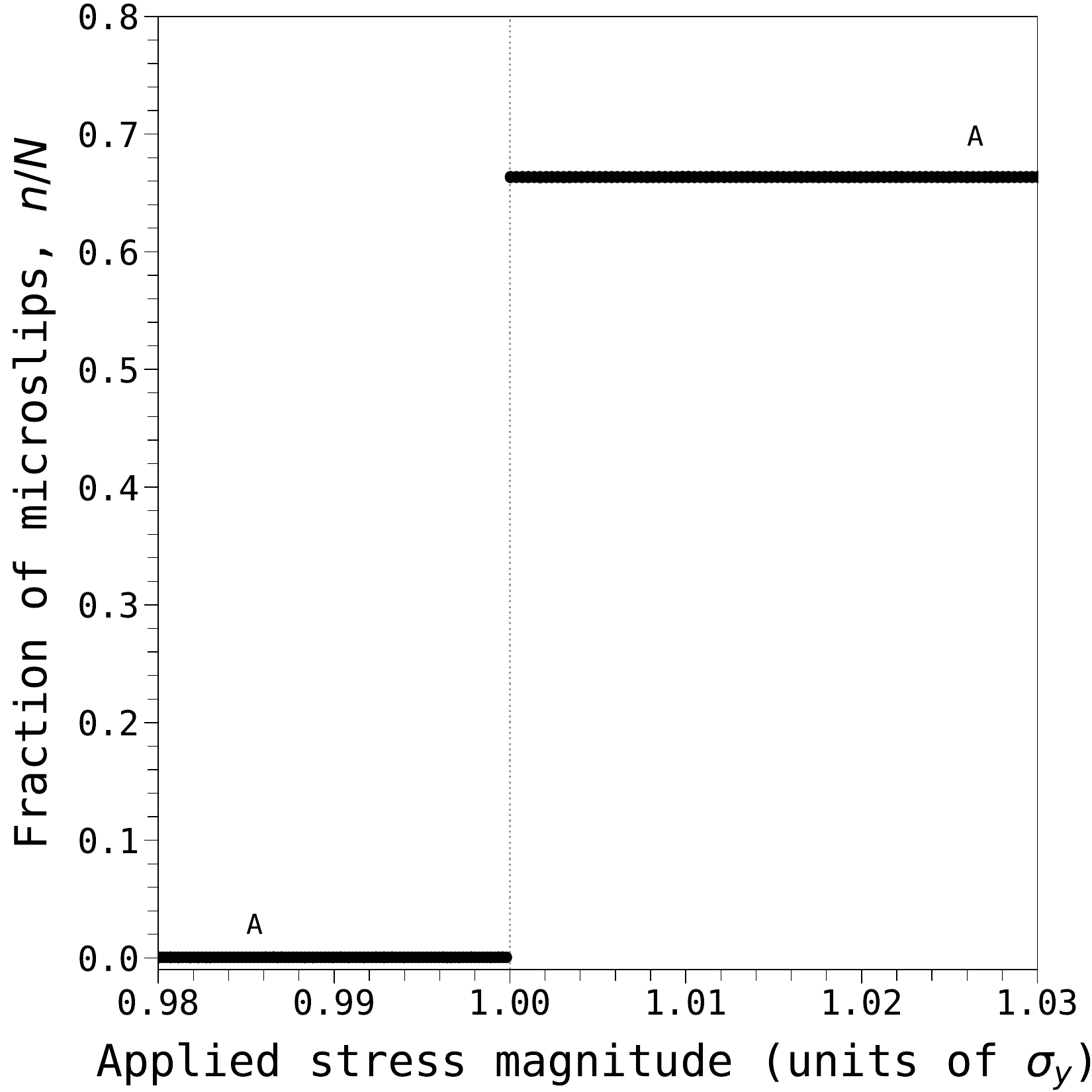}
         \caption{$\lambda = 0.01 \: \alpha$}
         \label{subfig: l0.01_fractions}
     \end{subfigure}
     \hfill
     \begin{subfigure}{0.329\textwidth}
         \centering
         \includegraphics[width=\textwidth]{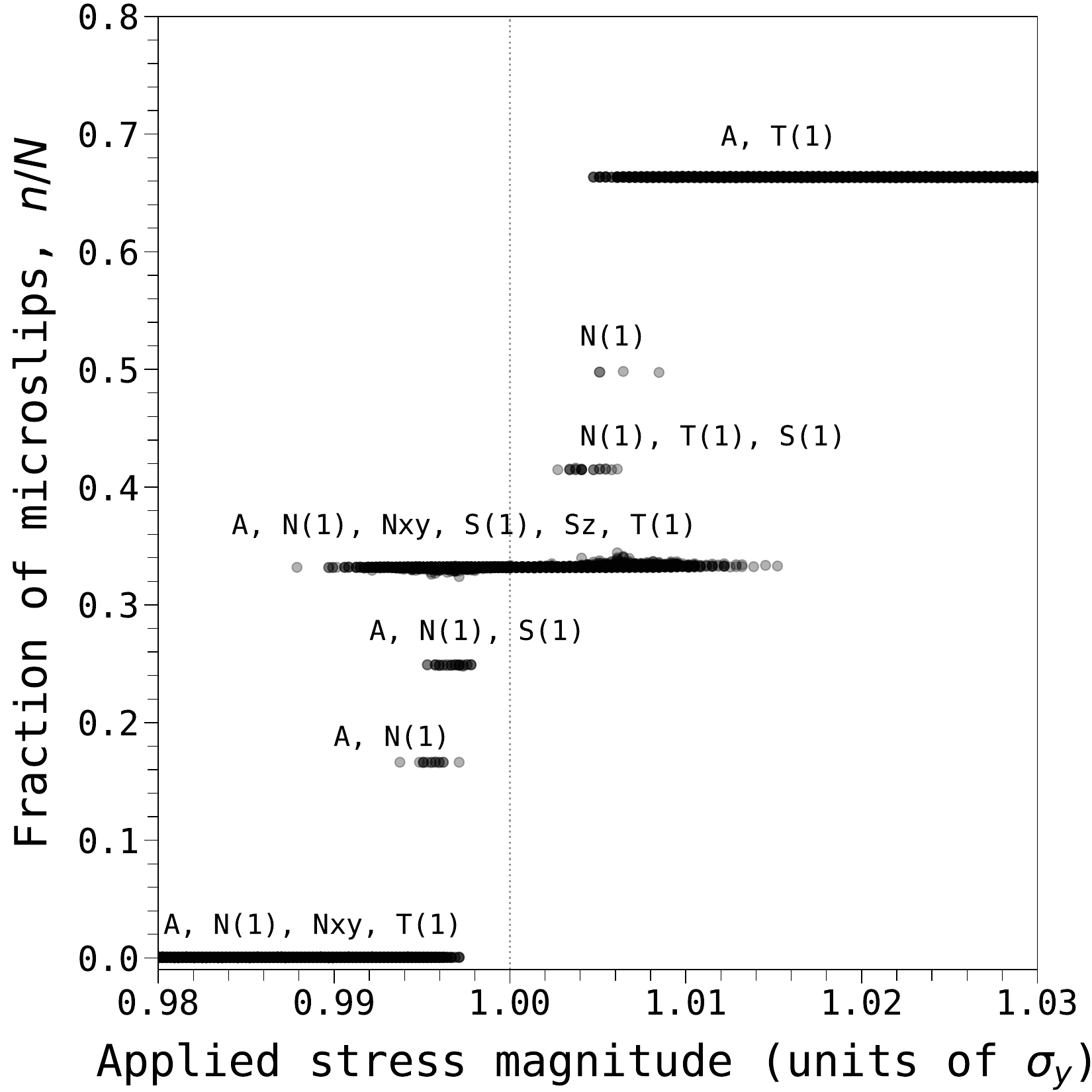}
         \caption{$\lambda = 4 \: \alpha$}
         \label{subfig: l4.0_fractions}
     \end{subfigure}
     \hfill
     \begin{subfigure}{0.329\textwidth}
         \centering
         \includegraphics[width=\textwidth]{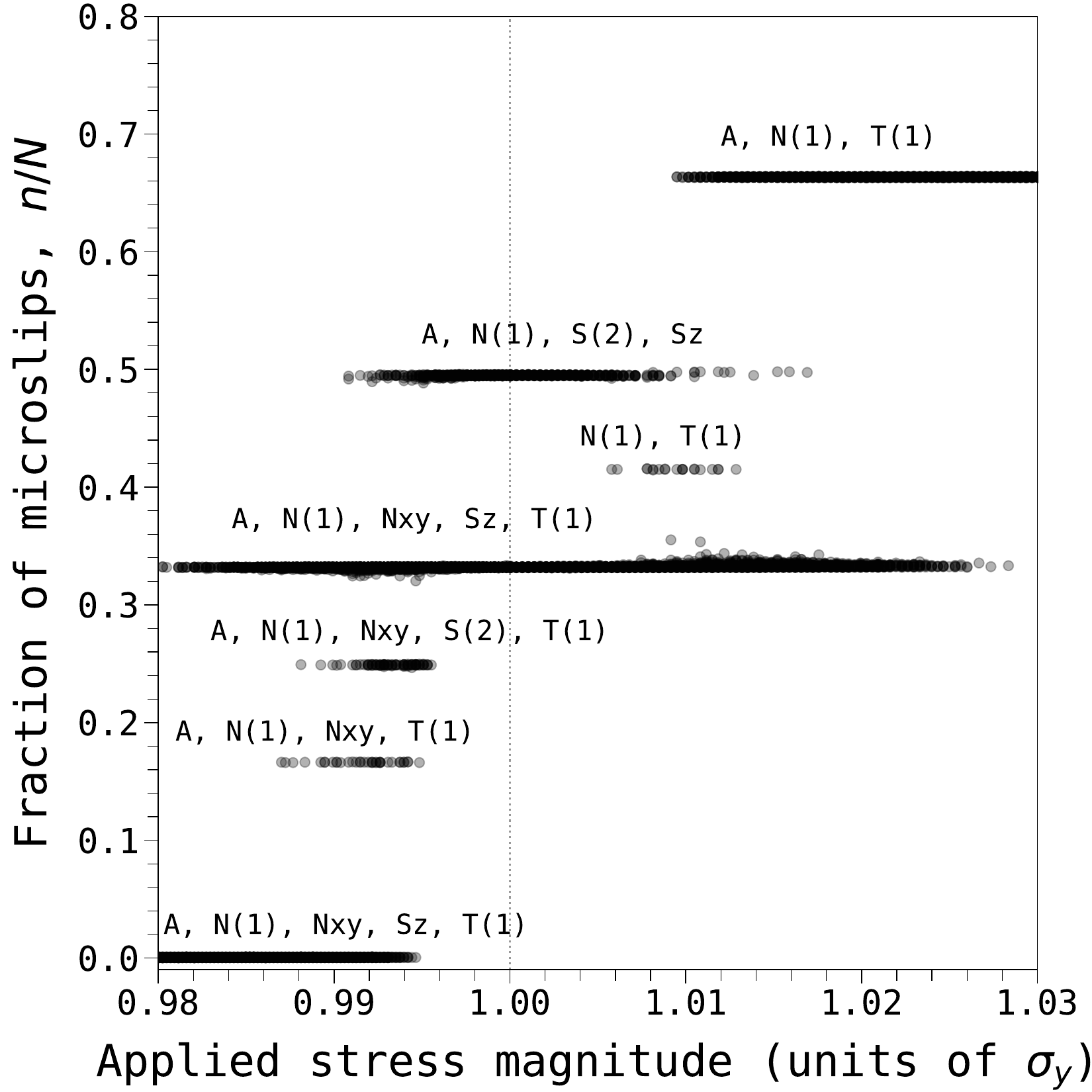}
         \caption{$\lambda = 8 \: \alpha$}
         \label{subfig: l8.0_fractions}
     \end{subfigure}
     \vfill
     \centering
     \begin{subfigure}{0.329\textwidth}
         \centering
         \includegraphics[width=\textwidth]{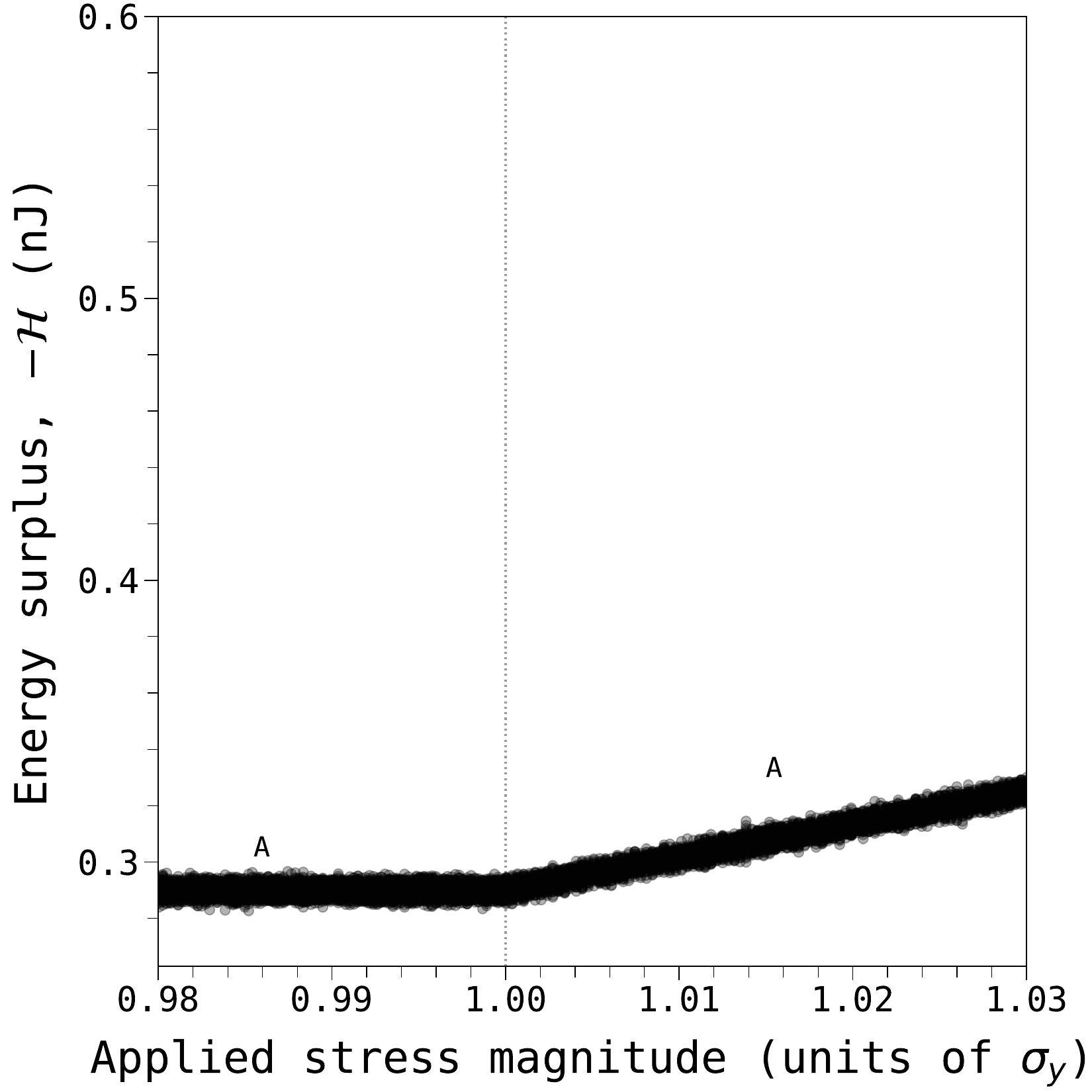}
         \caption{$\lambda = 0.01 \: \alpha$}
         \label{subfig: l0.01_energies}
     \end{subfigure}
     \hfill
     \begin{subfigure}{0.329\textwidth}
         \centering
         \includegraphics[width=\textwidth]{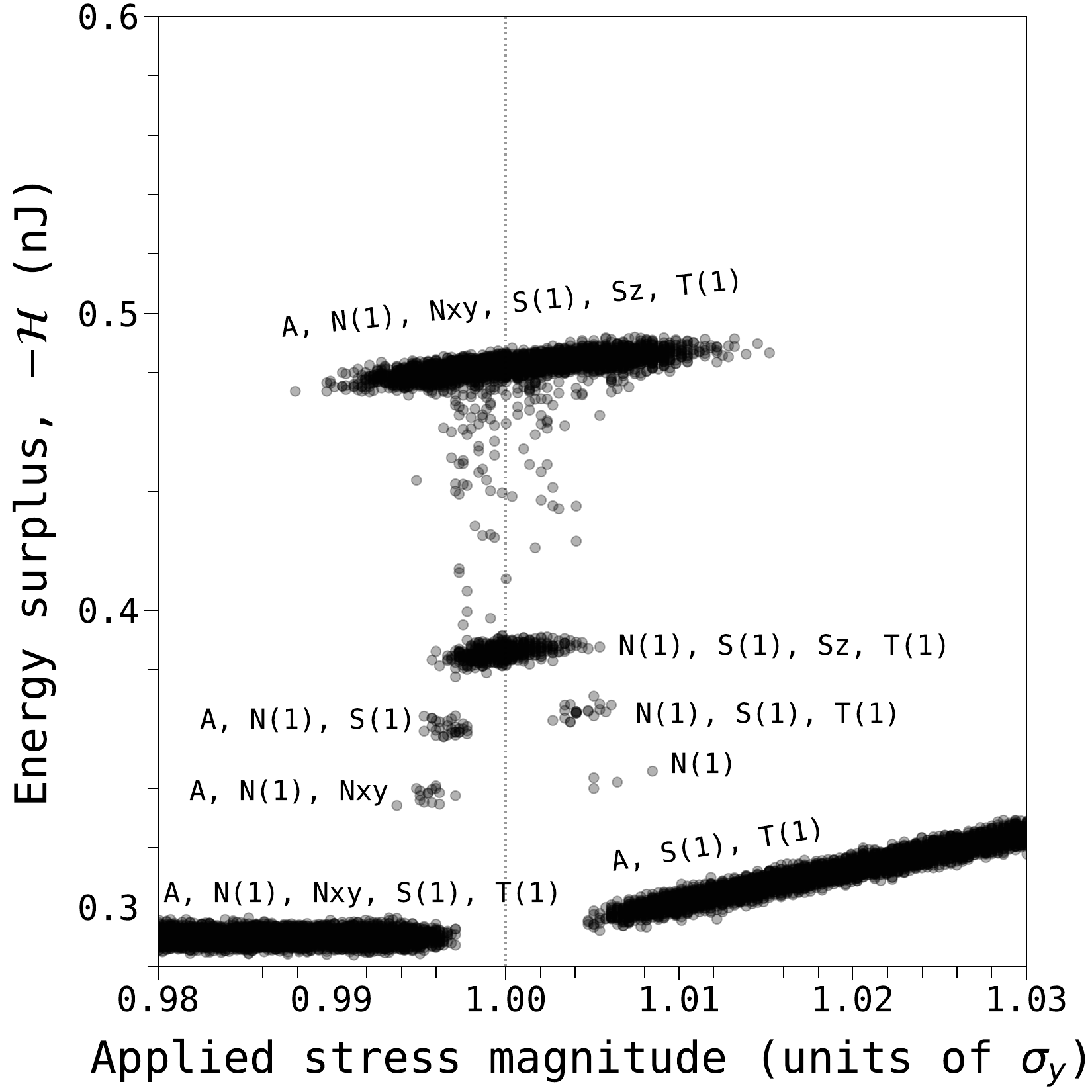}
         \caption{$\lambda = 4 \: \alpha$}
         \label{subfig: l4.0_energies}
     \end{subfigure}
     \hfill
     \begin{subfigure}{0.329\textwidth}
         \centering
         \includegraphics[width=\textwidth]{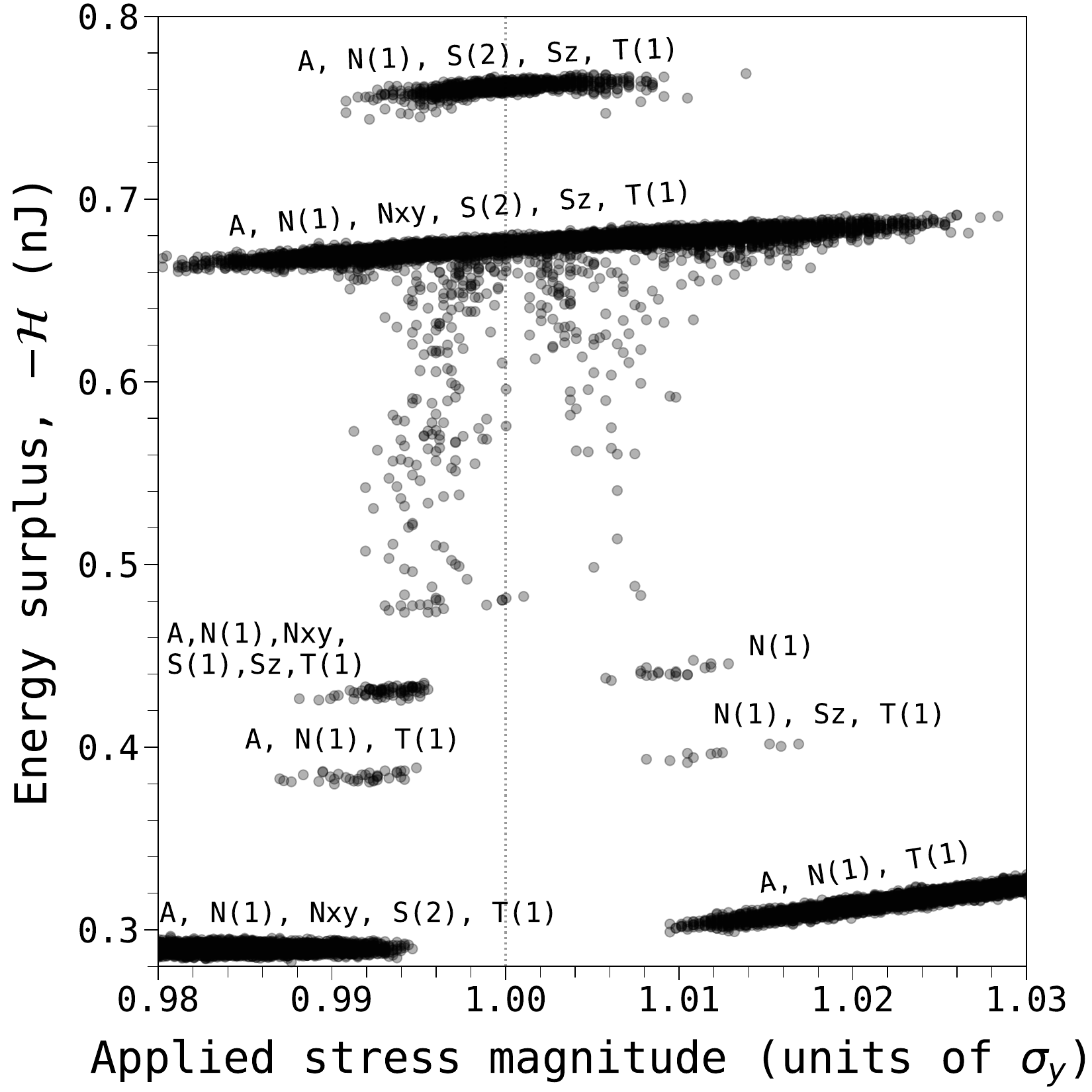}
         \caption{$\lambda = 8 \: \alpha$}
         \label{subfig: l8.0_energies}
     \end{subfigure}
     \vfill
     \centering
     \begin{subfigure}{0.329\textwidth}
         \centering
         \includegraphics[width=\textwidth]{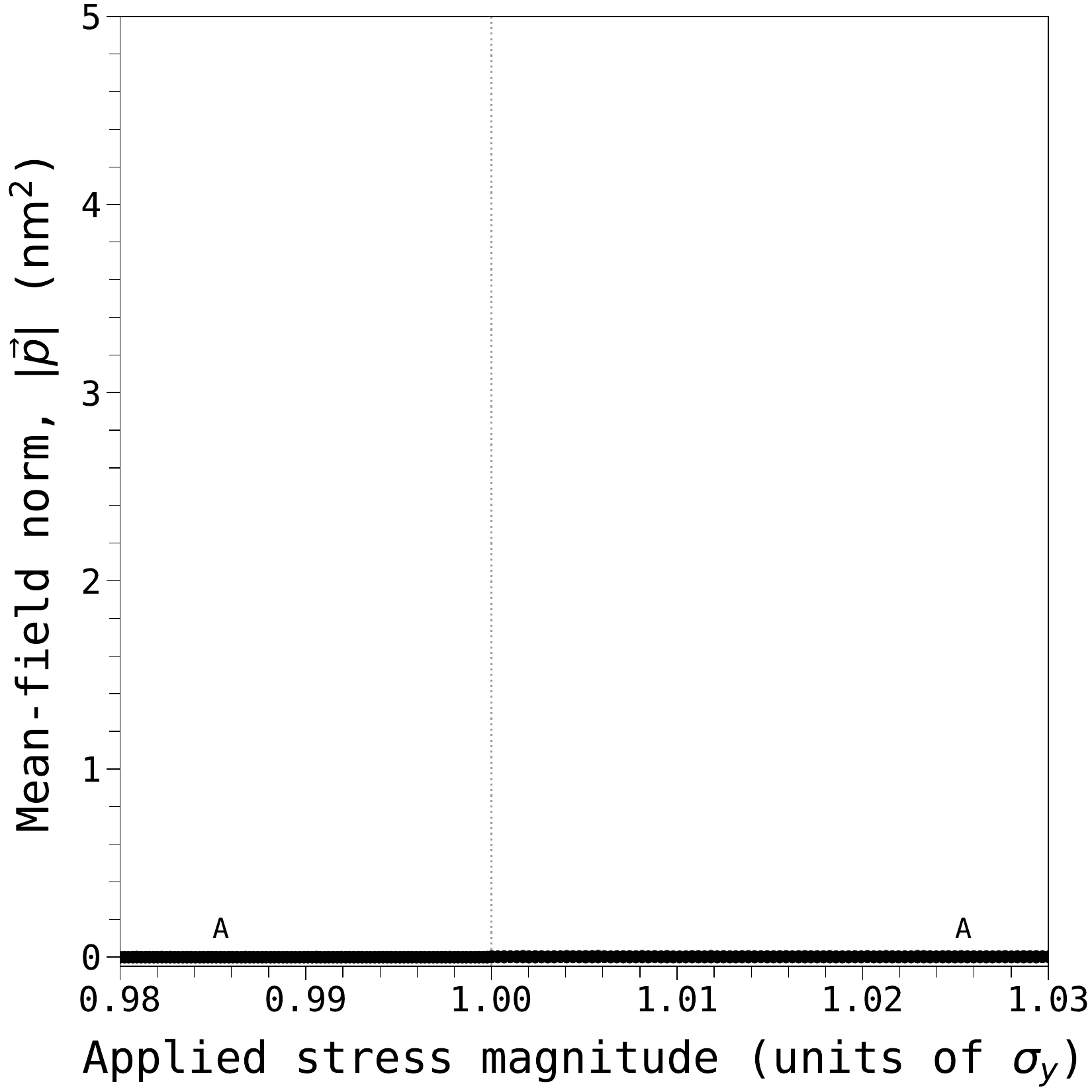}
         \caption{$\lambda = 0.01 \: \alpha$}
         \label{subfig: l0.01_meanfields}
     \end{subfigure}
     \hfill
     \begin{subfigure}{0.329\textwidth}
         \centering
         \includegraphics[width=\textwidth]{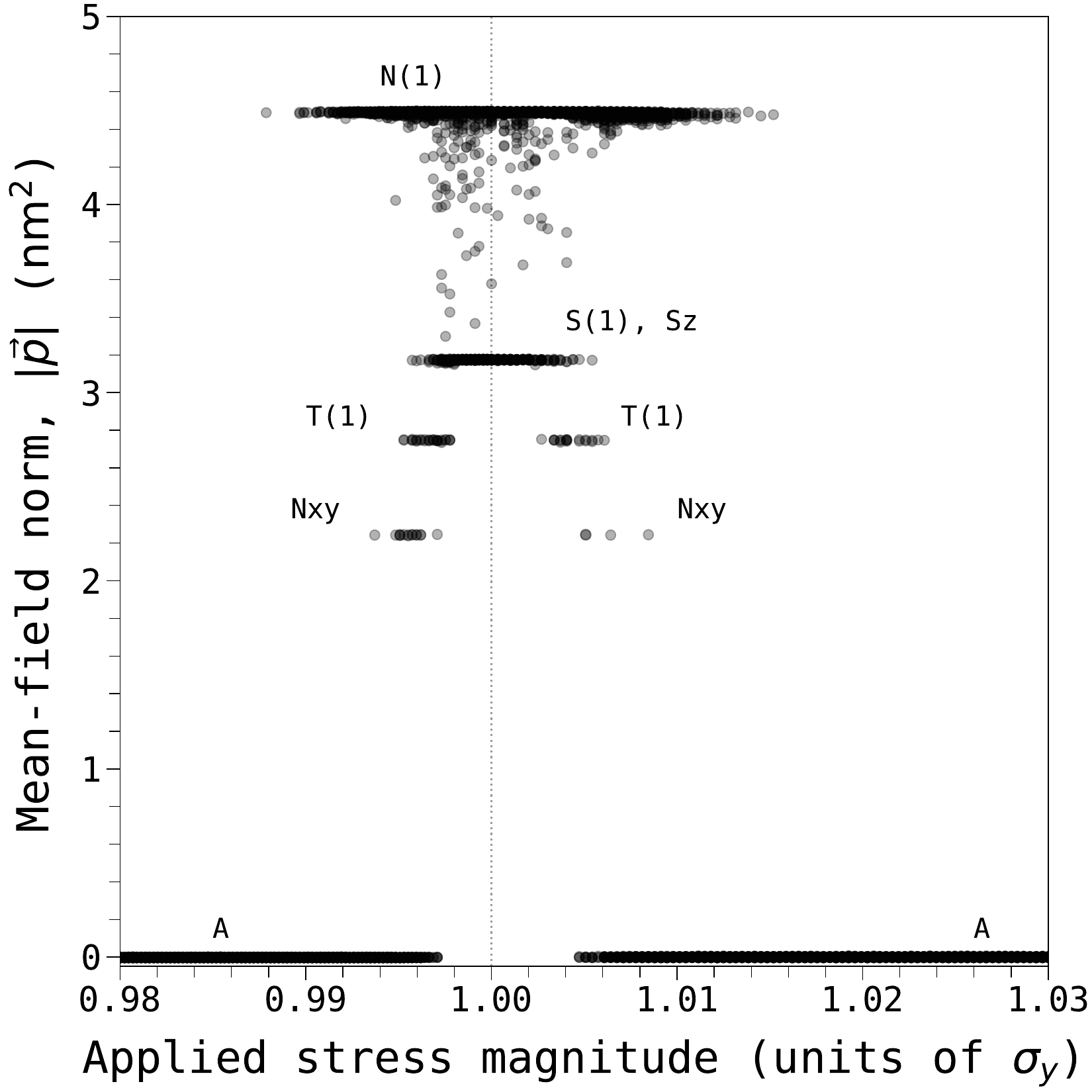}
         \caption{$\lambda = 4 \: \alpha$}
         \label{subfig: l4.0_meanfields}
     \end{subfigure}
     \hfill
     \begin{subfigure}{0.329\textwidth}
         \centering
         \includegraphics[width=\textwidth]{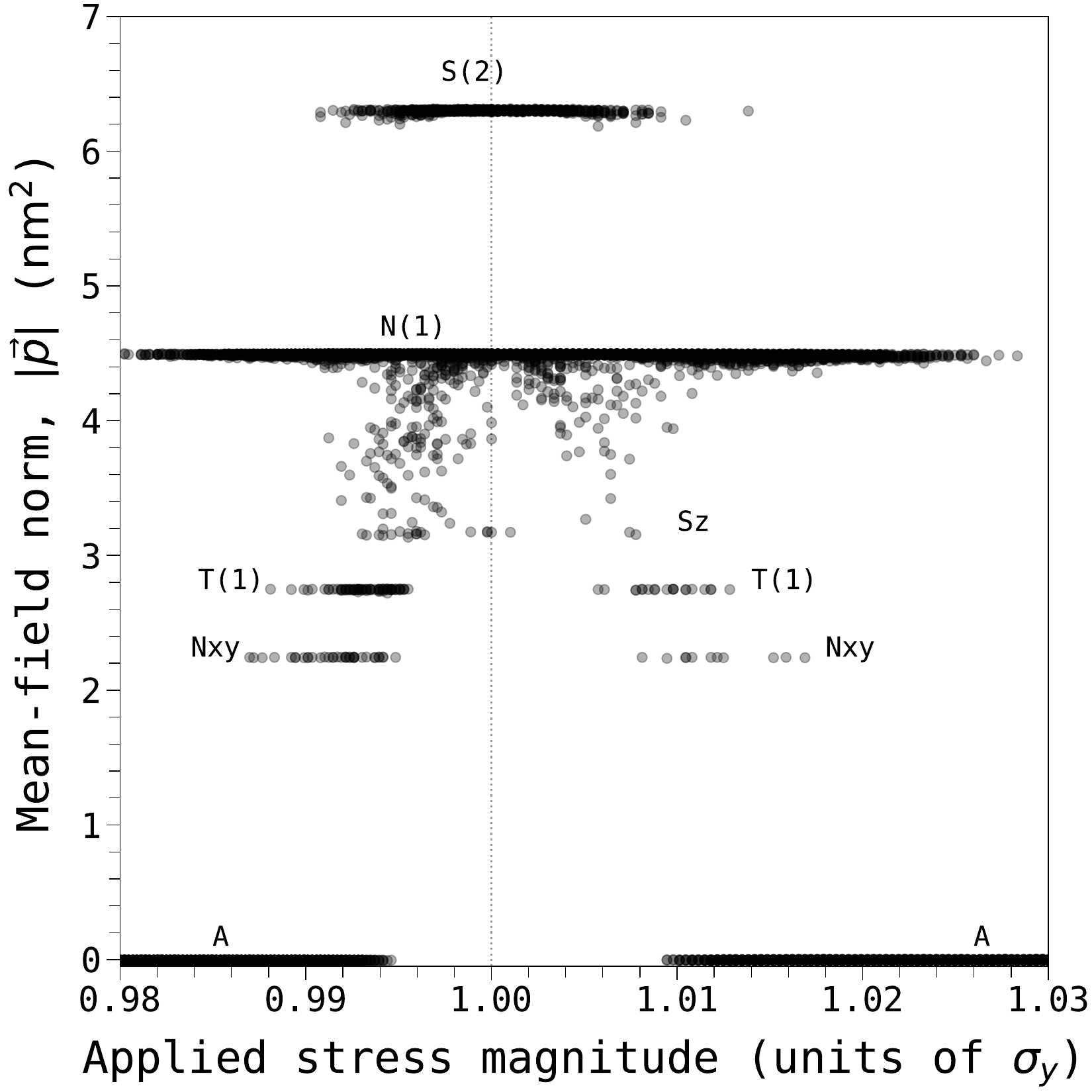}
         \caption{$\lambda = 8 \: \alpha$}
         \label{subfig: l8.0_meanfields}
     \end{subfigure}
        \caption{
(a)--(c) the fraction of microslip events, (d)--(f) the energy surplus and (g)--(i) the mean-field norm of the equilibrium states reached by the simulated system. Each data point seen here is the last data point taken from graphs such as Fig. \ref{fig: individual_simulations}. All simulations were run for 1 million steps. The self-energy parameter $\alpha$ was set to 465.15 GPa and the mean-field coupling parameter $\lambda$ was varied according to the captions on each individual plot. The text in each plot refers to the phases identified in Table \ref{table: phases} and indicate which phases were observed in the nearby collection of data points.  See the main text for more information.
}
        \label{fig: stress_range}
\end{figure}

To investigate further the effect of the applied stress magnitude on the behaviour of the system, systematic runs of the algorithm were conducted: Fig. \ref{fig: stress_range} shows the dependence of the fraction of activated microslips, the energy surplus and the mean-field norm on the magnitude of the applied uniaxial stress, at thermodynamic equilibrium, for increasing values of the mean-field coupling strength $\lambda$. The specific values of $\lambda$ were chosen to illustrate in a general sense how the behaviour of the system changes as $\lambda$ is increased. At each value of the applied stress magnitude, we considered the system to start at zero energy (i.e. $\mathcal{H}_0 = 0$) and set the initial defect network to be the same as described in Section \ref{subseq: Results_individualsimulations}. Then, we let the system reach equilibrium, i.e. each data point in the graphs of Fig. \ref{fig: stress_range} is the very last data point in the corresponding graphs of Fig. \ref{fig: individual_simulations}). Note that each data point is independent of the others.

With a very weak mean-field of $\lambda=0.01 \: \alpha$, the dependence of all three quantities was monotonically straightforward. The fraction of activated microslips (Fig. \ref{subfig: l0.01_fractions}) remained very close to zero until the applied stress magnitude reached the assumed yield stress, at which point it jumped to the maximum of about 0.66. This was in line with the observations in Section \ref{subseq: Results_individualsimulations}, where we noted that below yield the applied stress was too weak to maintain any defects in the system and above yield it was strong enough to produce every microslip. Such a weak mean-field simply took this case to the extreme. The energy surplus in the system (Fig. \ref{subfig: l0.01_energies}) remained constant at the minimum of about 0.289 nJ before yield, and as soon as the yield point was crossed it became a linear function of the applied stress magnitude. This linear trend is explained by looking at Eq. \eqref{eq: discrete_lagrangian} and noting that the inner product given by Eq. \eqref{eq: cochain_inner_product} is linear. The mean-field norm (Fig. \ref{subfig: l0.01_meanfields}) stayed very close to zero throughout the entire range of applied magnitudes because the contribution of all microslips was always circular, whether at a fraction of almost zero or a fraction of about 0.66. We thus identified in this case only one `state of deformation' which we labelled with the letter A. We refer to these `states of deformation', which were identifiable by the unique properties of the mean-field in each state, as phases. These phases are summarised in Table \ref{table: phases}. Phase A corresponded to a state of macroscopically perfect crystal structure, since either very few microslips had occurred (less than one hundred events) or they had occurred in such a way as to cancel each other out at the macroscopic scale, as per the comments that follow Eq. \eqref{eq: macrodefect_cochain}.

\setlength{\tabcolsep}{6pt} 
\renewcommand{\arraystretch}{1.35} 
\begin{table}
\centering
\begin{threeparttable}[t]
\caption{Averages (with standard deviations) of the mean-field vector norm $|\vec{p}|$, the absolute value of the mean-field $z$-component $|p_z|$ and the general form of the mean-field vector $\vec{p}$ given by Eq. \eqref{eq: macrodefect_cochain} for each phase as obtained from the simulations in Figs. \ref{fig: stress_range} and \ref{fig: lambda_range}.}
    \begin{tabular}{| m{1.5cm} | m{2.5cm} |  m{2.5cm} | m{7cm} |}
    \hline
    \textbf{Label} & $|\vec{p}|$ \textbf{(nm$\mathbf{^2}$)} & $|p_z|$ \textbf{(nm$\mathbf{^2}$)} & \textbf{General form of the normalised m.-f. vector $\vec{p}$}\\
    \hline
    A & $0.002 \pm 0.001$ & $0.001 \pm 0.001$ & (0, 0, 0) \\
    \hline
    N(1) & $4.489 \pm 0.012$ & $4.489 \pm 0.012$ & \multirow{2}{*}{$(0, 0, \pm 1)$} \\
    
    N(2) & $8.948 \pm 0.010$ & $8.948 \pm 0.010$ & \\
    \hline
    Nxy & $2.247 \pm 0.001$ & $0.001 \pm 0.001$ & ($\pm 1$, 0, 0) or (0, $\pm 1$, 0) \\
    \hline
    S(1) & $3.177 \pm 0.003$ & $0.002 \pm 0.003$ & \multirow{3}{*}{$(\pm 1, \: \pm 1, \: 0)/\sqrt{2}$} \\
    
    S(2) & $6.300 \pm 0.010$ & $0.002 \pm 0.002$ &  \\
    
    S(3) & $9.493 \pm 0.011$ & $0.007 \pm 0.008$ &  \\
    \hline
    Sz & $3.175 \pm 0.009$ & $2.231 \pm 0.012$ & $(\pm 1, \: 0, \: \pm 1)/\sqrt{2}$ or $(0, \: \pm 1, \: \pm 1)/\sqrt{2}$ \\
    \hline
    T(1) & $2.750 \pm 0.002$ & $1.121 \pm 0.003$ & $(\pm 2, \: \pm 1, \: \pm 1)/\sqrt{6}$ or $(\pm 1, \: \pm 2, \: \pm 1)/\sqrt{6}$ \\

    T(2) & $8.227 \pm 0.006$ & $2.210 \pm 0.006$ & $\approx (\pm 0.68, \: \pm 0.68, \: \pm 0.27)$ \\

    T(3) & $9.205 \pm 0.007$ & $2.289 \pm 0.003$ & $\approx (\pm 0.97, \: 0, \: \pm 0.28)$ or $(0, \: \pm 0.97, \: \pm 0.28)$ \\
    \hline
    \end{tabular}
\label{table: phases}
\end{threeparttable}
\end{table}

The behaviour of the system was markedly different for the stronger mean-field coupling strengths of $4 \: \alpha$ and $8 \: \alpha$, although Figs. \ref{subfig: l0.01_energies}, \ref{subfig: l4.0_energies} and \ref{subfig: l8.0_energies} only show the minima of the function $-\mathcal{H}(|\boldsymbol{\upsigma}|)$. In these two new cases, phase A was directly identifiable far from the yield point on either side, while the clear transition at yield seen in the previous case evolved into a complicated picture with multiple phases. These are labelled with uppercase letters and numbers in Fig. \ref{fig: stress_range} and summarised in Table \ref{table: phases}.


Comparing the cases $\lambda = 0.01 \: \alpha$ and $\lambda=4 \: \alpha$, the behaviour of the system mostly changed around the yield point --- roughly between 0.990 $\sigma_y$ and 1.014 $\sigma_y$ (Figs. \ref{subfig: l4.0_fractions}, \ref{subfig: l4.0_energies} and \ref{subfig: l4.0_meanfields}), where the new phases Nxy, T(1), S(1), Sz and N(1) appeared. From Figs. \ref{subfig: l4.0_fractions} and \ref{subfig: l4.0_energies} it is evident that multiple phases could be observed at different values of the microslip fraction and of the energy surplus, although the different phases were well-differentiated with respect to the general form of the mean-field vector $\vec{p}$ and its norm (Fig. \ref{subfig: l4.0_meanfields}). Although different phases populated the band of highest energy surplus, across the whole range of sampled applied stress magnitudes the most common phase after A was N(1) with a relative frequency of 23.8\% and the second most common phase was S(1) with a relative frequency of 2.8\%. Phase N(1) was observed at all values of the microslip fraction, except at the maximum of about 0.66 --- where only the phases A and T(1) were observed --- and was characterised by a mean-field vector in the $z$-direction with a norm of $4.489 \pm 0.012$, meaning that it provoked a net reduction in area in the $xy$-plane. Similarly, phase Nxy was characterised by a mean-field vector in the $x$-direction or $y$-direction with a norm of $2.247 \pm 0.001$, representing a net reduction in area in the $yz$- or $xz$-plane, respectively. Phase S(1) was characterised by a mean-field vector in the $xy$-plane (specifically along the $x=y$ direction) with a norm of $3.177 \pm 0.003$, meaning that it provoked a net reduction in area in a plane parallel to the $z$-direction. It also represents a net predominant activation of the $[110]$ slip direction. Phase Sz was characterised similarly to S(1) by a mean-field vector in the $xz$- or $yz$-plane (specifically along the $x=z$ or $y=z$ directions, respectively) with a norm of $3.175 \pm 0.009$, meaning that it provoked a net reduction in area in a plane parallel to the $y$-direction or $x$-direction, respectively. Like in the case of S(1), it represents a net predominant activation of the $[101]$ or $[011]$ slip directions, respectively. Lastly, phase T(1) was characterised by a mean-field vector of general form $(\pm 2, \: \pm 1, \: \pm 1)$ or $(\pm 1, \: \pm 2, \: \pm 1)$ (with any combination of the plus and minus signs) and norm $2.750 \pm 0.002$, indicating a net reduction in area in a plane perpendicular to it. All phases seemed to emerge discontinuously and vanish discontinuously.

Comparing the cases $\lambda=4 \: \alpha$ and $\lambda=8 \: \alpha$, the most significant changes were the reinforcement of phase S(1) into phase S(2) and the extension of the range of applied stress values over which phases other than A were observed. For example, phase N(1) was identified in a range of applied stresses from roughly 0.980 $\sigma_y$ to 1.028 $\sigma_y$, as seen in Fig. \ref{subfig: l8.0_meanfields}. Different phases were observed to be mixed at all values of the microslip fraction and of the surplus energy in Figs. \ref{subfig: l8.0_fractions} and \ref{subfig: l8.0_energies}. Across the entire range of applied stress magnitudes, the most common phase was N(1) with a relative frequency of 47.0\% and the second most common phase was A with a relative frequency of 46.2\%, followed by S(2) with 5.7\%. The new emergent phase S(2) was characterised similarly to S(1) by a mean-field vector in the $xy$-plane (specifically along the $x=y$ direction) but with a larger norm of $6.300 \pm 0.010$. As in the previous case, all phases seemed to emerge discontinuously and vanish discontinuously.

\begin{figure}[t!]
     \centering
     \begin{subfigure}{0.329\textwidth}
         \centering
         \includegraphics[width=\textwidth]{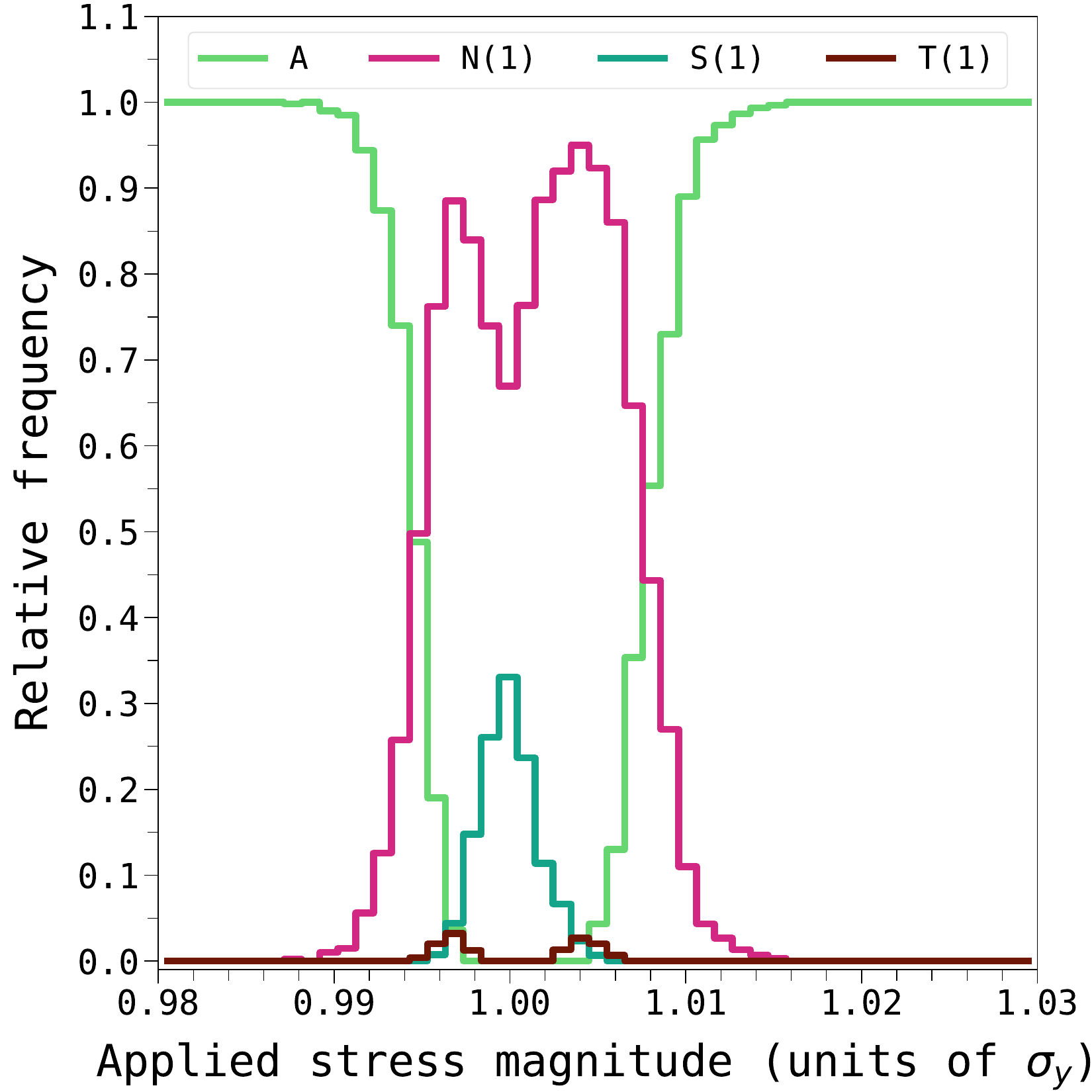}
         \caption{$\lambda = 4 \: \alpha$}
         \label{subfig: l4_histogram}
     \end{subfigure}
     \begin{subfigure}{0.329\textwidth}
         \centering
         \includegraphics[width=\textwidth]{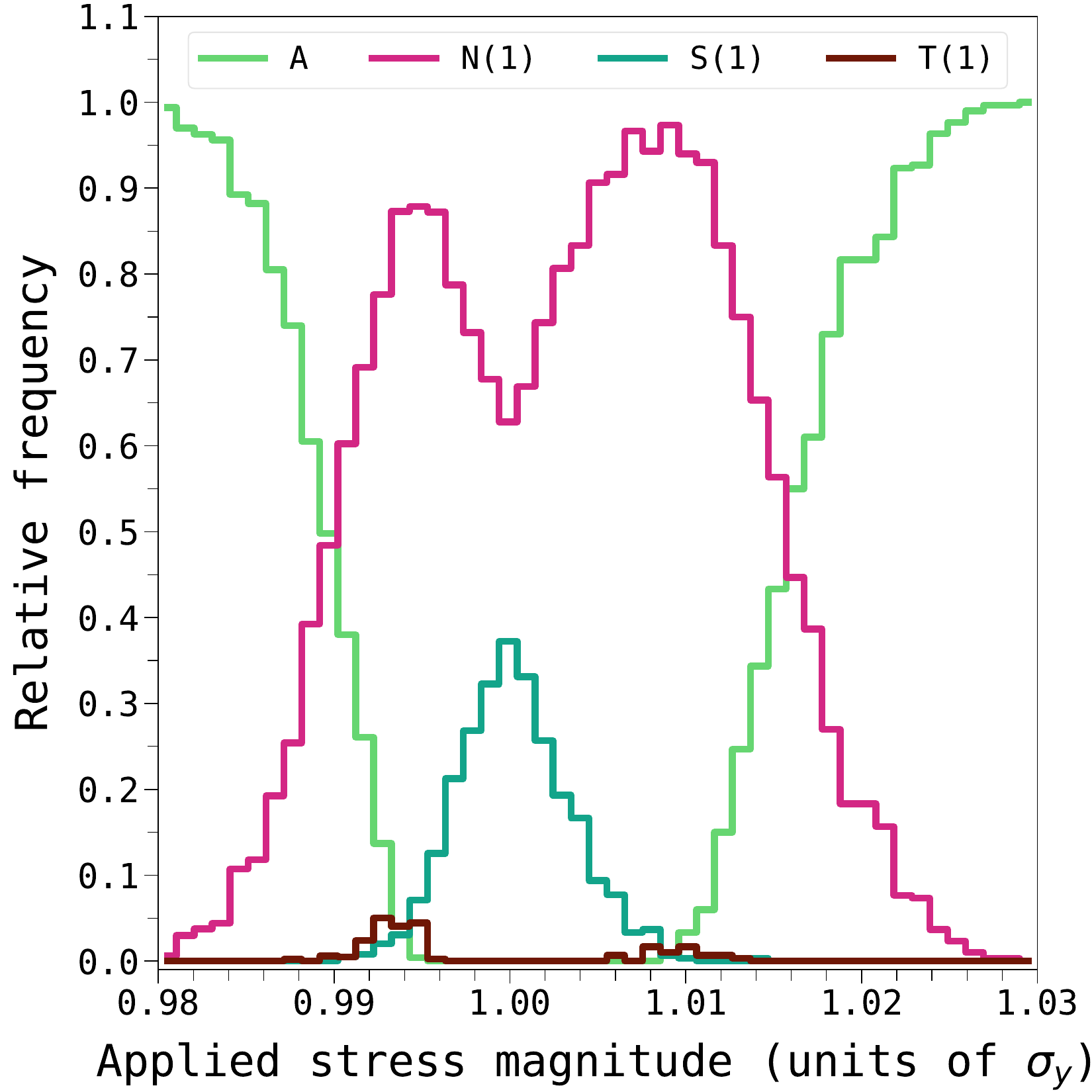}
         \caption{$\lambda = 8 \: \alpha$}
         \label{subfig: l8_histogram}
     \end{subfigure}
        \caption{
Histograms of the relative frequency of the phases observed (a) in Fig. \ref{subfig: l4.0_meanfields} and (b) in Fig. \ref{subfig: l8.0_meanfields}. The relative frequency is normalised so that it totals 1 at each bin in the horizontal axis. The bins have a width of $0.001\: \sigma_y$. Phase Nxy was excluded to enhance legibility of the plot.
}
        \label{fig: rel_freq}
\end{figure}

Upon further inspection, a probabilistic character was revealed in the discontinuous transitions between phases observed in Fig. \ref{fig: stress_range}. This is shown in Fig. \ref{fig: rel_freq}, where it is clear that the relative frequency of data points of a particular phase increased and decreased gradually. Phases A, N(1) and T(1) seemed to follow a bimodal probability distribution, while phases S(1) and S(2) seemed to follow a normal distribution about the yield point $\sigma=\sigma_y$.

Lastly, the spread of values in-between the horizontal lines $|\vec{p}| \approx 3.175$ and $|\vec{p}| \approx 4.489$ in Figs. \ref{subfig: l4.0_energies} and \ref{subfig: l8.0_energies} suggested that the number of simulation steps was insufficient for the system to reach equilibrium and that the simulated specimen may have transitioned from phase S(1), S(2) or Sz to phase N(1), as indeed was seen in Figs. \ref{subfig: med_energies} and \ref{subfig: hig_energies}. The reverse transition from N(1) to S(1), S(2) or Sz was never observed. Conversely, the lack of such intermediary data points between all the other discrete levels in Figs. \ref{subfig: l0.01_fractions}--\ref{subfig: l8.0_meanfields} may suggest that there was no viable trajectory in phase-space for the system to transition smoothly between the other phases.

\subsection{Phase transitions as a function of the mean-field coupling strength} \label{subseq: Results_lambdaranges}

\begin{figure}[!p]
     \centering
     \begin{subfigure}{0.329\textwidth}
         \centering
         \includegraphics[width=\textwidth]{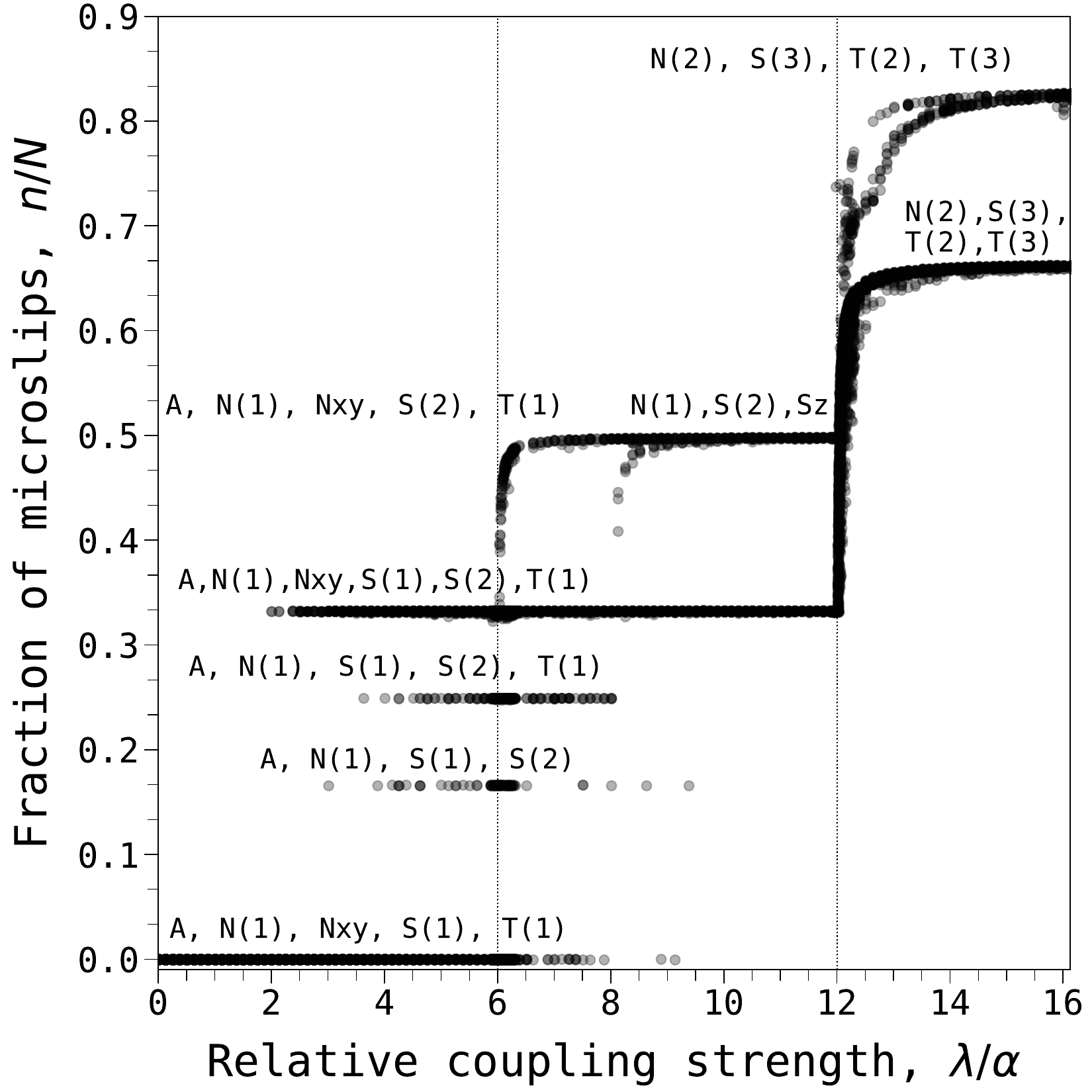}
         \caption{$\sigma= 0.995 \: \sigma_y$}
         \label{subfig: m0.995_fractions}
     \end{subfigure}
     \hfill
     \begin{subfigure}{0.329\textwidth}
         \centering
         \includegraphics[width=\textwidth]{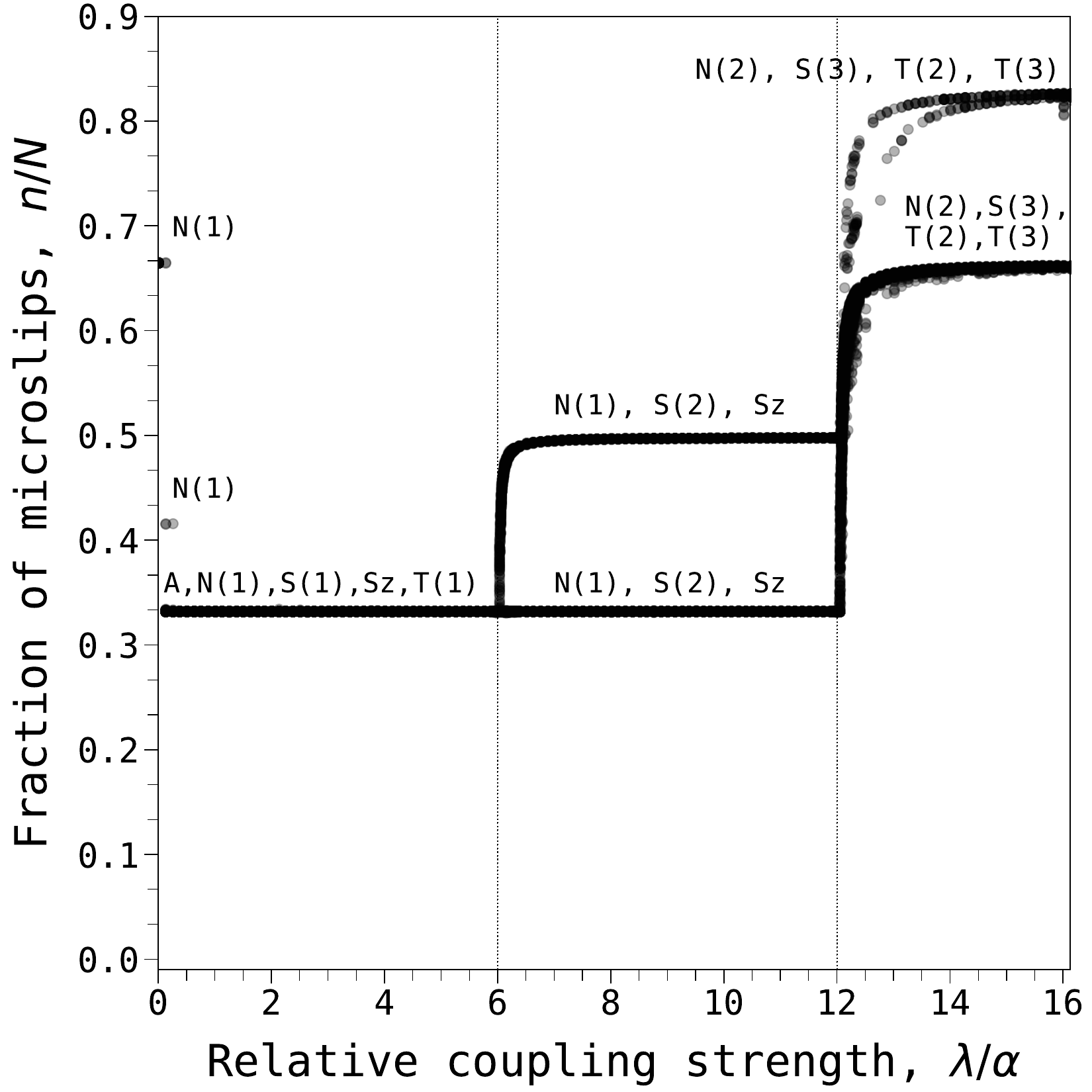}
         \caption{$\sigma= \sigma_y$}
         \label{subfig: m1.00_fractions}
     \end{subfigure}
     \hfill
     \begin{subfigure}{0.329\textwidth}
         \centering
         \includegraphics[width=\textwidth]{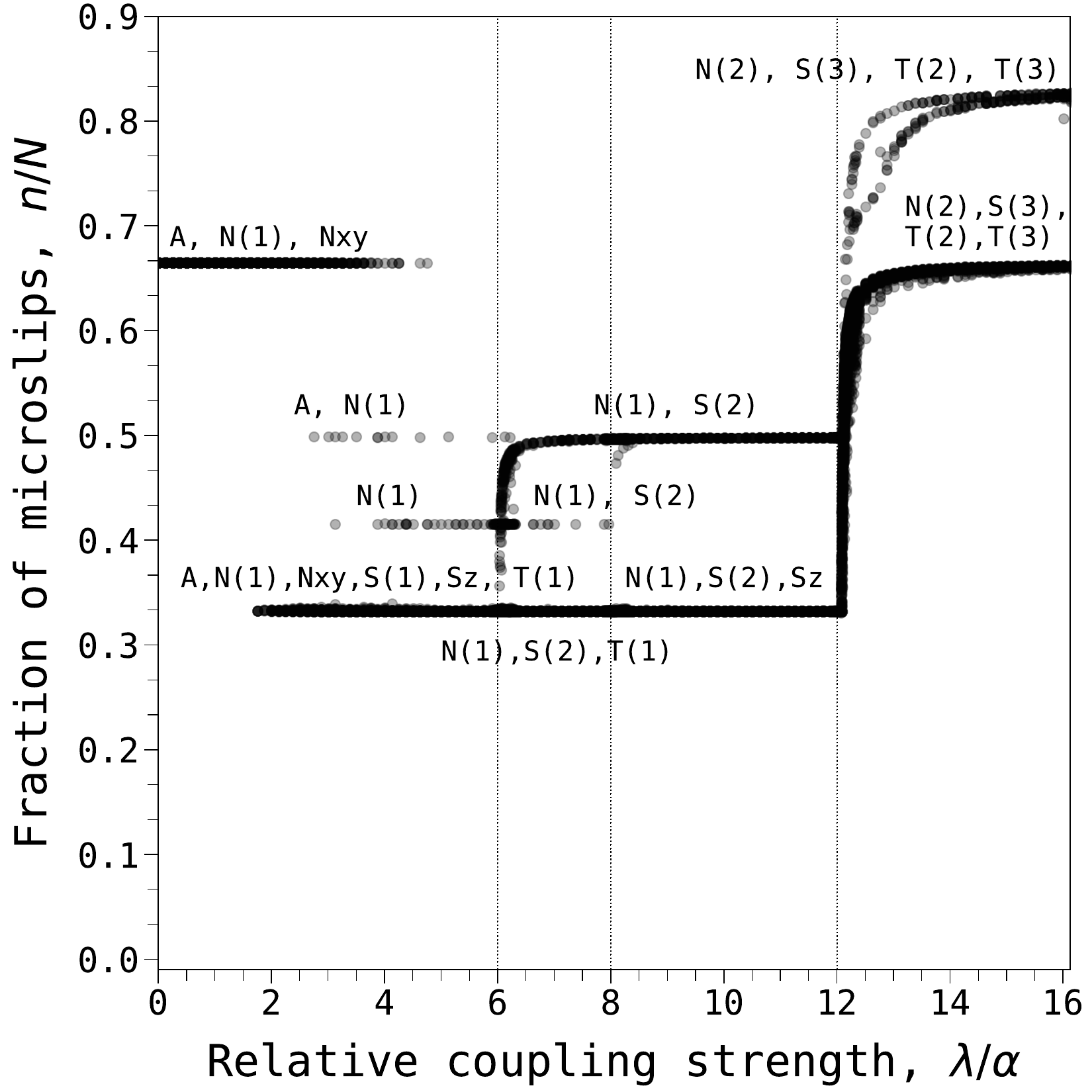}
         \caption{$\sigma= 1.005 \: \sigma_y$}
         \label{subfig: m1.005_fractions}
     \end{subfigure}
     \vfill
     \centering
     \begin{subfigure}{0.329\textwidth}
         \centering
         \includegraphics[width=\textwidth]{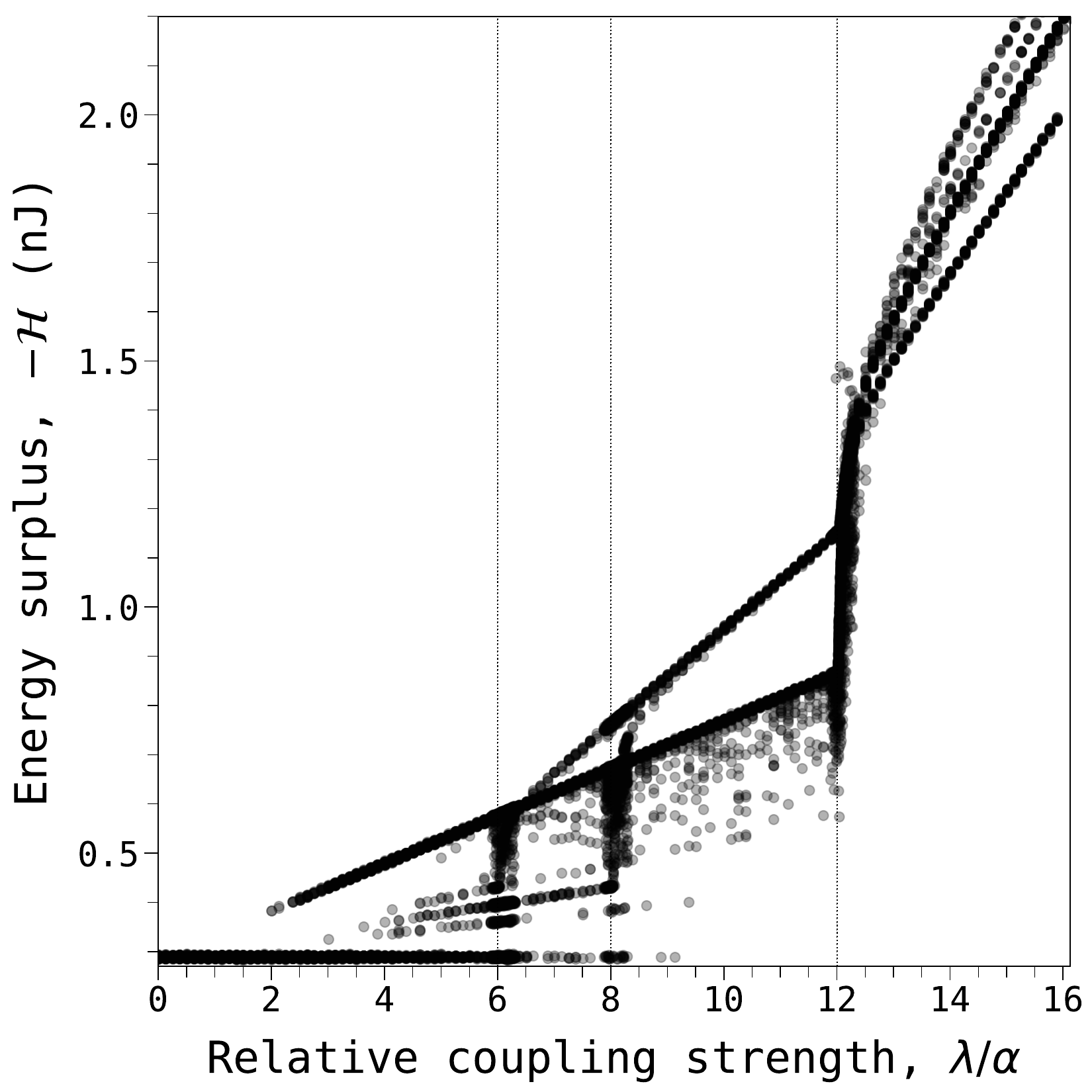}
         \caption{$\sigma= 0.995 \: \sigma_y$}
         \label{subfig: m0.995_energies}
     \end{subfigure}
     \hfill
     \begin{subfigure}{0.329\textwidth}
         \centering
         \includegraphics[width=\textwidth]{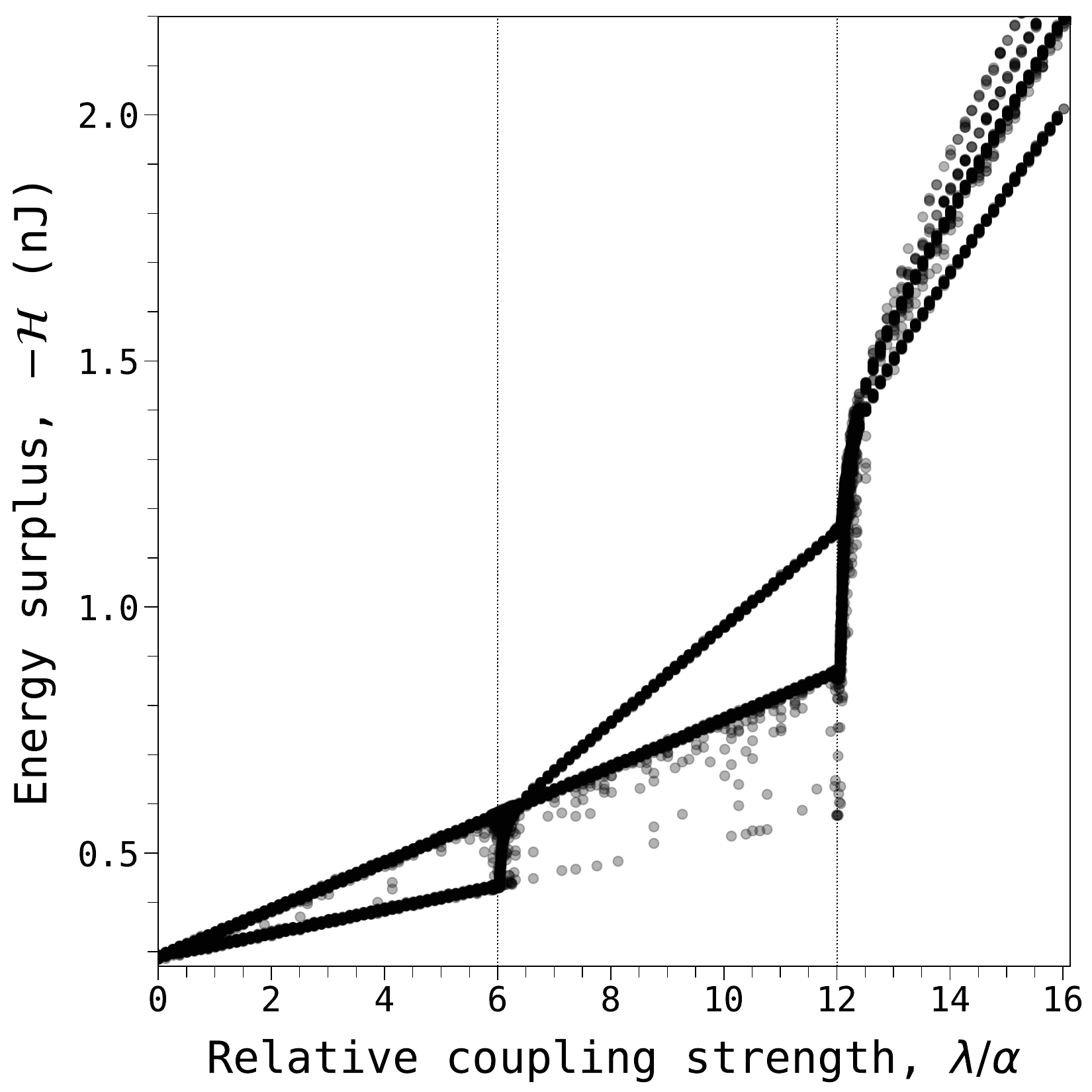}
         \caption{$\sigma= \sigma_y$}
         \label{subfig: m1.00_energies}
     \end{subfigure}
     \hfill
     \begin{subfigure}{0.329\textwidth}
         \centering
         \includegraphics[width=\textwidth]{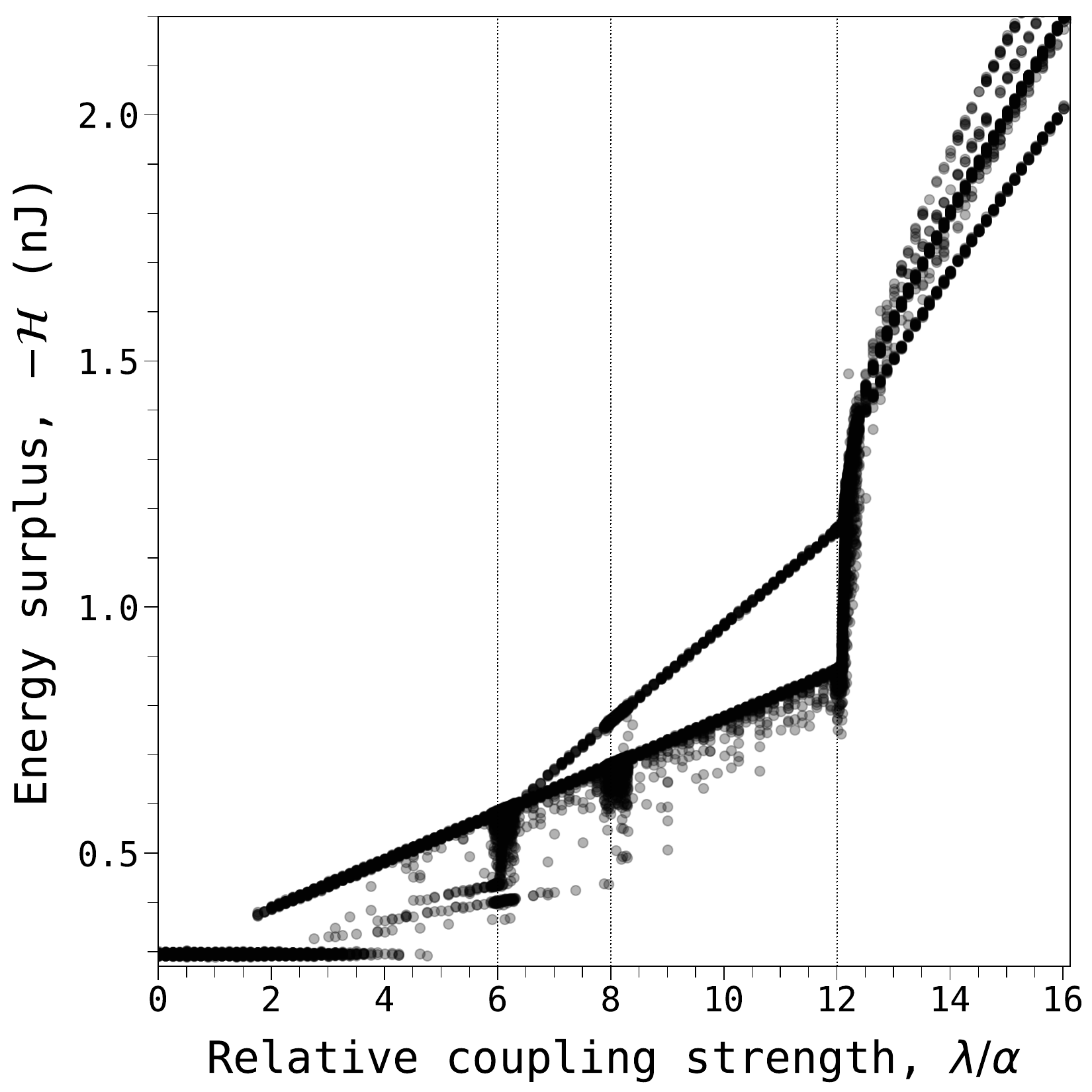}
         \caption{$\sigma= 1.005 \: \sigma_y$}
         \label{subfig: m1.005_energies}
     \end{subfigure}
     \vfill
     \centering
     \begin{subfigure}{0.329\textwidth}
         \centering
         \includegraphics[width=\textwidth]{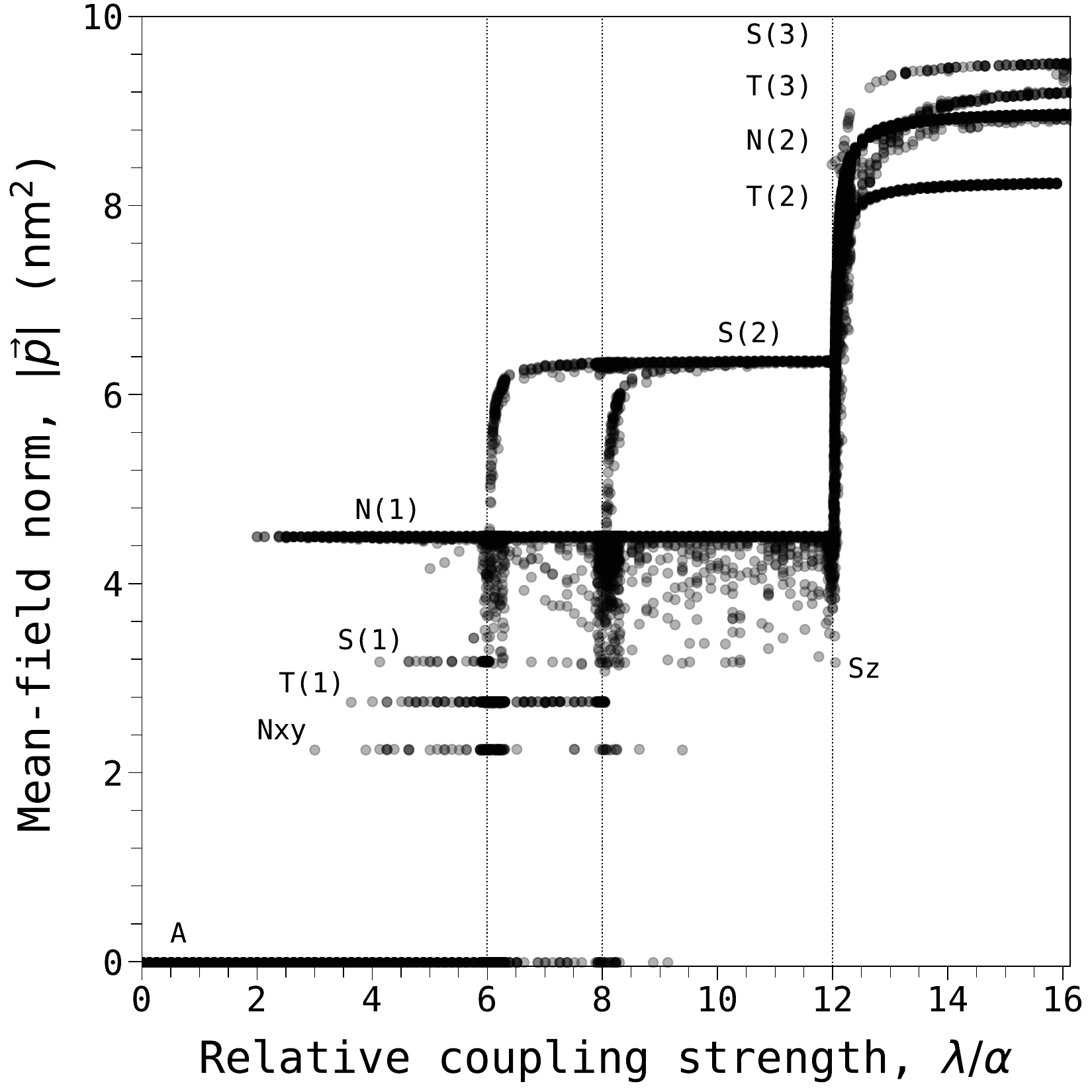}
         \caption{$\sigma= 0.995 \: \sigma_y$}
         \label{subfig: m0.995_meanfields}
     \end{subfigure}
     \hfill
     \begin{subfigure}{0.329\textwidth}
         \centering
         \includegraphics[width=\textwidth]{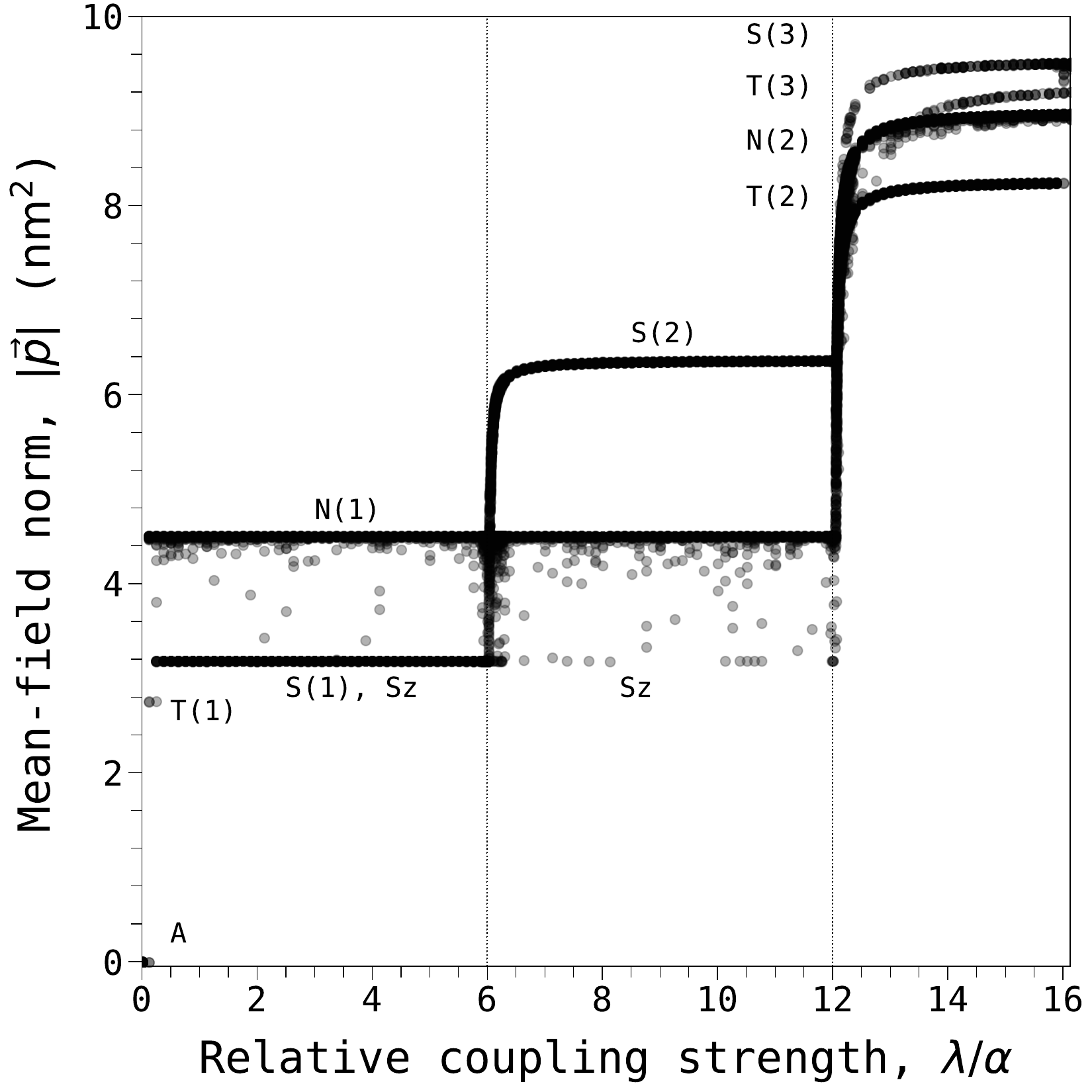}
         \caption{$\sigma= \sigma_y$}
         \label{subfig: m1.00_meanfields}
     \end{subfigure}
     \hfill
     \begin{subfigure}{0.329\textwidth}
         \centering
         \includegraphics[width=\textwidth]{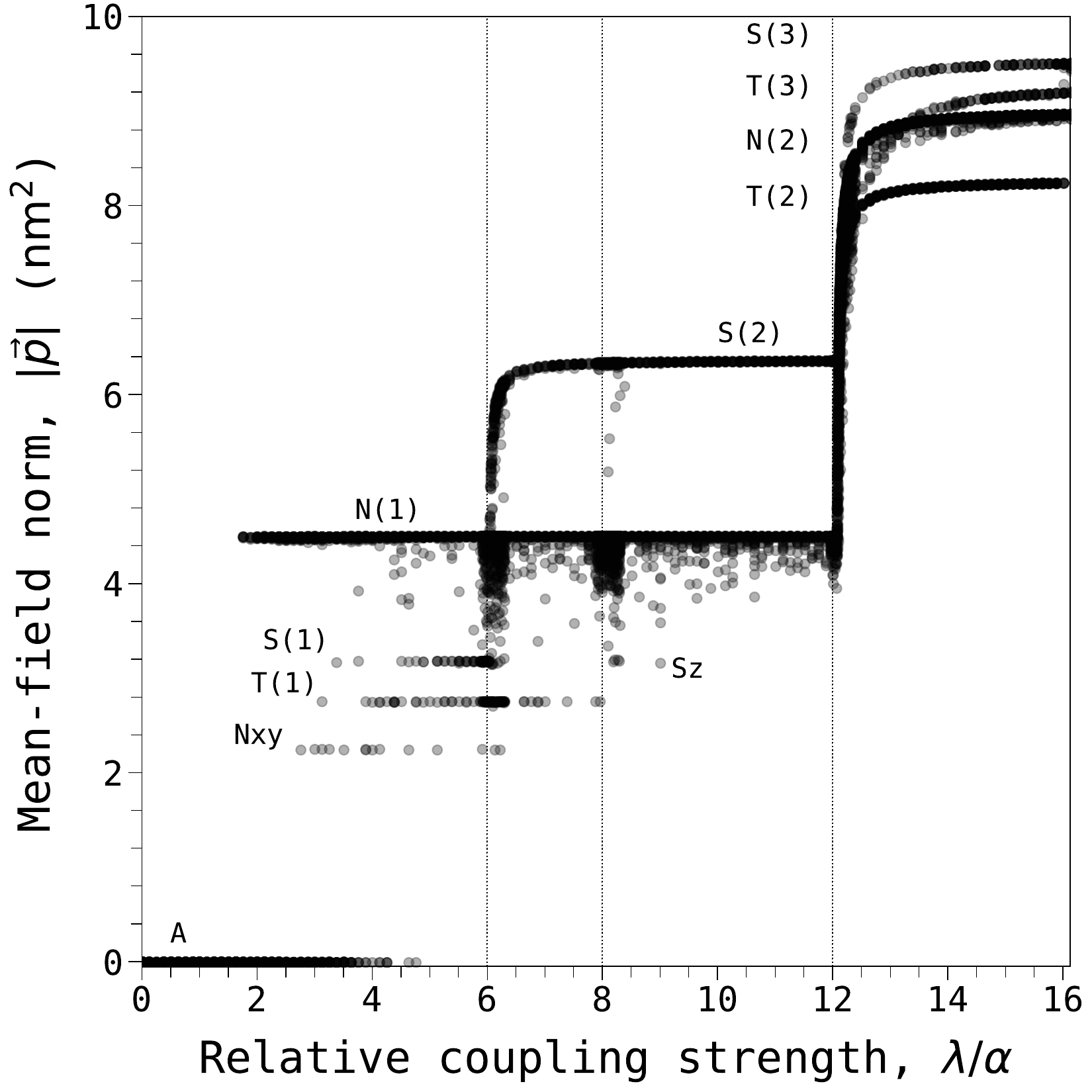}
         \caption{$\sigma= 1.005 \: \sigma_y$}
         \label{subfig: m1.005_meanfields}
     \end{subfigure}
        \caption{
(a)--(c) the fraction of microslip events, (d)--(f) the energy surplus and (g)--(i) the mean-field norm of the equilibrium states reached by the simulated system. Each data point seen here is the last data point taken from graphs such as Fig. \ref{fig: individual_simulations}. All simulations were run for 1.1 million steps, except those corresponding to most of the data points around $\lambda/ \alpha = 6,\: 8,\: 12 $ in (a), (c), (d), (f), (g) and (i) and $\lambda/\alpha = 6,\: 8$ in (b), (e) and (h), which were run for 1.3 million steps. The self-energy parameter $\alpha$ and the assumed yield stress $\sigma_y$ were set to 465.15 GPa and 10.09 GPa, respectively. The applied stress magnitude was varied according to the captions on each individual plot. The text in each plot refers to the phases identified in Table \ref{table: phases} and indicate which phases were observed in the nearby collection of data points. The labels for phases N(2), S(3), T(2) and T(3) refer to data points in the region $\lambda/\alpha>15$. See the main text for more information.
}
    \label{fig: lambda_range}
\end{figure}

Seeking to better understand the changes in the behaviour of the system as the mean-field coupling strength increases, Fig. \ref{fig: lambda_range} was plotted. An intricate picture arose of both discontinuous and continuous transitions as the mean-field coupling strength $\lambda$ increased. The continuous transitions, which can be identified as second-order phase transitions by their distinctive shape in Figs. \ref{subfig: m0.995_fractions}--\ref{subfig: m1.005_meanfields}, occurred at $\lambda/\alpha \approx 6$, at $\lambda/\alpha \approx 8$ and at $\lambda/\alpha \approx 12$. An initial sampling was conducted with data points spaced by $0.125 \: \lambda/\alpha$ on the horizontal axis, and afterward these data were supplemented in intervals around the transition points $\lambda/\alpha \approx 6$, $8$ and $12$ by data points spaced by $1/150 \: \lambda/\alpha$ on the horizontal axis.

Below the yield point, at an applied magnitude of $0.995 \: \sigma_y$, all three second-order phase transitions were identifiable. In the range $\lambda/\alpha \lesssim 6$, phases A, N(1), Nxy, S(1) and T(1) were observed at varying values of the microslip fraction, but never above $n/N \approx 0.33$, as labelled in Fig. \ref{subfig: m0.995_fractions}. In the range $\lambda/\alpha \lesssim 6$ all phases except A emerged discontinuously at varying points, which can be seen in Fig. \ref{subfig: m0.995_meanfields}. At $\lambda/\alpha \approx 6$, the first second-order transition occurred, whereupon phase S(1) transitioned into phase S(2), and the possibility for instances of phases A, N(1), Nxy, S(2) and T(1) to achieve a fraction $n/N \approx 0.50$ was unlocked. Phase Sz was registered for the first time in the range $6 \lesssim \lambda/\alpha \lesssim 8$.
The second second-order phase transition occurred at $\lambda/\alpha \approx 8$. This transition was not like the other two, where clear continuous changes could be observed from one phase to another. Additionally, it was not observed in the case at yield. As the mean-field coupling strength increased from $\lambda/\alpha < 8$ to $\lambda/\alpha > 8$, the only identifiable change in the behaviour of the system was the suppression of phase T(1). However, as can be seen in Fig. \ref{subfig: m0.995_meanfields}, this phase disappeared discontinuously. Instead, the actual second-order phase transition could be seen in terms of the microslip fraction or the energy surplus, in Figs. \ref{subfig: m0.995_fractions} and \ref{subfig: m0.995_energies}, where it corresponded to a transition from the level $n/N \approx 0.25$ to the level $n/N \approx 0.50$ (Fig. \ref{subfig: m0.995_fractions}) and where the energy surplus continuously jumped from about 0.41 nJ to about 0.85 nJ whence it continued on its linear trend (Fig. \ref{subfig: m0.995_energies}). In the interval $8 \lesssim \lambda/\alpha \lesssim 12$, instances of phases A, N(1), Nxy, S(2) and Sz were registered, occupying one of four discrete levels of the microslip fraction: 0, 0.25, 0.33 or 0.50. It was in this range that first-order phase transitions occurred as well, with the discontinuous suppression of phases A and Nxy. In the region $10 < \lambda/\alpha < 12$, phase N(1) had a relative frequency of approximately 81.0\% and phase S(2) had a relative frequency of around 18.0\%.
At $\lambda/\alpha \approx 12$, the last second-order phase transition took place, whereupon phases N(1), S(2) and Sz vanished and phases N(2), S(3), T(2) and T(3) emerged (Fig. \ref{subfig: m0.995_meanfields}). The microslip fraction at which instances of these new phases occurred stabilised at about 0.66 and 0.82 for $\lambda/\alpha > 15$, as seen in Fig. \ref{subfig: m0.995_fractions}. The value $n/N\approx0.82$ is higher than the theoretical maximum of 0.66 when only 8 of the 12 slips systems are activated, meaning that for some instances of the phases N(2), S(3), T(2) and T(3) the slip systems A6, B5, C5 and D6 were also activated. The relative frequencies of phases N(2), S(3), T(2) and T(3) in the range $14 < \lambda/\alpha < 16.125 $ were approximately 71.7\%, 5.9\%, 15.5\% and 6.9\%, respectively. The energy surplus of instances of these phases presented the steepest slopes, which can be directly seen in Fig. \ref{subfig: m0.995_energies}. The new phases N(2), S(3), T(2) and T(3) are described in Table \ref{table: phases}.

At the yield point, with an applied magnitude equal to $\sigma_y$, only two second-order phase transitions were identifiable: at $\lambda/\alpha \approx 6$ and at $\lambda/\alpha \approx 12$. For very low values of $\lambda/\alpha < 0.3$, phases A and T(1) were observed but quickly vanished discontinuously. Instances of phases N(1) and S(1) were the most frequently observed in the range $\lambda/\alpha \lesssim 6$ and almost always at a microslip fraction of about 0.33, although some instances of N(1) were registered at fractions of about 0.42 and 0.66 for values of $\lambda/\alpha < 0.3$ (Fig. \ref{subfig: m1.00_fractions}). At the threshold point $\lambda/\alpha \approx 6$ a second-order phase transition was observed where phase S(1) transitioned into phase S(2), like in the case discussed above. In the range $6 \lesssim \lambda/\alpha \lesssim 12$ only phases N(1), S(2) and Sz were observed at fractions $n/N$ of about 0.33 or 0.50, as labelled in Fig. \ref{subfig: m1.00_fractions}. In the region $10 < \lambda/\alpha < 12$, phase N(1) had a relative frequency of around 65.7\% and phase S(2) had a relative frequency of approximately 33.7\%. The second second-order phase transition occurred at $\lambda/\alpha \approx 12$, where phases N(1) and S(2) transitioned into phases N(2), S(3), T(2) and T(3). It was unclear from Fig. \ref{subfig: m1.00_meanfields} if phase Sz transitioned as well or if it vanished discontinuously at this point. The relative frequencies of phases N(2), S(3), T(2) and T(3) in the range $14 < \lambda/\alpha < 16.125$ were approximately 61.9\%, 9.4\%, 25.1\% and 3.5\%, respectively. Aside from the difference in relative frequencies, the behaviour of the system beyond the threshold $\lambda/\alpha \approx 12$ was the same as described at the end of the previous paragraph.

Lastly, in the case of an applied magnitude above the yield point, namely, $\sigma = 1.005 \: \sigma_y$, it is worthwhile to note how the behaviour of the system mirrored that of the case $\sigma = 0.995 \: \sigma_y$ in terms of the mean-field vector norm, by comparing Figs. \ref{subfig: m0.995_meanfields} and \ref{subfig: m1.005_meanfields}. The only three notable differences were the following: firstly, phase A vanished discontinuously in the range $\lambda/\alpha \lesssim 6$ in the case above yield rather than in the range $8 \lesssim \lambda/\alpha \lesssim 12$ as was the case below yield; secondly, and in a similar fashion, phase Nxy vanished sooner in the case above yield, disappearing discontinuously in the range $6 \lesssim \lambda/\alpha \lesssim 8$ rather than in the range $8 \lesssim \lambda/\alpha \lesssim 12$ as it had done in the case below yield; and thirdly, the transition at $\lambda/\alpha \approx 8$ was less pronounced in the case above yield than in the case below yield --- this was true whether one analysed the mean-field vector norm (Fig. \ref{subfig: m1.005_meanfields}), the microslip fraction (Fig. \ref{subfig: m1.005_fractions}) or the energy surplus (Fig. \ref{subfig: m1.005_energies}). By `less pronounced' we mean that the frequency of data points falling on the expected transition path was much reduced. The microslip fraction landscape was different between the two cases, as can be seen in Figs. \ref{subfig: m0.995_fractions} and \ref{subfig: m1.005_fractions}. In the case above yield, phases A, N(1), Nxy, S(1), Sz and T(1) were observed at $n/N \approx 0.33$ in the range $\lambda \lesssim 6$ and no instances were registered with a fraction lower than this. Additionally, phase N(1) was observed with microslip fractions of about 0.42, 0.50 and 0.66, phase A was observed at fractions of around 0.50 and 0.66, and phase Nxy was observed at a fraction of around 0.66. In the range $6 \lesssim \lambda/\alpha \lesssim 8$, only phases N(1), S(2) and T(1) were registered at a fraction of about 0.33 and instances of both N(1) and S(2) were also observed at microslip fractions of around 0.42 and 0.50. In the region $10 < \lambda/\alpha < 12$, phase N(1) had a relative frequency of around 83.5\% and phase S(2) had a relative frequency of about 16.5\%. In the range $8 \lesssim \lambda/\alpha \lesssim 12$, instances of the phases N(1) and S(2) were observed at fractions of around 0.33 and 0.50, with some instances of Sz also observed at $n/N \approx 0.33$. The relative frequencies of phases N(2), S(3), T(2) and T(3) in the range $14 < \lambda/\alpha < 16.125$ were approximately 72.4\%, 6.3\%, 14.9\% and 6.3\%, respectively. Aside from the difference in relative frequencies, the behaviour of the system beyond the threshold $\lambda/\alpha \approx 12$ was the same as in the case below yield.

\begin{figure}[t!]
     \centering
     \begin{subfigure}{0.329\textwidth}
         \centering
         \includegraphics[width=\textwidth]{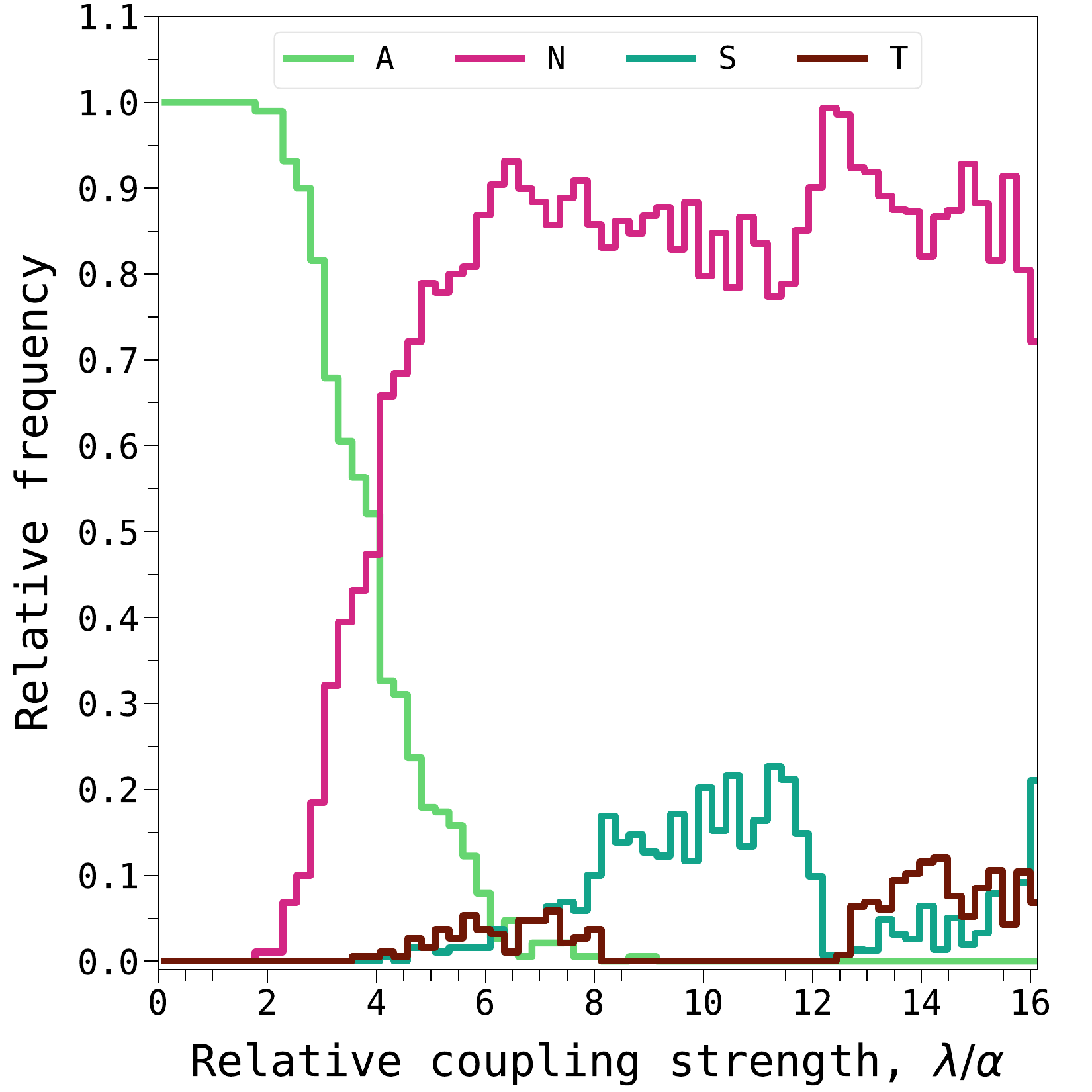}
         \caption{$\sigma=0.995 \: \sigma_y$}
         \label{subfig: m0.995_histogram}
     \end{subfigure}
     \hfill
     \begin{subfigure}{0.329\textwidth}
         \centering
         \includegraphics[width=\textwidth]{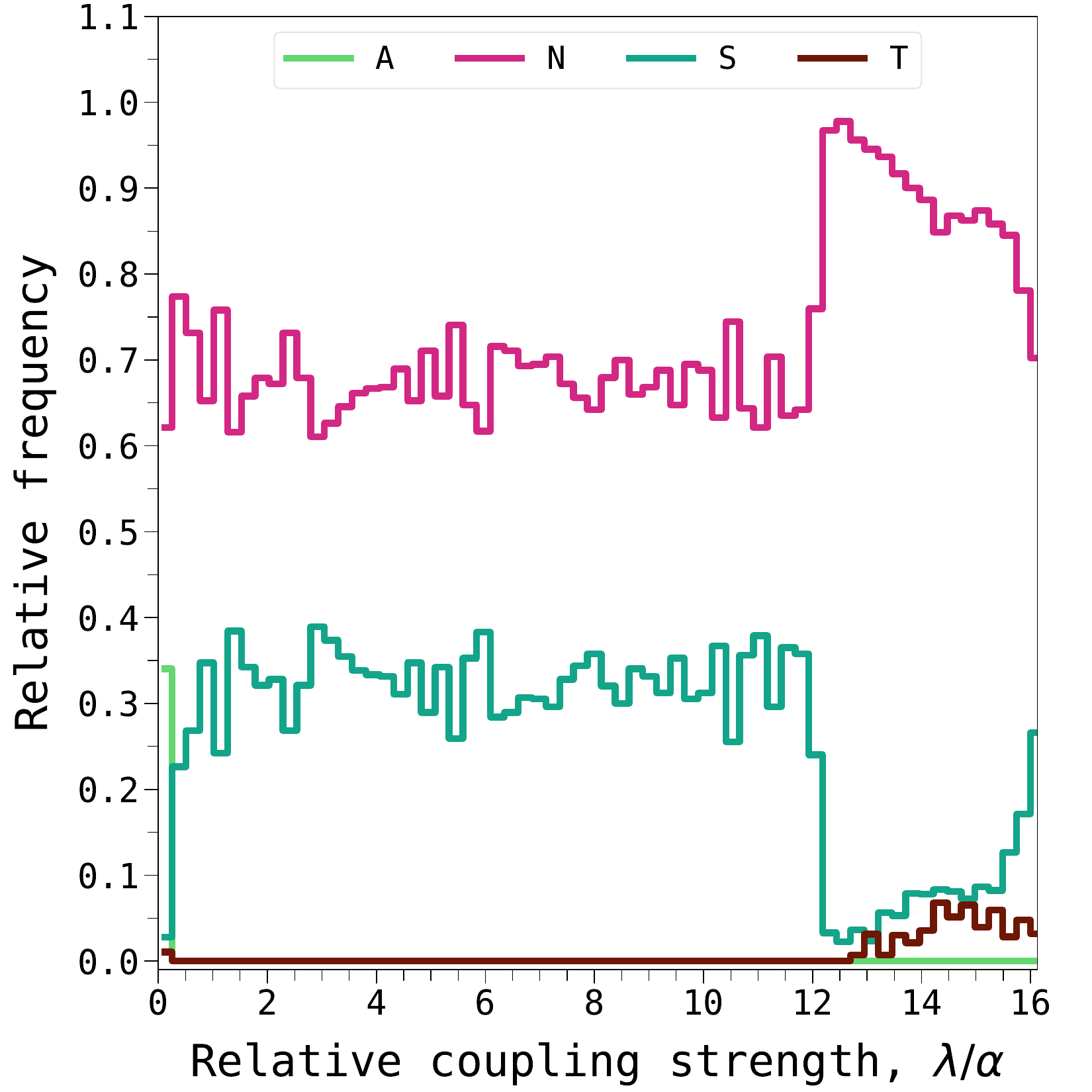}
         \caption{$\sigma=\sigma_y$}
         \label{subfig: m1.00_histogram}
     \end{subfigure}
     \hfill
     \begin{subfigure}{0.329\textwidth}
         \centering
         \includegraphics[width=\textwidth]{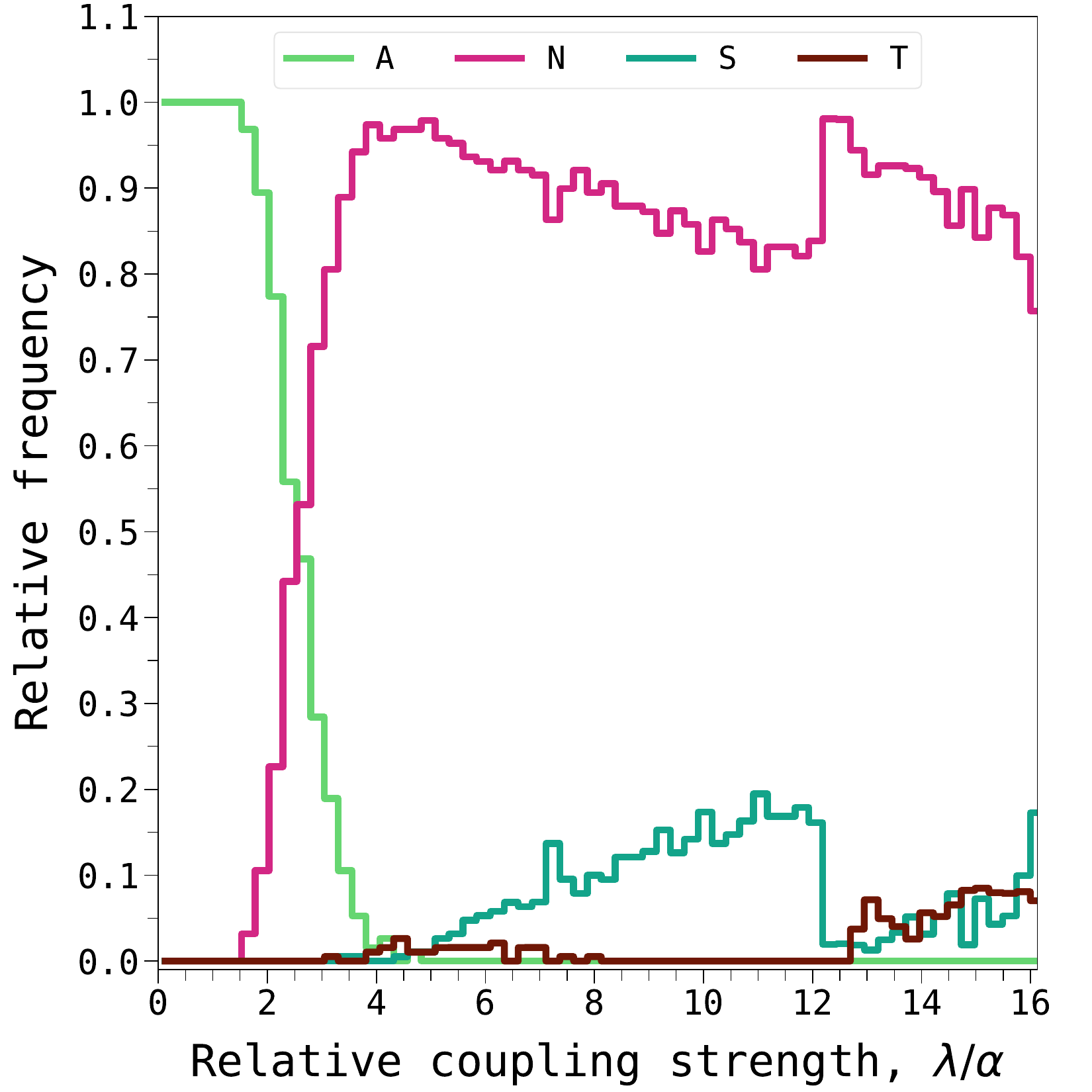}
         \caption{$\sigma=1.005 \: \sigma_y$}
         \label{subfig: m1.005_histogram}
     \end{subfigure}
        \caption{
Histograms of the relative frequency of the phases observed (a) in Fig. \ref{subfig: m0.995_meanfields}, (b) in Fig. \ref{subfig: m1.00_meanfields} and (c) in Fig. \ref{subfig: m1.005_meanfields}, bundled by type (so, N corresponds to N(1), N(2) and Nxy summed together). The relative frequency is normalised so that it totals 1 at each bin in the horizontal axis. The bins have a width of $0.250 \: \lambda/\alpha$.
}
        \label{fig: rel_freq_2}
\end{figure}

An analysis of the relative frequency of data points relative to each phase revealed the same behaviour that had been noted in Fig. \ref{fig: rel_freq}. This is shown in Fig. \ref{fig: rel_freq_2}. Specifically, phase T(1) was identifiable only in the range $3<\lambda/\alpha<8$ in Figs. \ref{subfig: m0.995_histogram} and \ref{subfig: m1.005_histogram} and in that range its relative frequency was seen to increase and then to decrease, showing how the transition is again characterised by a probability.

\section{Discussion} \label{seq: Discussion}

\subsection{Phases, macroscopic slip and localisation}

One might start by assuming that there is a distribution of $\vec{p}$-states of lowest energy --- possibly distributed in a spherical fashion around the origin $\vec{p}=(0,0,0)$ --- that the system is forbidden from accessing by the force of the self-energy term. The system is then coerced by the orientation and magnitude of the applied stress to adopt an optimal linear combination of these states, one which minimises the energy necessary to produce itself. These optimal states in $\vec{p}$-space are the ones highlighted in Table \ref{table: phases}. Due to the averaging mechanism of the mean-field, it is impossible to compare the final shape of the deformed simulated specimen with an actual specimen deformed under experimental conditions, since this shape depends not only on the activated slip systems but equally on where these systems are activated inside the material --- because the location will determine if the slip lines actually intersect. However, the phases identified in Table \ref{table: phases} carry physical meaning. As mentioned after Eq. \eqref{eq: macrodefect_cochain} was introduced, the mean-field vector $\vec{p}$ is perpendicular to the plane where the material suffers the most net reduction in area and its norm offers a measure of this reduction. The key word is `net' or, more to the point, this is a macroscopic feature.

Phase A at a fraction $n/N \approx 0$ is easily interpreted as that of complete order, but at a fraction of $n/N \approx 0.66$ could correspond to either an amorphisation deformation mode like the one reported by \cite{Liang2004}, or to a brittle fracture like the one reported by \cite{Lin2022}, both of which were observed in Molecular Dynamics simulations of copper nanowires at high strain rates and under adiabatic conditions. A phase like N(1), N(2) or Nxy, whose mean-field vector generally points in one of the Cartesian coordinates, cannot be attained without the activation of at least two non-collinear and non-coplanar slip systems, which could point to a number of physical phenomena such as activation of a secondary slip system \citep{Groma1993} or necking. For example, in the case of N(1) and N(2), possible combinations of active slip systems are: A3+B4, A2+C1, A2+D1 (second diagram in Fig. \ref{fig: necking_shearing}), B2+C1, B4+C3, B2+D1, C3+D4, etc.
Conversely, phases S(1), S(2), S(3) and Sz are characterised by a mean-field vector which is parallel to one of the allowed slip directions in face-centred cubic metals (as shown in Table \ref{table: Schmid&Boas}). Therefore, on the macroscopic scale, this results in the concerted activation of one slip direction above all others. This could point to another set of physical phenomena such as Lüders bands and localisation. For example, in the case of S(1), S(2) and S(3), possible combinations of active slip systems are: A6, B5, C5, D6, -A2+A3 (=A6), -B2+B4 (=B5), C1-C3 (=C5), D1-D4 (=D6) (third diagram in Fig. \ref{fig: necking_shearing}), -A2+B4, -B4+D1 (fourth diagram in Fig. \ref{fig: necking_shearing}), C1-A3, etc. The direct activation of slip directions 5 and 6 seems to have been less likely, since slip systems B5, C5, A6 and D6 were never observed to be activated in Fig. \ref{fig: individual_simulations} and the force applied on these systems by the external stress was null. Phase T(1) is characterised by a mean-field vector of the general form $(\pm 2, \: \pm 1, \: \pm 1)$ or $(\pm 1, \: \pm 2, \: \pm 1)$, for any combination of the plus and minus signs. Possible combinations of active slip systems are: A2-A6, B2-B5, C1+C5, D1+D6 (fifth diagram in Fig. \ref{fig: necking_shearing}), B2-D6, A2+A3-B4, B4+D1-A2, etc. The necessary activation of slip directions 5 and 6, which are orthogonal to the applied tension, or of three different slip systems may explain the low relative frequency of phase T(1). Finally, phases T(2) and T(3) do not seem to correspond to any such physical phenomenon and can be interpreted as superpositions of the other phases.

In multiscale simulations, \cite{Shehadeh2005} observed dislocations forming ``microbands'' on the \{111\} planes in shock-loaded copper single crystals when the strain rate was higher than $10^{6}$ s$^{-1}$. With an overwhelming activation of the slip systems A2, A3, B2, B4, C1, C3, D1 and D4, and using Kröner's interpretation of ``point stacking faults'', as highlighted in the Introduction, phases of the S type closely resemble the behaviour observed by Shehadeh et al.

In Figs. \ref{subfig: med_energies} and \ref{subfig: med_meanfields}, which refer to simulations at the yield point, some simulations initially developed on the path leading to an equilibrium at phase N(1) (the lower branch), but then changed to the path leading to phase S(1) (the upper branch). With the insight of the above paragraph, what this shows is an evolution from a phase characterised by the activation of non-collinear and non-coplanar slip systems to a phase characterised by the concerted activation of one slip direction resulting in strain localisation.

\begin{figure}[t!]
     \centering
     \begin{subfigure}{0.329\textwidth}
         \centering
         \includegraphics[width=\textwidth]{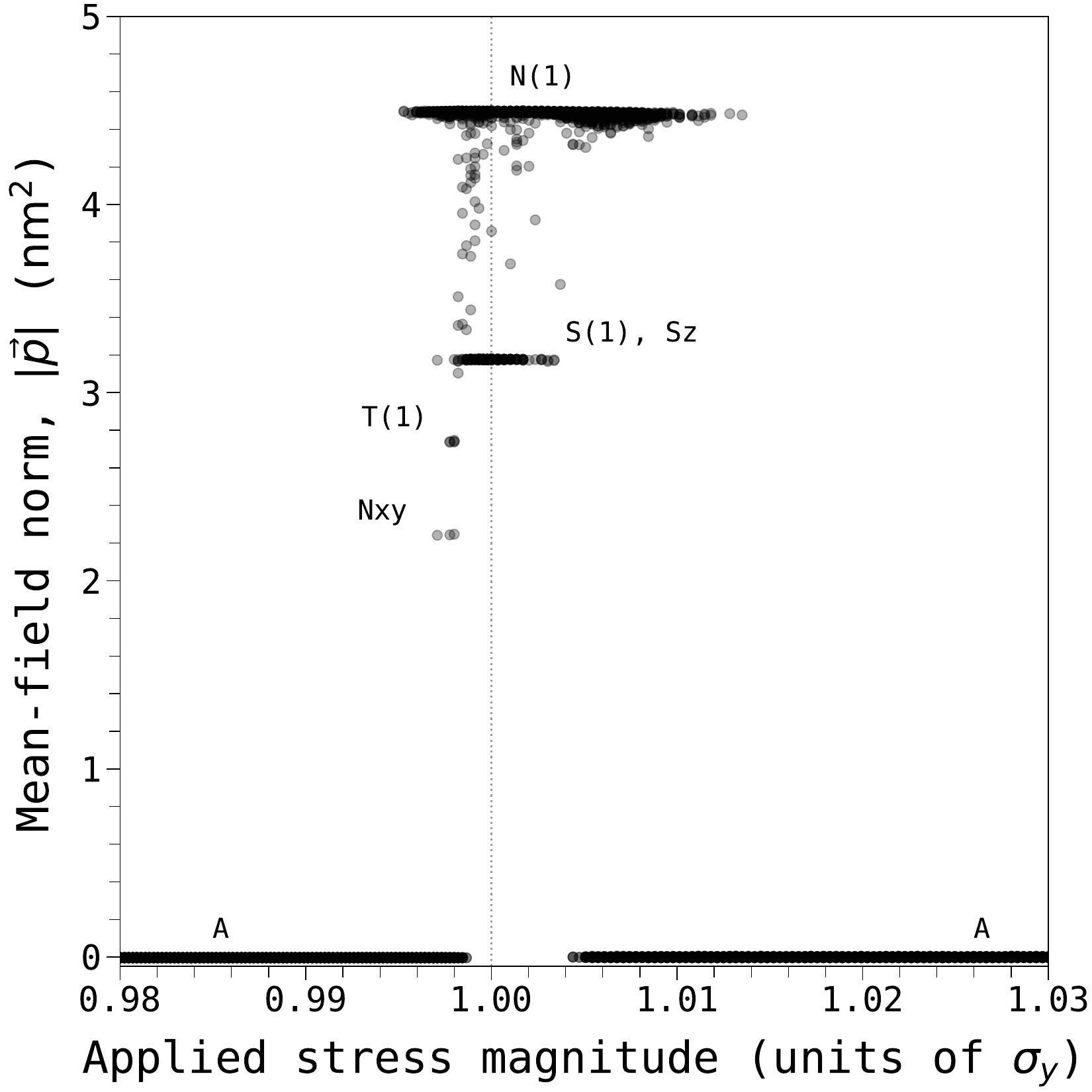}
         \caption{10\%}
         \label{subfig: f0.1}
     \end{subfigure}
     \hfill
     \begin{subfigure}{0.329\textwidth}
         \centering
         \includegraphics[width=\textwidth]{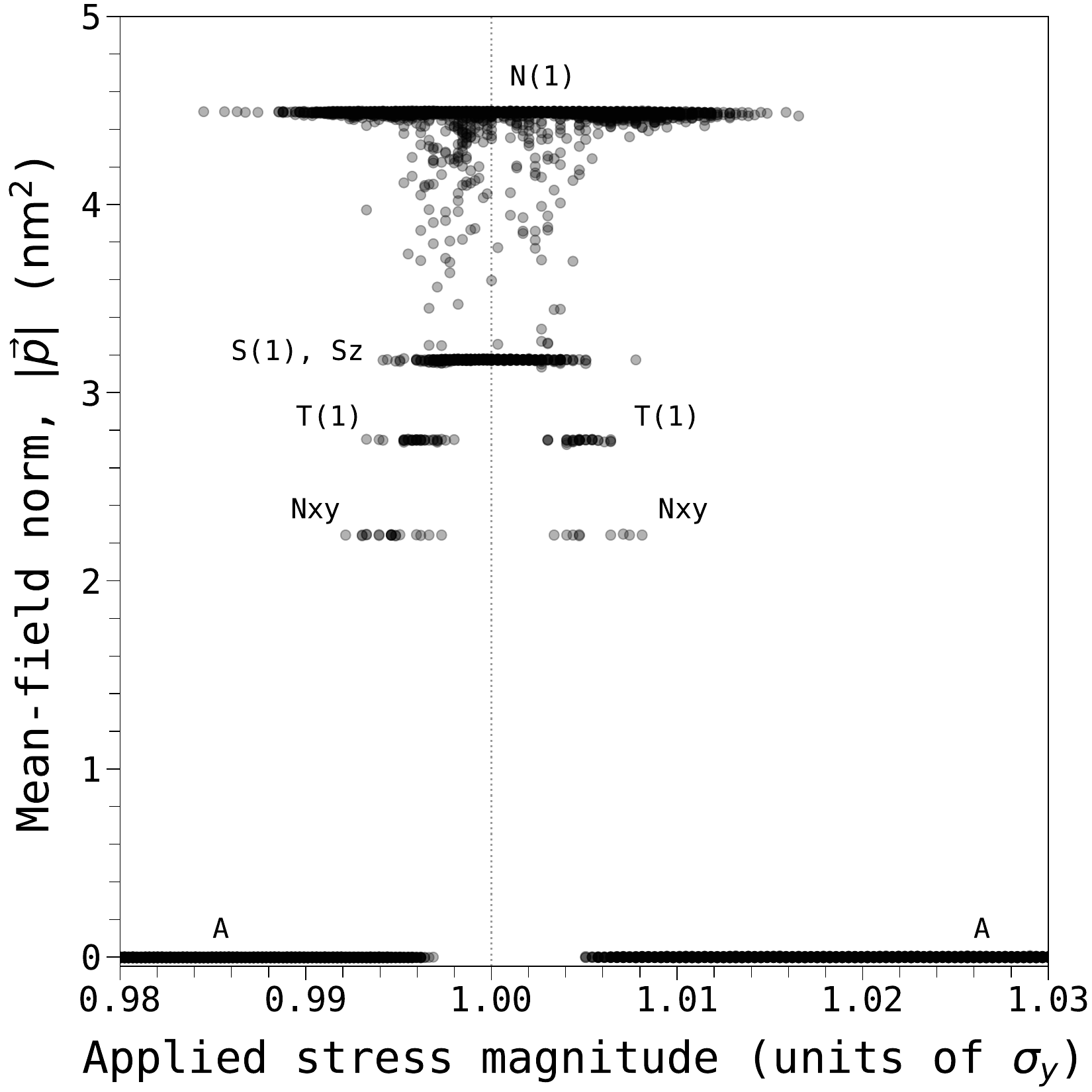}
         \caption{75\%}
         \label{subfig: f0.75}
     \end{subfigure}
     \hfill
     \begin{subfigure}{0.329\textwidth}
         \centering
         \includegraphics[width=\textwidth]{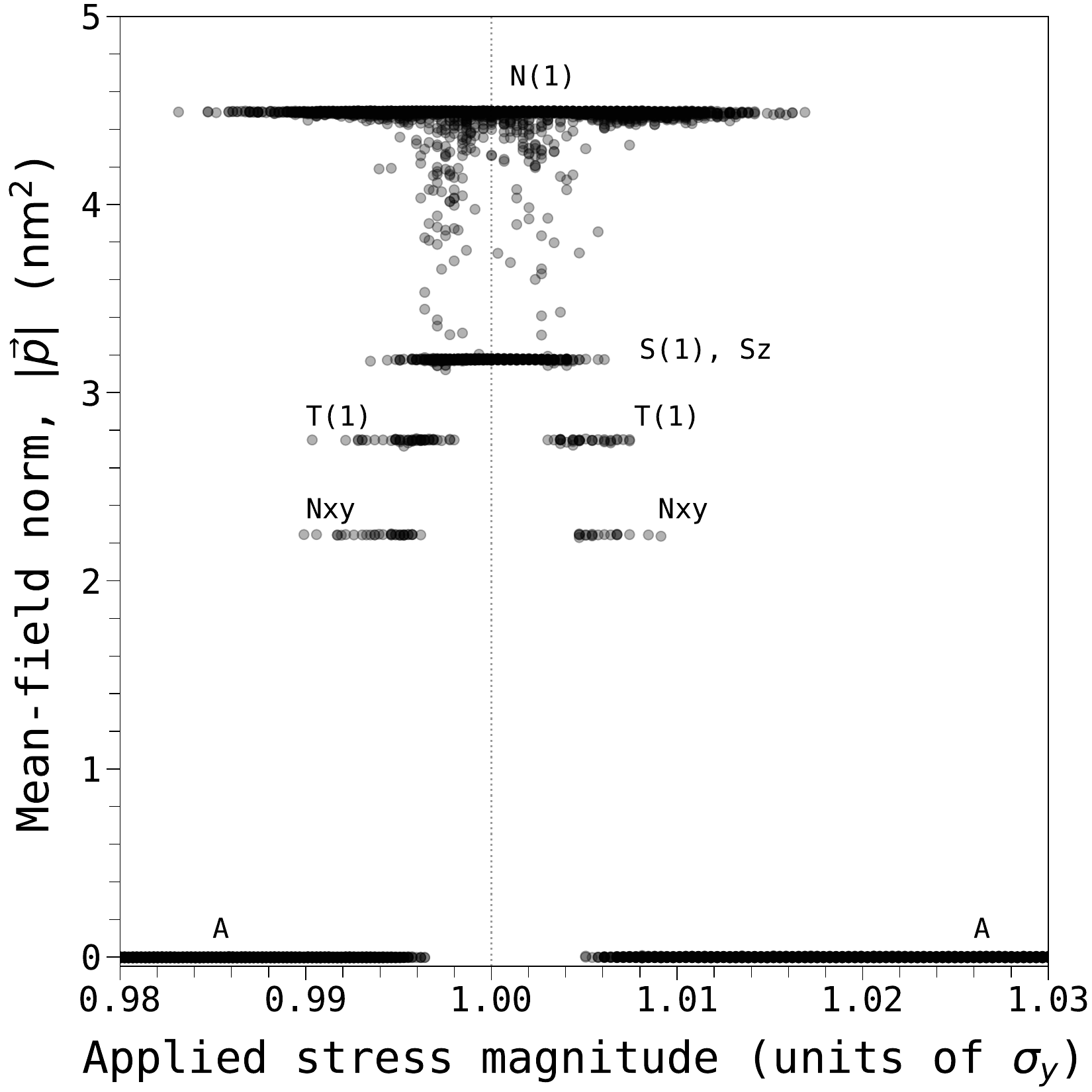}
         \caption{100\%}
         \label{subfig: f1}
     \end{subfigure}
        \caption{
The mean-field norm of the equilibrium states reached by the simulated system for different initial states. The initial state of the simulated material was determined by assigning one microslip event to a specific fraction of slip 2-cells, at random. That fraction varied according to the individual captions on each figure. Each data point seen here is the last data point taken from graphs such as Fig. \ref{fig: individual_simulations}. All simulations were run for 1 million steps. The self-energy parameter $\alpha$ was set to 465.15 GPa and the mean-field coupling parameter $\lambda$ was set to $4 \: \alpha$. The text labels refer to the phases identified in Table \ref{table: phases} and indicate which phases were observed in the nearby collection of data points.
}
        \label{fig: initial_state}
\end{figure}

\subsection{The effect of the applied stress}

As a function of the applied stress magnitude, first-order phase transitions appear between different configurations of the microslip structure, as seen in Fig. \ref{fig: stress_range}. The discontinuous jumps in the mean-field vector norm $|\vec{p}|$ are reminiscent of the first-order phase transition observed in the Ising model: at a fixed temperature below the Curie temperature, the magnetisation (average of the spins) of the system can be coerced to shift by a sufficiently strong applied magnetic field \citep[p. 10]{Kardar2007}. In the Ising model without an externally applied field, the magnetisation is larger (in absolute value) the lower the temperature of the system. In the model developed in this study, the self-energy term $\alpha \langle S, S \rangle$ precludes a non-zero mean-field in the absence of an applied stress. As pointed out in the Introduction, changes in the slip network due to the application of a stress can be associated with second-order phase transitions \citep{Schafler2005}.

Then, the first-order phase transitions obtained in this study might come from a failure to include the effect of the thermodynamics of slip in Eq. \eqref{eq: discrete_lagrangian}. The thermal energy dissipated into the system by the slipping mechanism is accounted for in the surplus energy term $-\mathcal{H}$, but it is unidentifiable, although one could approximate its value by considering the Taylor-Quinney coefficient \citep{Kositski2021}. Therefore, although the discontinuous jumps seen in Fig. \ref{fig: stress_range} tell us that there is a latent heat being used by the system to produce microstructural changes, we should take care not to assume that it is all of $-\mathcal{H}$ which is used in such a way \citep{Lubarda2016}. Accounting for the thermodynamics of slip could produce a butterfly effect in which the microstructural changes close to a phase transition are accelerated and the first-order phase transitions obtained with the present model are transformed into second-order phase transitions \citep{Nabarro2001}.

Furthermore, the self-energy term acts as a very steep barrier to the trajectory of the system along the minima of the free energy and the applied stress field carries the system across this discontinuous barrier. Another perspective is that in Figs. \ref{fig: stress_range} and \ref{fig: lambda_range} ``each data point is independent of the other data points'', as written in Sections \ref{subseq: Results_stressranges} and \ref{subseq: Results_lambdaranges}. A simulated specimen that reaches equilibrium in phase N(1) at a fraction of $n/N \approx 0.25$ under an applied stress magnitude $\sigma = 0.995 \: \sigma_y$ has no restriction on its trajectory in phase-space and so at yield $\sigma = \sigma_y$ can reach equilibrium in phase S(1) or T(1) --- because, in fact, it would not be the same simulated specimen at all. A real copper single crystal that equilibrates at phase N(1) at a fraction of about $0.25$ under an applied stress magnitude $\sigma = 0.995 \: \sigma_y$, when stretched further up to $\sigma = \sigma_y$ would have available to it only those trajectories in phase-space that start at phase N(1) at a fraction of about $0.25$ and end in a phase with lower energy $\mathcal{H}$. After all, hysteresis and energy dissipation follow from a reduction in accessible states (limited trajectories in phase space) \citep{Ortiz1999, Koslowski2002}.

Alternatively, the first-order phase transitions observed in Fig. \ref{fig: stress_range} might be a particularity of the assumptions made. With no pre-existing defects to smoothen the approach to the transition \citep{Dubrovskii1979}, the simulated specimen has no option but to react in an explosive fashion, with a fast multiplication of defects, giving rise to the first-order phase transitions observed. Fig. \ref{fig: initial_state} shows the effect of the initial state of the simulated specimen on the overall behaviour of the system. It can be discerned that, the more microslips are present in the system before the deformation starts, the more states are accessible. This is evident in the broadening of the bands from Fig. \ref{subfig: f0.1} to \ref{subfig: f1}. (Note that fig. \ref{subfig: l4.0_meanfields} represents the intermediary step of an initial state of one microslip assigned to 50\% of slip 2-cells at random.) This progression may be interpreted as something that is well-known from experimental research: it is easier to move dislocations than to create them.

Finally, Figs. \ref{fig: rel_freq} and \ref{fig: rel_freq_2} show that these stress-dependent first-order phase transitions are not well defined. Phases that showed first-order transitions did not get sharply suppressed at a certain value of the applied stress magnitude; rather, it was found that the probability of the system finishing the simulation in a certain phase was dependent on the applied stress magnitude, and this probability followed a normal or bimodal distribution. In this way, the first-order phase transitions identified occur not at single values of the applied stress, but have a probability of occurring in a range of values of the applied stress. Curiously, \cite{Dubrovskii1979} found that the presence of a network of defects can smoothen a second-order phase transition, broadening the range of temperature values over which it occurs. Perhaps there is a similar mechanism intertwining first-order phase transitions and the applied stress, although this is beyond the scope of our research. These transitions are still classified as first-order, however, because, from the perspective of the material, it will always be a discontinuous transition. For example, consider a system in phase T(1) with $\lambda=8 \: \alpha$ at $\sigma=0.993 \: \sigma_y$. If one increases the applied stress magnitude, the system might cling to that phase up until $\sigma\approx 0.996 \: \sigma_y$, as per Fig. \ref{subfig: l8.0_meanfields}, but the same plot shows that phase T(1) is not viable at $\sigma=\sigma_y$, so the system must transition to phase N(1) or S(2) at some point. Thus, it is not a matter of how, but of when.

\subsection{The effect of the mean-field coupling}

The continuum model predicts three typical responses of the system in the ranges $n\lambda/\alpha > 1$, $1/1.3 < n\lambda/\alpha < 1$ and $n\lambda/\alpha < 1/1.3$, corresponding to quasi-brittle, ductile and fine-grain behaviours; the boundaries $n\lambda/\alpha=1$ and $n\lambda/\alpha=1/1.3$ correspond to bifurcation points \citep{Bayandin2019}. These predictions are based on a dimensional analysis that is not straightforwardly reproduced with the discrete model developed here. Nevertheless, it is clear from Fig. \ref{fig: lambda_range} that second-order phase transitions arise as a functions of the ratio $\lambda/\alpha$.

As a function of the mean-field coupling strength, both first-order and second-order phase transitions were observed. In the case of a very weak mean-field, each microslip event is an independent event. This can be interpreted as a material where the dislocations are highly screened and their stress and strain fields do not extend far beyond their location. A stronger mean-field can be achieved in two separate ways: the mean-field coupling strength $\lambda$ can be increased directly, or the number of microslips in alignment with the mean-field can be increased. Mathematically, looking at Eq. \eqref{eq: macrodefect_cochain}, these two changes are equivalent, for one can either double $\lambda$ or double $n$ to achieve the same result. A material with a higher mean-field coupling strength $\lambda$ can be interpreted as a material with a stronger coupling between the stress fields of the dislocations, or as a material where the slipped regions, on average, are not as screened and their stress fields extend farther beyond their position. A material with a stronger mean-field will become more easily `locked' into a certain deformation state and so we see the discontinuous disappearance of the microstructureless phase A for large values of $\lambda/\alpha$ in Fig. \ref{fig: lambda_range}. Furthermore, the second-order phase transitions that occur at $\lambda/\alpha \approx 6$, $\lambda/\alpha \approx 8$ and $\lambda/\alpha \approx 12$ reinforce the localisation of strain. This can be interpreted as a refining of dislocation structures organised along a particular plane, in the case of the transition from phase S(1) to S(2), from S(2) to S(3), or from N(1) to N(2), or as a reorganisation of the network which differentially activates non-coplanar slip systems, in the case of the suppression of phase T(1). The former is analogous to the experimentally observed second-order phase transition from stage III of deformation, which is characterised by square dislocation cells, to stage IV, characterised by rectangular dislocation cells \citep{Schafler2005}. The latter may be analogous to the flipping of domains in the Ising model or to the nucleation and growth of ice domains in the freezing of water, thus creating the context for a first-order phase transition. Aside from A and T(1), the other phase that disappears discontinuously as $\lambda$ increases is Nxy. Its suppression shows a preferred deformation state that aligns more closely with the applied stress.

\subsection{The yield point}

It has been observed experimentally that, in copper microwhiskers or micrometre-sized single copper crystals under tension, plastic deformation starts before the yield point is achieved (e.g. see \cite{Gotoh1974}). This phenomenon occurs by a single slip mechanism and it is only when a secondary slip system is activated that the specimen shows a sharp drop in the stress necessary to deform it further; thus, in these cases it is common to speak of a higher yield point and a lower yield point. The results shown in Fig. \ref{fig: stress_range} reproduce this effect and even suggest that the yielding of microscopic single crystals can occur in multiple stages, not just two. However, since every equilibrium state discussed in Sections \ref{subseq: Results_individualsimulations} and \ref{subseq: Results_stressranges} exhibited more than one activated slip systems, our model has failed to capture the specific effect of the secondary slip system's activation. This failure can be explained by the lack of a hardening mechanism in Eq. \eqref{eq: discrete_lagrangian}.

\subsection{The mean-field coupling parameter $\lambda$, revisited}

Having studied the possible configurations of the deformation state of the simulated crystal, Eq. \eqref{eq: lambda} could now be used. However, this would require the modeller to choose which state (phase) they considered optimal and hence deserving of being preserved after the applied stress was removed. The two notable contenders are phases of the N and S type. With a phase of the N type and magnitude $|\vec{p}|=p$ in nm$^2$, Eq. \eqref{eq: lambda} gives a maximal $\lambda/\alpha \approx 6.94/p$. If, instead, a phase of the S type is considered, then it gives a maximal $\lambda/\alpha \approx 13.88/p$. With the results obtained in our research, this places $\lambda$ in the interval $\lambda/\alpha < 4.4$, which is strictly below the first second-order phase transition at $\lambda/\alpha \approx 6$. If the behavioural regimes identified by the continuuum theory \citep{Bayandin2019} could be applied to the results obtained in this work, then, considering that the fraction $n/N<1$, we would find that defect-free microscopic copper single crystals behave according to ductile behaviour or to fine-grain behaviour.

Alternatively, it is likely that the mean-field coupling strength ought to be a function of an internal variable such as the number of defects $n$ or the energy surplus $-\mathcal{H}$, instead of taking a fixed value throughout an entire simulation. Namely, it is tentative to infer that $\lambda$ should decrease with a higher number of defects in the system, since each individual defect's stress field would be screened more and affect a smaller region of the material.

\section{Conclusion} \label{seq: Conclusion}

Metallic specimens of the smallest dimensions (nano to micro-scale) exhibit plastic properties that differ from what is generally observed in bulk-sized specimens \citep{Brenner1956, Brenner1957, Brenner1958, Kamada1974, Tschopp2007}. In the limit of a featureless microstructure, an applied stress with increasing magnitude imparts on the system an increasing elastic strain energy, which accumulates until the first plastic slip event occurs, upon which the energy is released \citep{Cao2008}. This results in a propagating wave of plastic slip and energy release \citep{Shehadeh2005}.

Furthermore, experimental observations have shown that changes in the dislocation network of a plastically deforming metal are in some contexts associated with second-order phase transitions \citep{Schafler2005, Friedman2012}. This has been supported by theoretical work on the basis of percolation theory \citep{Dubrovskii1979} and statistical-mechanical models \citep{Ortiz1999}.

Nevertheless, popular continuum models fail to describe the strange behaviour of these very small objects. This is due to a fundamental failure of the continuum assumption, as already \cite{Kroner1995, Kroner2001} had recognised. The very nature of a continuum is a homogeneous featureless structure which is antithetical to a plastically induced microstructure. Some models and techniques, such as discrete dislocation dynamics (DDD), try to overcome this discrepancy by discretising some of the objects involved in the theory. However, the kind of discretisation they attempt is insufficient, for the domain where dislocations, grain boundaries, strain fields and stress fields exist is still itself left as a continuum (e.g. see \cite{Gurrutxaga-Lerma2013}).

Therefore, the model proposed in this study attempts to take this discretisation procedure to its conclusion. Firstly, a continuum mean-field model designed for shock-like plastic deformation \citep{Naimark1992, Naimark1998, Naimark2016, Naimark2017, Bayandin2019} was rewritten and adapted to the language of chains and cochains on a cell complex (Section \ref{seq: Theory}). Then, a three-dimensional tessellation of a cubical space was developed (Fig. \ref{fig: FCC_complex}) which reproduced the network of primary slip systems (Table \ref{table: Schmid&Boas}). The discrete model was applied to this construction by assigning microscopic slip (microslip) events to pairs of one 2-cell (face) and one 1-cell (edge) on its boundary in the cell complex (Section \ref{seq: Methodology}). The application of an external uniaxial stress was simulated via a Metropolis-Hastings algorithm that tried to maximise the energy surplus of the total deformation process, which was the energy input into the system but not used in the production of the defect itself. The simulated specimen was a perfect copper single crystal with volume 1 $\upmu$m$^3$ and the applied stress was assumed constant and homogeneous.

The main results can be summarised as follows:
\begin{itemize}
    \item At the end of each simulation, the microslip network reliably arranged itself into one of eleven phases that were characterised by a definite mean-field vector (these are summarised in Table \ref{table: phases}) and different combinations of collinear, coplanar, non-collinear and non-coplanar slip systems. The mean-field vector was perpendicular to the plane of greatest net area reduction and its norm gave a measure of this area. Some of these phases can be interpreted physically to reproduce activation of secondary slip systems (N type), strain localisation (S type), and amorphisation or fracture (phase A), although the model did not allow for the last two explicitly. Some simulations showed an evolution from a phase of N type to a phase of S type, revealing an allowed phase-space trajectory from an instance of multiple active slip systems to a spontaneous strain localisation (Fig. \ref{subfig: med_meanfields}). Throughout all simulations, the most common phases aside from A were those of N type, followed by those of S type.
    \item As a function of the applied stress magnitude and in terms of the observed phases, the behaviour of the system was almost symmetrical on either side of the yield point, with the main difference being the number of activated microslips, which increased with increasing magnitude (Fig. \ref{fig: stress_range}). The final states of the simulations aggregated into discrete bands (ranges of stress values) with no smooth transitions between them, which meant that the microslip network went through first-order phase transitions imposed by the external stress. However, these transitions did not happen at a specific value of the applied stress magnitude, but instead had a probability of occurring within a range of values (Fig. \ref{fig: rel_freq}).
    \item As a function of the mean-field coupling strength, the behaviour of the system went through two bifurcations at yield and three bifurcations above and below the yield point (Fig. \ref{fig: lambda_range}). These bifurcations corresponded to second-order phase transitions \citep[pp. 37-52]{Sole2011}. In order of increasing mean-field coupling strength, which translates to decreasingly screened dislocations: the first second-order phase transition was characterised by a refinement of the phase of S type, suggesting an increased strain localisation, which matched experimental observations \citep{Schafler2005}; the second second-order phase transition was accompanied by a first-order phase transition that suggested a change from active coplanar slip systems to non-coplanar, possibly collinear, systems; and the third second-order phase transition produced another refinement of the slip microstructure (N or S type), along with the emergence of more exotic phases.
\end{itemize}

Furthermore, the simulations conducted during this study support the idea that yield is not a single event but a process of events localised in time (Fig. \ref{fig: stress_range}). As the applied stress magnitude approaches the yield `point', there is a successive number of changes in the microstructure that start as early as at 98\% of `the' yield stress, reach a peak of activity at `the' yield stress and continue until as late as 103\% of `the' yield stress.

Lastly, the model that has been developed here suffers from many drawbacks. Firstly, the model is constrained by the tessellation of the simulation space, which, while it does not determine the plasticity process, it also does not allow for the explicit activation of less common slip systems, including partial dislocations. Secondly, strain hardening and thermodynamics were not included in the model. Had they been accounted for, the picture obtained in this study of very definite phases and transitions between them might have been severely altered. Including strain hardening would impede the development of secondary slip systems, while including thermodynamics would accelerate changes by virtue of increased kinetic energy. Failure to include these two fundamental processes essentially restricts our model to the case of adiabatic deformation under very high strain rates. These limitations are currently being addressed by considering local interactions instead of a mean-field approach.


\section*{Acknowledgments}
Authors acknowledge the financial support from EPSRC, UK via grants EP/V022687/1 (PRISB) and EP/N026136/1 (GEMS).

The authors confirm that the data supporting the findings of this study is available within the article.


\section*{Declaration of generative AI and AI-assisted technologies in the writing process}

No AI tool was used in the writing of this article.

\bibliographystyle{apalike}
\bibliography{references}


\end{document}